\documentclass{article}
\usepackage{epubtk}   
\usepackage{booktabs}  
\usepackage{graphicx}  
\usepackage{graphicx}
\usepackage{lscape}
\usepackage{indentfirst}
\usepackage{latexsym}
\usepackage{multirow}
\usepackage{tabls}
\usepackage{epsfig}
\usepackage{multirow}
\usepackage{caption}
\usepackage{subcaption}
\usepackage{hyperref}

\usepackage{color}
\usepackage{amssymb}
\usepackage{amsfonts}
\usepackage{amsmath}
\usepackage{bm}
\usepackage{mathrsfs}

\usepackage{algorithm}
\usepackage{algpseudocode} 
\usepackage{setspace} 
\usepackage[]{units}
\usepackage{siunitx}
\usepackage{color}
\usepackage{titlesec}
\titleclass{\part}{top}
\titleformat{\part}[display]
  {\normalfont\huge\bfseries}{\centering\partname\ \thepart}{20pt}{\Huge\centering}
\titlespacing*{\part}{0pt}{50pt}{40pt}

\newcommand{\msun}{M_{\odot}} 

\newcommand{\bea}{\begin{eqnarray}}
\newcommand{\eea}{\end{eqnarray}}

\newcounter{comment}
\newenvironment{comment}[1][]{\refstepcounter{comment}\par\medskip\noindent%
   \textbf{Comment~\thecomment. #1} \rmfamily}{\medskip}

\newcounter{discussion}
\newenvironment{discussion}[1][]{\refstepcounter{discussion}\par\medskip\noindent%
   \textbf{Discussion~\thediscussion. #1} \rmfamily}{\medskip}

\newcounter{example}
\newenvironment{example}[1][]{\refstepcounter{example}\par\medskip\noindent%
   \textbf{Example~\theexample. #1} \rmfamily}{\medskip}

\newcounter{result}
\newenvironment{result}[1][]{\refstepcounter{result}\par\medskip\noindent%
   \textbf{Result~\theresult. #1} \rmfamily}{\medskip}

\def\cE{ {\cal E} }

\def\cF{ {\cal F} }
\def\cH{ {\cal H} }
\def\cK{ {\cal K} }
\def\cP{ {\cal P} }

\def\cT{ {\cal T} }

\def\cO{ {\cal O} }

\def\bA{ {\bf A} }

\def\bH{ {\bf H} }
\def\bI{ {\bf I} }

\def\integer{\mathbb{Z}}

\def\complex{\mathbb{C}}

\def\DoL{ {{\Delta}_{1\text{L}}} }
\def\DtL{ {{\Delta}_{2\text{L}}} }
\def\Dot{ {{\Delta}_{12}} }

\def\MMm{ {\mathrm{MM}}} 

\newcommand{\cI}{{\cal I}}
\newcommand{\be}{\begin{equation}}
\newcommand{\ee}{\end{equation}}
\newcommand{\blank}{\bigskip}
\newcommand{\lam}{\lambda}
\newcommand{\Lam}{\Lambda}
\newcommand{\ft}{\footnote}

\def\lsim{\mathrel{\rlap{\lower3pt\hbox{\hskip1pt$\sim$}}
    \raise1pt\hbox{$<$}}}              
\def\gsim{\mathrel{\rlap{\lower3pt\hbox{\hskip1pt$\sim$}}
    \raise1pt\hbox{$>$}}}         
\def\coordeq{ \, \mathrel{ \rlap{\hbox{\hskip-2.5pt$=$} }
    \raise4pt\hbox{$\cdot$}} \, } 
\def\ceq{ \mathrel{\mathop:}=}

\begin{document}

\title{Reduced Order and Surrogate Models for Gravitational Waves}

\author{Manuel Tiglio\ft{email: mtiglio@unc.edu.ar} \and Aar\'on Villanueva\ft{email: aaron.villanueva@unc.edu.ar}}
\date{%
    {\it Facultad de Matem\'atica, Astronom\'ia, F\'isica y Computaci\'on, \\
              Universidad Nacional de C\'ordoba, C\'ordoba (5000), Argentina}\\[2ex]
\today
}
\maketitle
\begin{abstract}

We present an introduction to some of the state of the art in reduced order and surrogate modeling in gravitational wave (GW) science. Approaches that we cover include Principal Component Analysis, Proper Orthogonal Decomposition, the Reduced Basis approach, the Empirical Interpolation Method, Reduced Order Quadratures, and Compressed Likelihood evaluations. We divide the review into three parts: representation/compression of known data, predictive models, and data analysis. The targeted audience is that one of practitioners in GW science, a field in which building predictive models and data analysis tools that are both accurate and fast to evaluate, especially when dealing with large amounts of data and intensive computations, are necessary yet can be challenging.  As such, practical presentations and, sometimes, heuristic approaches are here preferred over rigor when the latter is not available. This review aims to be self-contained, within reasonable page limits, with little previous knowledge (at the undergraduate level) requirements in mathematics, scientific computing, and other disciplines. Emphasis is placed on optimality, as well as the curse of dimensionality and approaches that might have the promise of beating it. We also review most of the state of the art of GW surrogates.  Some numerical algorithms, conditioning details, scalability, parallelization and other practical points are discussed. The approaches presented are to large extent non-intrusive and data-driven and can therefore be applicable to other disciplines. We close with open challenges in high dimension surrogates, which are not unique to GW science. 

\end{abstract}

\newpage
\tableofcontents

\newpage
\subsection*{Notation}
\label{sec:notation}
\begin{itemize}
\item $h_{\lam}$: a function (for the purposes of this article, a GW waveform) associated with parameter value $\lam$. It can be, for example, a time series $h_{\lam} = h_{\lam}(t)  $, or a frequency domain function $h_{\lam} = h_{\lam} (f)$.  
\item $\lambda$: a (usually multi-dimensional) parameter for $h_{\lam}$.  
\item $\tt dim$: number of parameters of $\lam$. 
\item $L$: number of time/frequency samples.
\item $\cF$: an abstract space of functions of interest. Typically, $\cF:= \{ h_{\lam}\,|\,\lam\in\Phi \}$, with $\lam$ in some compact region (continuous or discrete) $\Phi$. We also refer to $\cF$ as the {\it fiducial} or {\it underlying} model. 
\item $N$: size of the training set. Also, the number of points in standard quadratures, polynomial interpolation, etc.
\item $\cT$: training set of parameter points $\cT := \{ \lam_i \}_{i=1}^N$. It is assumed to be compact.
\item $\cK$: training set of functions in $\cF$, $\cK := \{ f_i \}_{i=1}^N =  \{f(\lam_i , \cdot) \}_{i=1}^N\subset\cF$.
\item RB: Reduced Basis as a framework.
\item $\Lam_i$: selected (typically, by a greedy algorithm) parameter values.
\item {\tt rb}: a specific reduced basis.
\item $t,f$: time, frequency domains. 
\item EIM: The Empirical Interpolation Method. 
\item $T, F$: selected (typically, by the EIM) time/frequency interpolation nodes.
\item $n$: number of basis elements and of EIM nodes (they are equal by construction). The goal of ROM, if the problem is amenable to dimensional reduction, is to find an accurate basis such that $n \ll N$. 
\item $\Lambda_n$: the Lebesgue constant for an approximation with $n$ basis elements/interpolation points. This involves some ambiguity: we use the same symbol for greedy selected points, but it should be clear from context which one we refer to.
\item $h_{\tt s}$: a surrogate function for $h$.
\item Boldfaces are used for matrices, for example $\bf A$ has elements $A_{ij}$.  
\item $\langle a,b \rangle$: scalar product between two vectors or functions $a,b$.
\end{itemize}

\newpage
\section{Introduction} \label{sec:intro}
Gravitational wave (GW) science has reached a level of maturity 
in which many tools from areas such as modern approximation theory, data science, machine learning, and artificial intelligence are being incorporated into the field. These attempt to address challenges such as dealing with complex modeling, analysis, and handling of big data. A common feature of these challenges is the computational cost involved, which in many cases can be prohibitive, to the point that it cannot be solely overcome with larger or faster (super)computers, specialized hardware such as GPUs, or software optimization. This is particularly the case for parametrized problems, where each query depends on multiple input parameters that might only be known at run time. This is exacerbated as the number of parameters grow, usually resulting in the curse of dimensionality. This refers to the complexity of the problem (here leaving the term  {\it complexity} ambiguous on purpose) growing fast, sometimes exponentially, with the number of parameters. In the case of gravitational waves from binary systems, parameters can be intrinsic or extrinsic. The former relate to parameters such as the mass and spin of the binary components, the initial separation and eccentricity of the system, and equations of state if matter is present. Extrinsic parameters include distance of the source to Earth, sky position, orientation, time and phase of arrival. 

One of the first challenges of a parametrized problem is sampling it. With standard methods (for example, equally spaced, using the {\it metric} approach in GWs, or stochastic sampling) the accuracy of such catalogs for GW detection increases in many cases at best linearly with the number of samples, and their sizes typically increase exponentially with the number of parameter dimensions {\tt dim} as ${\cal O}(N^{\tt dim})$.  In fact, for the metric approach, the number of templates grows as $(1-\text{MM})^{-\text{dim}/2}$, with $\text{MM}$ the minimal match. Producing such catalogs, as well as their storage and analysis, can become challenging and, again, even not tractable through raw computational power and storage. 

One approach to this problem is decreased fidelity: the modeling of the problem is simplified by approximations to Einstein's General Relativity equations which are cheaper to solve for and analyze. Decreased fidelity is a delicate approach, though, since the accuracy of the approximations might not be known without access to the high fidelity models for an {\em arbitrary} query, which if not available could become an issue when attempting to assign error bars and a given precision needs to be guaranteed for statistical purposes. Decreased fidelity models can also lead to missed signals or biases in parameter estimation of detected GWs. 

Thus, another challenge is to simulate the GWs emitted by, for example, the coalescence of compact binary objects in real time, without any physical approximation. Here the gold standard is high accuracy numerical relativity (NR) solutions to the full Einstein field equations. It might appear unattainable to replace online (that is, not evaluated in advance), for an arbitrary query, a supercomputer simulation which might take per query hundreds to tens of thousands of hours of computing time with a substitute or surrogate model of equal accuracy but which can be evaluated in real time on a commodity laptop. As we show throughout this review, this can be -- and is being -- done, though there are still outstanding challenges left and the problem is not completely solved.

Finally, another challenge of equal importance is to perform parameter estimation on the sources of any detected GW signal in quasi real time. Meaning, fast enough so as to process the large number of detections by modern GW laser interferometers. And, most important, in the case of sources with electromagnetic counterparts, fast enough to allow for rapid telescope followups. This requires both online fast evaluation of the GW waveforms, as well as rapid likelihood evaluations.   

In this review we discuss some state of the art approaches which attempt to (and in some, {\em but not all}, cases manage to) obtain accurate and fast-to-evaluate and analyze surrogate models of gravitational waves emitted by binary systems, accomplishing some of the aforementioned challenges.  The common aspect underlying all these methods is reduced order modeling (ROM). The approaches here reviewed are Principal Component Analysis (PCA), Singular Value Decompositions (SVD), Reduced Basis (RB), the Empirical Interpolation Method (EIM), and derivatives, such as Reduced Order Quadratures (ROQ). The goal is intendedly not to be a definite survey, since it is a very active field. Instead, we attempt to provide an introduction to some of the approaches that are being used in  practice and do deliver on some or several of the above challenges, along with some basic theory underlying each of them.  The review is divided into three parts, each of which builds upon the previous one: 
\begin{enumerate}
\item Representation

One example is the generation of compact catalogs or banks of gravitational waves for searches. Another one is to analyze a system and look for redundancies.  
\item Predictive Models

The most ambitious goal here is to build surrogate models that can be evaluated in real time and are indistinguishable from numerical relativity supercomputer simulations of the Einstein equations, without any physical approximation. The target is the evaluation of a waveform in the order of  milliseconds per angular mode; that is, a speed up of at least $\sim 10^8$ with respect to NR and without loss of accuracy. 
\item Data Analysis

One of the main goals of ROM and other efforts in GW science is to achieve very fast parameter estimations, in particular so that real time alerts can be sent for searches of electromagnetic counterparts. From an astrophysical point of view, the target is from months using standard methods to around $10$ minutes using Focused Reduced Order Quadratures (discussed in Section~\ref{sec:PE}), including millions to tens of millions waveform evaluations and likelihood evaluations.    
\end{enumerate}
We encourage the reader to provide us with feedback, including topics to cover in future versions of this Living Review article. For briefness we have to skip many references, we apologize in advance for any and all omissions in such an active field. 

\section{Reduced Order Modeling and Gravitational Waves}
\label{sec:rom-gw}

This article reviews approaches which attempt to solve many of the aforementioned problems in GW science through Reduced Order Modeling (ROM), also known as Dimensional or Complexity Reduction, and the related field of Surrogate Modeling. ROM as a field has been around for a long time, but over the last decade and a half there have been major theoretical advances followed by a rapid raise in the number of applications and pace at which they both take place, in many areas of science, engineering and technology. 

This is in part, again, due to powerful new approaches and results, from approximation theory to numerical analysis and scientific computing, but also due to the recognition of the power of dimensional reduction in many important problems, some of them long-standing. That is, problems which are only now being seen in the light of the latest developments of ROM. On top of that, fields such as data science (DS), Machine Learning (ML) and Artificial Intelligence (AI) are considerably benefiting from ROM to either eliminate redundancies in big data, making them more amenable to analysis, and/or identifying relevant features for further studies using techniques from these disciplines. In fact, depending on the definitions of these fields, some include ROM as a subdiscipline. 

There is a big difference, though. There are many books on established DS, ML, and AI practices and theory, while the literature on modern approaches to ROM for parametrized problems, with some important exceptions that we mention in this review, is largely composed of technical papers, which could be difficult to grasp for practitioners as introductory material. Also, these notable exceptions usually focus on time-independent partial differential equations (PDEs), and even as introductory material they might be hard to absorb by non-mathematicians. In contrast, this review covers approaches which are purely data-driven and attempts to be amenable as an introduction for GW science practitioners, keeping theory to the minimum necessary to build intuition on why a given approach works as it does, what are the challenges left, and possible approaches to them. 

One of the reasons why ROM for parametrized system has been largely devoted to time-independent PDEs is because in that case many rigorous results, such as a priori error bounds on the reduced model, can be proved; these are called {\it certified approaches}.  Here we move beyond that particular arena to ROM for general parametrized systems. This includes problems which might not involve any differential equation, be time-dependent, to purely data-driven problems such as analysis and handling of large amounts of data. This broadening is at the expense, in many cases, of more heuristics and less rigor, such as a posteriori validation as opposed to rigorous a priori error bounds. The rationale for this is simply that many problems of interest in GW science are far too complex for existing detailed rigorous theorems. On the other hand, GW problems share many similarities with others for which approaches with proven properties have been developed and therefore certain algorithms can be adapted to cases of interest in GWs. In other instances it is quite the opposite: available rigorous results are quite generic, and can be abstracted from any previous application in which they were introduced. 

Half-rigorous, half-heuristic approaches are not as pessimistic as it might seem. Many widely used techniques in DS, ML and AI are somewhat heuristic in nature and require a posteriori validation. Similarly with half-rigorous, half-heuristic cutting-edge ROM approaches as those discussed here: they can (and should) always be a posteriori validated. 

Our presentation keeps technicalities and analyses to the minimum necessary for building intuition of why a technique works as it does, under which conditions it does so, and what can be expected given the properties of the problem of interest. At the same time, it is not a survey of all the work done in the field. 

There are two important books about Reduced Basis; though focused on partial differential equations there is a lot of valuable information (well beyond the scope of this short introduction): \cite{Jan-RB} and \cite{quarteroni2015reduced}. A website for ROM with several resources is \cite{morepas}, while \cite{rom-gr} has the speakers slides from a workshop on ROM in General Relativity; even though the latter is several years old and there has been much progress since then, the topics covered are still highly relevant. A very recent and lengthier, year-long program at ICERM (Brown University) focused on ROM in GR can be found on the website \cite{brown_sp-s20}, including  many of the slides and talks of each workshop. 

\section{Mathematical Preliminaries}
This section serves as a brief recap of some basic mathematics and to introduce some notation used hereon.  Since in this review we deal with model reduction for parametrized problems, we next consider an example amenable to it.

\begin{example}

Consider the model
\be
h(t; \lambda) = \exp(\lambda t) \, .  \label{eq:exp_1}
\ee
where $\lambda$ is a complex parameter and $\operatorname{Im}({\lam}) > 0$. This represents oscillatory, exponentially damped functions; it can be seen as a toy model for the ringdown of a black hole (discussed in Section~\ref{sec:ringdown}). 

In the notation that we use throughout this review, $h$ is a function of the physical variable $t$ parametrized by $\omega$ and for compactness we often use the following type of shortcuts:
$$
h = h_{\lambda}  = h(\lam) =  h(t; \lam)  \,.  
$$
Suppose now that we have a set of $N$ samples of $\lambda$, leading to a training set 
$$
{\cal K} = \{ h_i \} = \{ h_{\lambda_i} \}  = \{ h(\lam_i) \} = \{ h(t; \lam_i) \}\, , \quad i=1\ldots N. 
$$
Since the functions of interest (\ref{eq:exp_1}) are known in closed-form, building such a training set is straightforward, though in general this {\it is not} the case (as when solving the Einstein equations is required). 

One of the goals of ROM is to find a {\it reduced basis}, that is, a subset of $\cK$ with number of elements $n \leq N$ -- with, hopefully, $n \ll N$ --, such that its span (the span of a set of vectors is its set of linear combinations) represents  ${\cal K}$ with arbitrarily high accuracy. The compression rate is then
$$
C_r := N/n \, .
$$
Furthermore, in practice one also needs to discretize time (or frequency). So let's sample these training set elements at an arbitrary set of times $\{ t_i \}_{i=1}^L $, 
where $L$ stands for {\it Length}. Using, for example, the Empirical Interpolation Method (EIM, discussed in Section~\ref{sec:eim}), one can subsample these time samples to a subset $l \leq L$. In fact, in EIM, if the number of reduced basis is $n$, then by construction $l=n$. 

That is, the initial set being of size $N\times L$ can now be reduced to $n^2$, with a double compression rate (both in parameter and time domains), 
$$
C_r := \frac{N \times L} {n^2} \, . 
$$

\end{example}

\subsection*{Inner products and norms for functions or vectors}

Let $\Omega$ denote the physical domain. Throughout this review it represents time or frequency interval, $[t_{\rm min},t_{\rm max}]$ or $[f_{\rm min},f_{\rm max}]$ respectively, though in general it could be space, space-time, or some more abstract arena. For the sake of discussion we consider time intervals. For any two complex-valued functions, $h(t)$ and $g(t)$, we consider inner/scalar/dot products and their corresponding norms of the form 
\be \label{eq:dotproduct}
\langle h_1,  h_2 \rangle := \int_{\Omega} \bar{h}_1(t) h_2(t) \omega (x) dx , \quad  \| f \|^2 := \langle f , f \rangle \, ,
\ee
where the bar over $h_1$ denotes complex conjugation and $ \omega(x)$ is a generic weight function. These are referred to as weighted $L_2$ scalar products. In data analysis/signal processing and usually in the frequency domain, $S:= \omega^{-1}$ characterizes the sensitivity of the detector and is referred to as the power spectral density (PSD). 

We consider also discrete inner products and norms of the form 
\be \label{eq:dotproduct_d}
\langle h_1,  h_2\rangle := \sum_{i=1}^L \bar{h}_1(t_i) h_2(t_i) \omega_i  \,  \quad  \| h \|^2 := \langle h , h \rangle \, .
\ee
This discrete version of an $L_2$ scalar product is usually denoted as $\ell_2$, with a lowercase to distinguish it from the continuum case. 

Whenever \eqref{eq:dotproduct_d} is the discrete approximation of an integral \eqref{eq:dotproduct} the values $t_i \in \Omega$ and $\omega_i$ are respectively referred to as quadrature nodes and weights. Together, $\{t_i,\omega_i\}_{i=1}^L$ is called a quadrature rule. When 
$\omega_i =1$, (\ref{eq:dotproduct_d}) is referred to as the Euclidean scalar product. Throughout this review we also use the infinity, or max, norm
$$
\| h \|_{\infty} := \max_{1 \leq i \leq L}|h (t_i)| \, .
$$
of a vector $f \in \complex^{L}$. This norm is not induced by an inner product. That is, it can be shown that there is no scalar product $\langle \cdot, \cdot \rangle$ for which $\| h \| _{\infty} =\langle h, h \rangle$.   

Formally speaking, \eqref{eq:dotproduct} is defined for functions while \eqref{eq:dotproduct_d} is for vectors. We use the same notation $\langle \cdot , \cdot \rangle$ for both cases; with the hope that the distinction is clear from the context. Some proofs are sometimes more convenient in a continuous setting while numerical computations are restricted to discrete values. Depending on the context, we sometimes switch between these two settings using the same notations for both of them. 

\begin{example} Polynomials.

Consider the space of degree $n$ polynomials defined on $\Omega = [-1,1]$. Any element 
\be
f(t) = \sum_{i=0}^n c_i t^i \, ,
\ee
can be written as a sum of $(n+1)$ terms and such space is a linear one of dimension $(n+1)$. The first $(n+1)$ normalized Legendre polynomials $\{P_i(t)\}_{i=0}^{n}$ form an orthonormal basis with respect to the scalar product \eqref{eq:dotproduct} with $\omega(t)\equiv 1$.
\end{example}

\subsection*{Matrices}

Consider a complex $L \times N$ matrix $\textbf{H} \in \complex^{L \times N}$, where $L$ is the number of rows (``length of the time series'') of $\textbf{H}$ and $N$ is the number of columns (``number of parameter samples''), 
\be
\bH:= 
 \begin{pmatrix}
  h_{11} & h_{12} & \cdots & h_{1N}\\
  h_{21} & h_{22} & \cdots & h_{2N} \\
  \vdots  & \vdots  & \ddots & \vdots  \\
  h_{L1} & h_{L2} & \cdots & h_{LN} 
 \end{pmatrix} \, . 
\ee
As one might imagine, for gravitational waves $h_{ij} = h_j(t_i)$. Each column is referred to as a {\it snapshot}. A matrix $\bH$ of shape $L\times N$ appears often throughout this review, where $N$ is the number of samples in the training set (or $n$, the dimensionality of the reduced basis) and $L$ is the length of each time series, so it is a very concrete example. 

A square matrix $\bH$ is said to be non-singular if its inverse -- denoted by $\bH^{-1}$ -- exists, i.e. $\bH\bH^{-1}=\bH^{-1}\bH = \bI$, where 
$$
\bI = \begin{pmatrix}
  1 & 0 & \cdots & 0 \\
 0  & 1 & \cdots & 0 \\
  \vdots  & \vdots  & \ddots & \vdots  \\
0 & 0 & \cdots & 1 
 \end{pmatrix} =: diag (1,1,\ldots  ,1) \, .
$$
The (Hermitian) transpose of $\bH$, denoted by $H^{\dagger}$ is defined as the matrix with elements
$$
\left(\bH\right)^{\dagger}_{ij} = \bar{\bH}_{ji} \, , 
$$
the bar again indicates complex conjugation, and we say that $\bH$ is symmetric (hermitian or self-adjoint, in the current context), if $\bH=\bH^{\dagger}$. \\

\noindent{\bf Eigenvectors and eigenvalues:} \\

\noindent A vector $x\neq 0$ is an eigenvector of $\bH$ with eigenvalue $\alpha$ if 
$$
\bH x = \alpha x \,.
$$
Eigenvectors are defined up to a normalization constant; that is, if $x$ is an eigenvector with eigenvalue $\alpha$ so is $a x$, for any 
non-zero $a$. 

\subsubsection*{Matrix Norms}

\noindent{\bf The Frobenius norm}. Imagine ``unpacking" $\bH$ into a long vector of size $L \times N$. Measuring this vector with the Euclidean norm defines the Frobenius norm
\be \label{eq:Frobenius}
\| { \bH} \|^{2}_{F} := \sum_i^L \sum_j^N \, \left| h_{ij} \right|^{2}
\ee
of a matrix $\bH$.

Later we will see that the Frobenius norm plays a key role when discussing approximation by proper orthogonal decompositions.

\blank

\noindent{\bf Induced norm}. 

The matrix can be viewed as a linear operator mapping vectors from $\complex^N$ to $\complex^L$. Given norms for both $\complex^L$ and $\complex^N$, which need not be the same, the {\em induced norm} of a matrix is defined as 
\be \label{eq:matrix2Norm}
\| { \bH} \| =  \max_{ \| x \| = 1} \| { \bH } x \| \, .
\ee
This norm characterizes the maximal possible ``amplification" from the application of ${ \bf H}$ to $x$. More precisely: the value of $\| { \bH} \|$ is the smallest positive number $c$ such that 
\be
\| { \bH} x \| \leq  c \| x \| \, .
\ee

Notice that, following standard practice, we use the same notation -- namely, $|| \cdot ||$ -- for both the norm of a vector ($\| x \|$ and $\| \bH x \|$ in the above definition) and a matrix ($\| \bH \|$). It should be clear from the context, though, which one we are referring to. 

\subsubsection*{Rank and Kernel} \label{sec:rank}
The range of a complex $L\times N$ matrix $\bH$ is
$$
\mbox{range}(\bH) = \{ y \in\complex^L |\, y = \bH x \;\;  \mbox{for some } x \in\complex^N \} \, , 
$$ 
its rank 
$$
\mbox{rank}(\bH) = \dim \left(  \mbox{range} (\bH)  \right) \,,
$$
its kernel 
$$
\mbox{ker}(\bH) = \{  x \in\complex^L |\, \bH x =0  \} \,,
$$
and its nullity
$$
\text{nul}(\bH)=\text{dim}(\text{ker}(\bH))\,.
$$
As a linear algebra exercise it can be shown that
$$
\text{rank}(\bH) +\text{nul}(\bH) = N \,.
$$

For square $N\times N$ matrices, the following properties are equivalent:
\begin{enumerate}
\item $\bA$ is non-singular
\item $\mbox{det}(\bA) \neq 0$
\item $\mbox{ker}(\bA) = \{ {\bf 0}\in\complex^N \}$
\item $\mbox{rank}(\bA)=N$
\item $\bA$ has linearly independent columns and vectors. 
\end{enumerate}

\subsection*{Approximation by projection}\label{sec:Projection}

\subsubsection*{The least squares problem}\label{sec:ls_problem}

It is rather easy to approximate one function by other functions, such as Fourier, wavelets, polynomial expansions, or somewhat physically inspired bases. In this article we focus on representing gravitational waves by themselves. Intuitively, this should be (and is) more efficient. In this subsection we briefly introduce projection-based approximations, with more details given in Section~\ref{sec:spec}. 

To motivate the problem consider an $n$-dimensional vector space $W_n$ which is itself a subspace of a Hilbert's one ${\cal H}$ (for the purposes of this article, this means that a scalar product is assumed to exist). A common approximation criteria is a least squares (LS) one. That is, one seeks to approximate $h \in \cH$ by $h^{(n)} \in W_n$ which is the solution to 
\be \label{eq:orthopt}
h^{(n)} = {\tt argmin}_{\tilde{h}} \| h - \tilde{h} \|^2 \, ,
\ee
where $\tilde{h} \in W_n$. Equation~(\ref{eq:orthopt}) means finding the element of $W_n$ that minimizes the squared norm of the representation error.\\
If $\{e_i\}_{i=1}^n$ is an orthonormal basis of $W_n$ and $\langle.,. \rangle$ the scalar product associated with $\cH$, the following is the (unique) solution to the LS problem, 
\be
h^{(n)} = \sum_{i=1}^n c_i e_i \, , \quad \text{with} \;c_i = \langle e_i, h\rangle.  \label{eq:projection-LS}
\ee 
The solution $h^{(n)}$ is the orthogonal projection of $h$ onto $W_n$, therefore denoted as 
$$
{\cal P}_n h := h^{(n)} \, . 
$$
What it means is that if $h \in W_n$ then
\be
 {\cal P}_n h = h \, , \label{eq:proj1}
\ee
and the residual $\delta h = {\cal P}_n h - h $ satisfies 
\be
\langle {\cal P}_n h , \delta h \rangle = 0 \, .\label{eq:proj2}
\ee
The solution in \ref{eq:projection-LS} is basis-independent and $h^{(n)}$ is uniquely defined. The orthogonal projection onto any linear space is a geometric construction, so it is independent of the basis used to represent it: If $\{ e_i\}_{i=1}^n$  and $\{ \tilde{e}_i\}_{i=1}^n$ are any two orthonormal bases, and the projection coefficients $c_i$ and $\tilde{c}_i$ are computed according to  (\ref{eq:projection-LS}) (replacing in the second case $e_j$ by $\tilde{e}_j$), then
$$
{\cal P}_n h = \sum_{i=1}^n c_i e_i  = \sum_{i=1}^n \tilde{c}_i \tilde{e}_i \, . 
$$ 
\be
\sum_{i=1}^n \langle e_i, v\rangle e_i = \sum_{i=1}^n \langle \tilde{e}_i, v\rangle \tilde{e}_i \, ,  \label{eq:LS2}
\ee
leading to the same solution of the LS problem. One might ask why the emphasis on orthogonal or orthonormal bases; it is due to a conditioning issue which we discuss this in Section~\ref{sec:spec}. For the time being the summary is that one should always use orthogonal or orthonormal bases. An exception is when building  predictive models, as discussed in Section \ref{sec:surrogates}. 

\subsubsection*{Representing a set of elements: collective error} \label{sec:best_space} 

Suppose we seek to approximate not one function but a set of them:  
\be \label{eq:SetF}
{\cal F} = \{h_1, h_2, \dots , h_N | h_i \in {\cal H} \} 
\ee
From the above discussion in Section~\ref{sec:ls_problem} we know that ${\cal P}_n h_i$ is the the best representation of any particular $h_i$. The question of how one measures the approximation error of the collective set naturally arises. Two reasonable and standard criteria are 
\be
\frac{1}{N} \sum_{i=1}^N \| h_i - {\cal P}_n h_i \|^2  \, , \label{eq:avg-error} 
\ee
and 
\be
\max_i \| h_i - {\cal P}_n h_i \| ^2 \, , \label{eq:max-error}
\ee
which have the interpretations of the mean and maximum errors, respectively. Whether the notion of error ${\rm (1)}$ or ${\rm (2)}$ is chosen for an error minimization criteria will lead to the main two ROM approaches discussed in this review. Namely, Proper Orthogonal/Singular Value Decompositions, or the Reduced Basis-greedy approach. 

\subsection*{Further reading}
The material discussed can be found in standard linear algebra/numerical analysis books. There are many introductory good ones, we suggest  \cite{stewart1996afternotes} and \cite{stewart1998afternotes}.

\part{Representation and Compression}

\newpage

\section{Principal Component Analysis} \label{sec:PCA}

Principal Component Analysis (PCA) is perhaps one of the most used tools when seeking for redundancy or a hierarchy of  importance of variables in statistical analysis. It has a close relationship with Proper Orthogonal Decomposition (POD) as we discuss in Section~\ref{sec:pca-pod}. With this approach one seeks to determine the most statistically relevant variables and to potentially dimensionally discard from the problem the least relevant ones. In order to do this, we recall the definition of the covariance between two stochastic variables $X,Y$ as given by 
\begin{align}
	{\rm Cov} (X,Y) & = \big\langle ( X - \langle X \rangle ) ( Y - \langle Y \rangle ) \big\rangle \nonumber \\ 
		& = \langle X Y \rangle - \langle X \rangle \langle Y \rangle \, , 
			\label{covmx_1}\,
\end{align}	
where the brackets denote expectation values. The covariance between two variables provides a measure of the degree to which their fluctuations are correlated. A smaller (larger) covariance implies lower (higher) correlation. In particular, the covariance of a variable with itself is its variance (i.e., the standard deviation squared) and measures deviations from the mean value. 

When there are multiple  stochastic variables $X_i$ ($i=1, \ldots, n$) one can construct their associated covariance matrix $\bf{C}$ with components $C_{ij} = {\rm Cov} (X_i,X_j)$. This matrix is symmetric, non-negative definite and can therefore be diagonalized with an orthogonal transformation. Consider the $i^{\rm th}$ normalized eigenvector $\hat{{\bf V}}_i$. If we set ${\vec {\bf X}}:=(X_1,\ldots,X_n)$ then the {\it principal components} (PCs) are the 
associated eigenmodes, 
\begin{equation}
	\cE_i  = {\vec {\bf X}}\cdot\hat{\bf V}_{i} \, .
\end{equation}
The PCs are new variables representing directions in which the $X_i$ have different levels of variance. They are uncorrelated, a consequence of the orthogonality of the eigenvectors, and their associated eigenvalues $\lambda$ are their variances,
\begin{equation}
	{\rm Cov}(\cE_i, \cE_j ) = \lambda_i \delta_{ij} \, .
\end{equation}
The fact that, by construction, principal components are uncorrelated with each other is important since they provide independent pieces of statistical information. 

The smaller an eigenvalue $\lambda_i$, the more likely that the corresponding linear combination $\cE_i$ will not deviate from its average value for a randomly set of variables. Therefore, if there exist small eigenvalues then the associated PCs are largely conserved in a statistical sense. Conversely, the larger an eigenvalue is then the more relevant the associated PC is in describing the dynamics and variations in the problem.

There are two related but different senses in which for a parametrized time series a principal component with small variance can be semi-conserved. The first is being constant as a function of time for an arbitrary but fixed set of parameters. The second one is in the statistical sense that deviations of a principal component from the mean value are small for arbitrary but fixed initial and final times over a set of runs with the initial configurations. A small variance automatically implies approximate conservation in the second sense but not necessarily in the first one. The interest is not only in those PCs which have the smallest variances (and thus identify semi-conserved quantities in the second sense) but also in those with the largest variances, which encode the most information about the system dynamics. This should be clarified through the following example. 

\begin{example}{\bf PCA for spin dynamics.}

One may wonder if given a uniform initial spin orientation distribution, after a while those orientations turn into some preferred orientation, such as being aligned into some preferred direction. This was studied in \cite{Galley:2010rc} through massive numerical simulations. We only briefly review some of the results of that reference. The interest at the time was motivated by the unexpectedly large kicks found in numerical relativity simulations and whether they were of a generic nature in a statistical sense; for a review on recent comprehensive studies on this topic see Section~\ref{sec:kicks}. In order to perform a large enough analysis the Post-Newtonian equations were used, up to 3.5PN order in the angular frequency and 2PN with the covariant spin supplementary condition. 

Next, in any PCA study one has to define which variables to analyze. In order for the  quantities to be invariant under rotations of the system of reference, at least in a Newtonian sense, the first quantities chosen were scalar products between the normalized spins of each binary component  and the orbital angular momentum of the system. Moreover, their differences between initial and final values:
\begin{eqnarray}
	\Delta ({\bf\hat S}_1 \cdot {\bf\hat L})   &=& {\bf\hat S}_1 \cdot {\bf\hat L} |_f - {\bf\hat S}_1 \cdot {\bf\hat L} |_i \label{delta1L}  =: \DoL \\
	\Delta ({\bf\hat S}_2 \cdot {\bf\hat L})  & = & {\bf\hat S}_2 \cdot {\bf\hat L} |_f - {\bf\hat S}_2 \cdot {\bf\hat L} |_i  =: \DtL  \label{delta2L} \\
	\Delta ({\bf\hat S}_1 \cdot {\bf\hat S}_2) &=& {\bf\hat S}_1 \cdot {\bf\hat S}_2 |_f - {\bf\hat S}_1 \cdot {\bf\hat S}_2 |_i \label{delta12} =: \Dot  \, , \label{delta} 
\end{eqnarray}
where hats stand for unitary vectors. The choice of initial and final values is something to define and is discussed in~\cite{Galley:2010rc}, the summary is that the main results do not depend on these choices, which gives insight into the fact that the binary problem in GR is highly redundant. 

The orbital angular momentum and spin orientations naturally become correlated due to spin-orbit and 
spin-spin interactions as each of these binary black hole configurations evolve in time. However, at least within the PN approximation here considered, the orbital angular momentum and spin vectors remain perfectly uniformly distributed~\cite{Herrmann:2009mr}. For example, a Kolmogorov-Smirnov test for a representative configuration returns a p-value of $\sim 10^{-5}$ when testing for lack of uniformness~\cite{Herrmann:2009mr}. Higher PN expansions might introduce small biases~\cite{Lousto:2009ka} but if so they appear to be at a level in which approximating the mean of the above scalar products at any instant of time by zero is a very good approximation. 

Ref.~\cite{Galley:2010rc} starts with a simple case and makes contact with previous conservation results. The authors start building towards the more general case by first doing a PCA using only the two spin-orbit (SO) variables in (\ref{delta1L}) and (\ref{delta2L}), 
\begin{equation}
\DoL = \Delta ({\bf\hat S}_1 \cdot {\bf\hat L})\, , \quad \DtL \ceq \Delta ({\bf\hat S}_2 \cdot {\bf\hat L}) \, .  \label{SOPCA}
\end{equation}
However, spin-spin interactions in both the numerical simulations and in the analytical calculations {\em are} included in the PN equations of motion used when solving for the evolution of each configuration.

For mass and spin magnitudes ($m_j,\chi_j$) of each black hole the covariance matrix for the variables (\ref{SOPCA}) is
\begin{equation}
{\bf C}  = 
\left(
\begin{array}{cc}
{\rm Cov} ( \DoL, \DoL ) &  {\rm Cov} ( \DoL, \DtL ) \\
{\rm Cov} ( \DtL, \DoL ) &  {\rm Cov} ( \DtL, \DtL ) 
\end{array}
\right)  \, , 
\label{covmxSO}
\end{equation}
where the entries can come either from numerical simulations or from what the authors call the {\it instantaneous approximation}. 
The matrix $\bf C$ is then diagonalized to find the principal components.

From numerical simulations the authors find that, sampling across many random initial spin orientations, each of the principal components has zero mean over time (to numerical accuracy), $\langle \Delta \cE_j^{\text{SO}} \rangle = 0$, a consequence of the spin orientation distributions remaining highly uniform during the inspiral. 

Furthermore, they find $\lambda_2$ to be in the range $\sim 10^{-9} - 10^{-4}$ for the parameters sampled and that it grows with both spin magnitudes, which is expected from physical intuition, but also that it {\em increases} as the equal mass case is approached, which was unexpected.  

As an example with $(m_1,m_2, \chi_1,\chi_2) = (0.4, 0.6, 1.0, 1.0)$, Figure \ref{fig:PC_2d} shows a graphical representation of the principal components overlaid on a scatter plot of the $\DoL$ and $\DtL$ data from $1,\!000$ out of $100,\!000$ numerical simulations using random initial spin orientations. Notice that the first PC, which points along the direction of the eigenvector ${\bf\hat V}_1$ with the largest eigenvalue $\lambda_1$, captures the largest variation in the data while the second PC, pointing along ${\bf\hat V}_2$, indicates that there is very little spread in the data in that direction, which is also implied by the smallness of $\lambda_2$ relative to $\lambda_1$. Therefore, for the time interval considered, the second PC is largely irrelevant. This figure is almost an ideal example of dimensionality reduction through PCA. For a detailed analysis see \cite{Galley:2010rc}.

\begin{figure}
\begin{center}
\includegraphics[width=0.5\columnwidth]{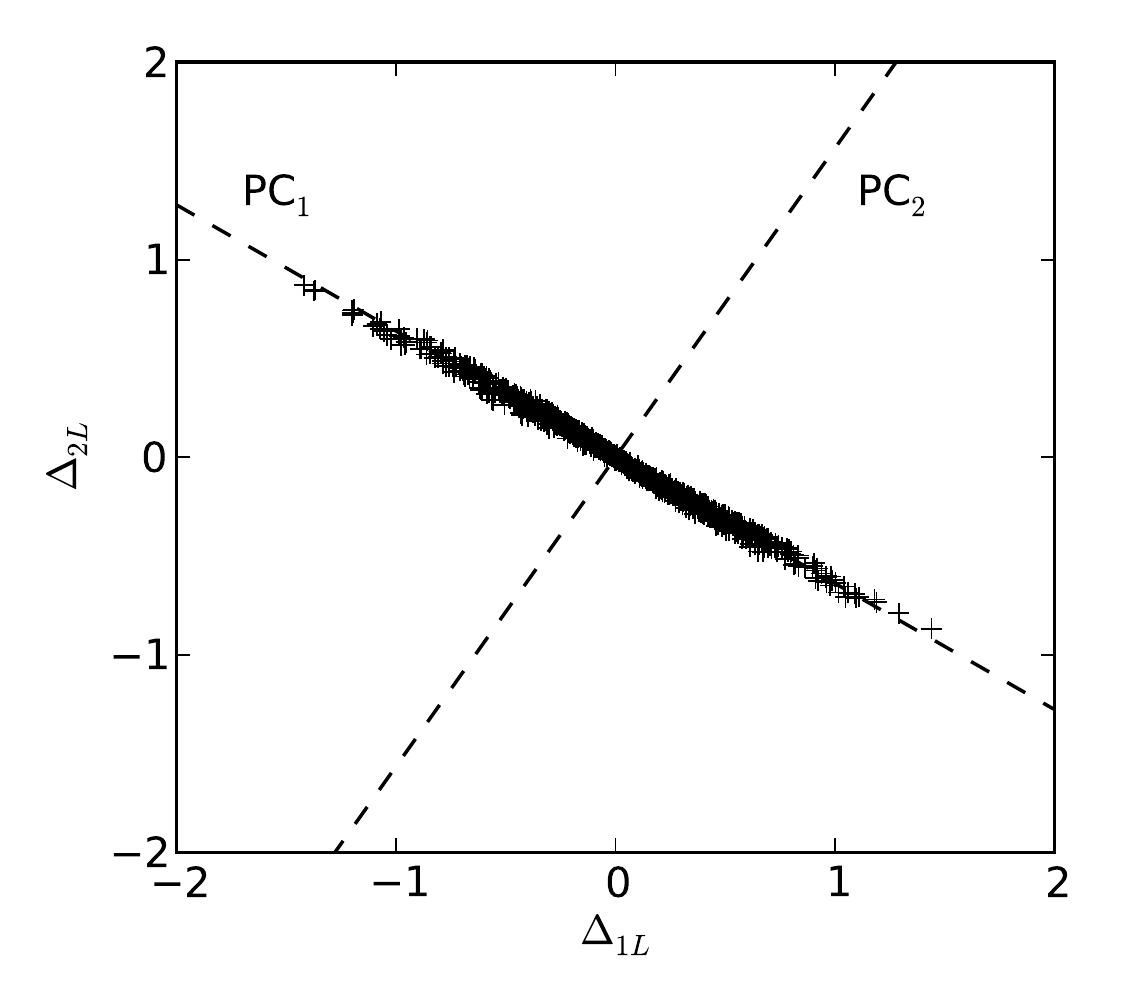}
\caption{A graphical representation of the principal components for the spin-orbit variables and the numerical data of $\DoL$ and $\DtL$ for a binary black hole system with, as illustration, $m_1 = 0.4$ and maximal spin magnitudes. One can see that PC2 is largely irrelevant. For details see \cite{Galley:2010rc}.}
\label{fig:PC_2d}
\end{center}
\end{figure}

\end{example}
\subsection*{Further Reading} \label{sec:PCA-fr}

Principal Component Analysis is a very well known tool to look for redundancies or categorize the relevance of different variables. One of its weaknesses is that it depends on which variables to look for. The case study that we chose is due to the fact that if looking in more detail at the original reference reviewed, there are strong indications that there are at least three (out of eight) redundant quantities in the problem, and therefore the problem is amenable to an unsupervised dimensional reduction approach, as presented throughout this review.  Reference \cite{JolliffePCA} is a very pedagogical book solely on PCA, with many case studies and modern developments.

\section{Proper Orthogonal Decomposition}\label{sec:low_rankd}
Given the snapshots $H_1, \ldots, H_N \in \complex^{L}$ (the training samples), the snapshot matrix is 
$$
{\bf H} = [H_1, \ldots H_N] \in \complex^{L \times N}
$$
with, in general,  $L\geq N$. The goal here is to find $n \leq N$ orthonormal vectors $\{ \phi_i \}_{i=1}^n$ in $\complex^{L}$ minimizing the average approximation error (\ref{eq:avg-error}) 
\be
\frac{1}{N}\sum_{i=1}^N \| H_i -  {\cal P}_n  H_i \|^2\,, \label{eq:pod}
\ee
where ${\cal P}_n$ is the orthogonal projector onto $W_n:= \text{span}\{ \phi_i \}_{i=1}^n$ and $\| \cdot \|$ is the Euclidean norm. The minimizer subspace of (\ref{eq:pod}) is called a Proper Orthogonal Decomposition (POD) of rank $n$, and its generating basis the POD one.

Notice that in general the POD basis elements $\phi_i$ are not members of $\{H_i\}_{i=1}^N$. In the gravitational wave case, this means that the reduced basis is not a subset of waveforms, neither does the method provide a set of the ``most representative'' points in the space of waveforms for building a posteriori bases through numerical relativity simulations. 

Another disadvantage is that the problem, involving a potentially very large linear algebra system, is not easily parallelizable in distributed memory architectures. Still, the method is straightforward to implement and can, for example, quickly provide insight into whether a problem is amenable to dimensionality reduction: after performing a POD decomposition of the snapshot matrix, one can look at the decay rate of singular values as a function of $n$ in order to know if the problem can be codified in a space of lower dimensionality than the original one. This can be turned into an actual strategy: a POD for a small subset of the problem can provide a quick insight, and if the singular values decay fast enough with $n$, the larger problem of interest can be tackled with a greedy approach, as discussed in Section~\ref{sec:greedy}.

The optimal solution to the minimization problem defined by Eq.~(\ref{eq:pod}) can be accomplished by means of a Singular Value Decomposition (SVD) procedure. In this framework the POD basis is given by the first $n$ left singular vectors of the snapshot matrix $\bf{H}$ (this can be proved by considering the first-order optimality conditions for the minimization problem).

Consider then the following singular value decomposition of {\bf H}:
\be
\bf{H}=\bf{U\Sigma V}^\dagger \, ,    \label{eq:SVD}
\ee
where ${\bf U}=[u_1, \ldots, u_L]\in\complex^{L\times L}$, ${\bf V}=[v_1, \ldots, v_N]\in\complex^{N\times N}$ are orthogonal matrices and
\be
{\bf \Sigma}=\begin{bmatrix} {\bf D}\\ {\bf 0} \end{bmatrix}\in\complex^{L\times N}, {\bf D}=diag(\sigma_1, \ldots, \sigma_N)\in\complex^{N\times N}, \sigma_1\geq\ldots\geq\sigma_N\geq 0.
\ee
Then, the rank-$n$ POD basis is given by $\phi_i=u_i,\,i=1, \ldots, n$. This basis provides a low-rank approximation to {\bf H} and represents the minimizer basis for the minimization problem stated in (\ref{eq:pod}). To see this, consider from (\ref{eq:SVD}) the relation
$$
u_i^\dagger{\bf H}=\sigma_i v_i^\dagger\,,
$$
valid for all column vectors of ${\bf U}$. Next, multiply by $u_i$ on the left and sum
\be
\sum_{i=1}^n u_i u_i^\dagger{\bf H}=\sum_{i=1}^n\sigma_i u_i v_i^\dagger\,.\label{eq:proj}
\ee
The l.h.s. of (\ref{eq:proj}) is exactly the orthogonal projector associated to the basis $\{u_i\}_{i=1}^n$ acting on $\bf H$ and the r.h.s. represents its rank-$n$ approximation ${\bf H}_n$. Therefore we can rewrite (\ref{eq:proj}) as
\be
\cP_n{\bf H}={\bf H}_n\,.
\ee

\noindent{\bf Remarks on the SVD decomposition}

\begin{itemize}
\item The numbers $\sigma_i$ are known as the {\it singular values} of the matrix {\bf H} and correspond to the positive square roots of the eigenvalues of the associated matrix $\bf{K}=\bf{H}^{\dagger}\bf{H}$. They are usually chosen in descending order in practice to facilitate the recognition of the most relevant principal components.
\item For the matrices {\bf U} and {\bf V}, only the first $\mathrm{rank}(\mathbf{H})$  columns are unique, whereas the remaining ones are arbitrarily extended such that orthogonality is maintained. Since $\sigma_i = 0$ whenever $i > \mathrm{rank}(\mathbf{H})$ the factorization \eqref{eq:SVD} does not depend on the choice of this extension, since such extension is annihilated by null entries of $\bf\Sigma$.
\item The first $N$ columns of {\bf U} and {\bf V} are known as the left- and right-singular vectors of {\bf H} respectively. The right-singular vectors, $v_i$, are the normalized eigenvectors of ${\bf H}^{\dagger}{\bf H}$ and the left-singular vectors, $u_i$, are the normalized eigenvectors of ${\bf H}{\bf H}^{\dagger}$.
\item The number $K$ of non-zero elements in $\mathbf{\Sigma}$ is exactly $\text{rank}(\mathbf{H})$. Consequently, Eq.~(\eqref{eq:SVD}) is sometimes called a {\em rank-revealing} factorization.
\end{itemize}

\noindent{\bf Error of a POD approximation}

The accuracy of the low rank-$n$ approximation ${\bf H}_n={\cal P}_n{\bf H}=[{\cal P}_n H_1,\ldots, {\cal P}_n H_N]$ is given by the following lemma.

\noindent{\bf Lemma 1.}  Define the approximation error by

\be
\epsilon= \frac{1}{N} \sum_{i=1}^N \|H_i-{\cal P}_n H_i \|^2.\label{eq:errSVD}
\ee
It can be shown that  $\epsilon$ satisfies
\be
\epsilon= \frac{1}{N} \sum_{i=n+1}^N\sigma_i^2.
\ee
In terms of the Frobenius norm,
\be
\epsilon=\frac{1}{N}\|{\bf H}-{\bf H}_n\|^2_F.\label{eq:errorF}
\ee
Intuitively, the square of the Frobenius norm of the difference between a matrix and its low-rank approximation represents the total squared difference of the rows of ${\bf H}$ due to omitting the last $(N-n)$ singular values when forming ${\bf H}$. As discussed above, the projection ${\bf H}_n$ represents the best rank-$n$ approximation of {\bf H} in the Frobenius norm. 

In summary, going back to section \ref{sec:best_space}, one can see that the minimization problem (1) related to the average error of a set of functions ${\cal F}$ is optimally solved by the POD/SVD decomposition of the matrix associated to those functions.

In the context of a gravitational-wave template bank the rows of the snapshot matrix are the waveforms evaluated at different time/frequency values, and the different rows correspond to different intrinsic parameters. The inner product of rows with themselves, in turn, correspond to the total power in the bank. Thus, the squared Frobenius norm of a template bank is the total power contained within all the templates in it. For normalized templates, this is simply the size of the bank. It can be seen how Eq.~(\ref{eq:errorF})
directly corresponds to the total power in a template bank lost from a low-rank approximation to it. More precisely, the error measure is directly related to the average fractional SNR loss 
which, up to a constant, is the squared Frobenius norm of the difference 
between the full template bank and its low-rank approximation. 
Fixing the average fractional SNR loss thus determines the total number of non-zero singular values which must be retained to guarantee that the rank-reduced bank remains effective. This is discussed in practical terms through the following case study.  

\begin{example}{\bf SVD for gravitational waves.}

In Ref.~\cite{PhysRevD.82.044025} the authors studied the application of SVD to gravitational wave templates to reduce the redundancies in the bank and build an orthogonal basis to represent the whole set. As a proof of concept, the authors applied a SVD approach to a set of CBC waveforms corresponding to a sliver of the BNS parameter space. Next, we summarize some of the results of this reference, closely following its notation, which might be different from the rest of this review.

In order to detect a GW signal, the common choice of the minimal match between an arbitrary point in parameter space and its nearest point of the template bank is $97\%$.  

In order to compare and filter the data against the entire template bank, an approximation to the matched filtering $\rho_\alpha=\langle \bar{h}_{\alpha}, s\rangle$ is sought for, where $h_{\alpha}$ is a complex waveform vector and $s$ is the data vector (the presumed signal). In this way the number of evaluation of inner products can be reduced as well as its computational cost. Let's define the $N\times L$, where $N=2M$ and $M$ stands for the number of complex waveform (there is no obvious reason why to perform an SVD on real and imaginary parts of a waveform, given that the POD approach can handle complex snapshots) template matrix
\be
{\bf H}=\{h_1^R,h_1^I,\ldots,h_M^R,h_M^I\}\,,
\ee
where $h_i^{R,I}$ are the real and imaginary part of the $i$-template waveform, each one corresponding to  the rows of $\bf H$. Applying an SVD decomposition to ${\bf H}$ and writing it in component form (here we follow the notation of \cite{PhysRevD.82.044025})
\be
{\bf H}_{\mu\nu}=\sum_{\kappa=1}^N v_{\mu\kappa}\sigma_\kappa u_{\kappa\nu}\,,
\ee
one can define the truncated sum
\be
{\bf H}'_{\mu\nu}=\sum_{\kappa=1}^{N'} v_{\nu\kappa}\sigma_\kappa u_{\kappa\nu}\,,
\ee
and approximate $\rho_\alpha$ by
\be
\rho'_\alpha= \langle H'_{2\alpha-1}-iH'_{2\alpha}, s\rangle =\sum_{\nu=1}^{N'} (v_{{2\alpha-1}\,\nu} \sigma_\nu - i v_{{2\alpha}\,\nu}\sigma_\nu) \langle u_\nu, s\rangle\,,
\ee
where the $H'_j$ are the vector rows of ${\bf H}'$. 

Figure \ref{fig:svd} shows a representation of the matrix of waveforms {\bf H} and its associated SVD-basis. One disadvantage of this kind of dimensional reduction is that the remaining basis barely resembles the structure of the original template. This can be fixed with the RB-greedy approach that is presented in Section~\ref{sec:rb}. 
\begin{figure}[h!]
\begin{center}
\includegraphics[width=0.4\linewidth]{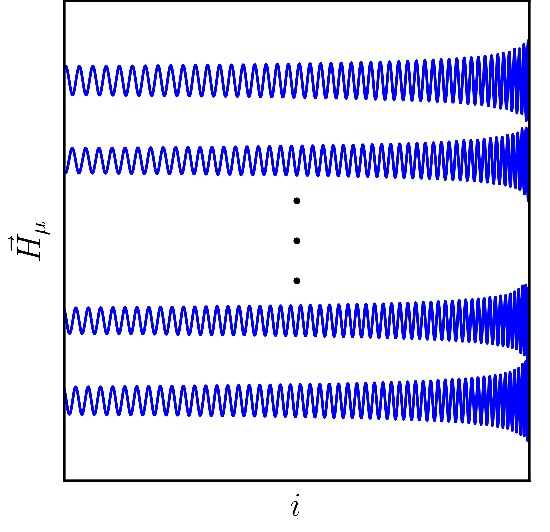}
\includegraphics[width=0.4\linewidth]{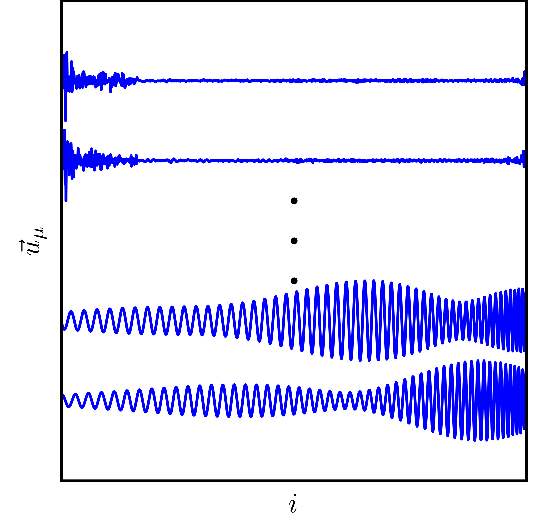}
\caption{{\bf Left}: Representation of template waveforms corresponding to $\bf H$ against time. {\bf Right}: Representation of the first four SVD-basis.  Notice that since the SVD basis functions do not correspond to actual waveforms, they display non-physical behavior which might be difficult to fit for when building a surrogate model as discussed in Section~\ref{sec:surrogates}. Figures taken from~\cite{PhysRevD.82.044025}.}
\label{fig:svd}
\end{center} 
\end{figure}

\noindent {\bf Accuracy}. An approximation to the data snapshot matrix is sought such that $\|{\bf H}_\mu - {\bf H}'_\mu \|\sim 1\%$. Following this requirement, the fractional SNR
\be\label{eq:SNR}
\Big\langle\frac{\delta\rho}{\rho}\Big\rangle:=\frac{1}{M}\sum_{\alpha=1}^M \frac{\delta\rho_\alpha}{\rho_\alpha}
\ee
can be approximated by:
\be
\Big\langle\frac{\delta\rho}{\rho}\Big\rangle=\frac{1}{2N}\sum_{\mu=N'+1}^N \sigma_\mu^2\,,
\ee
resulting from a Taylor expansion valid in the range $\langle\delta\rho/\rho\rangle<10\%$. Note that this approximation is proportional to the squared Frobenius norm of the truncation error of ${\bf H}$,
\be
\Big\langle\frac{\delta\rho}{\rho}\Big\rangle=\frac{1}{2N}\|{\bf H}-{\bf H}' \|_F^2\,.
\ee

As a case of application, \cite{PhysRevD.82.044025} shows an SVD analysis to gravitational waves emitted by a CBC-BNS, with chirp masses $1.125 M_\odot\leq M_c<1.240M_\odot$ (that is, a rather small sliver in parameter space) and component masses $1M_\odot\leq m_1,m_2<3M_\odot$. In order to satisfy a minimal match of $96.8\%$, a number of templates $M=456$ ($N=912$) to cover the parameter space was found. In Fig.~\ref{fig:bns-svd} (plot of $\langle\delta\rho/\rho\rangle$ vs. \# of SVD-basis elements) it can be seen that, to obtain $\langle\delta\rho/\rho\rangle=10^{-3}$, the number of basis elements needed to reconstruct the whole template bank to that accuracy can be reduced from $N=912$ to $N'=118$. Though POD/SVD is a good starting point for ROM, in following sections we will develop a more modern framework for modeling reduction.

\begin{figure}[h!]
\begin{center}
\includegraphics[width=0.4\linewidth]{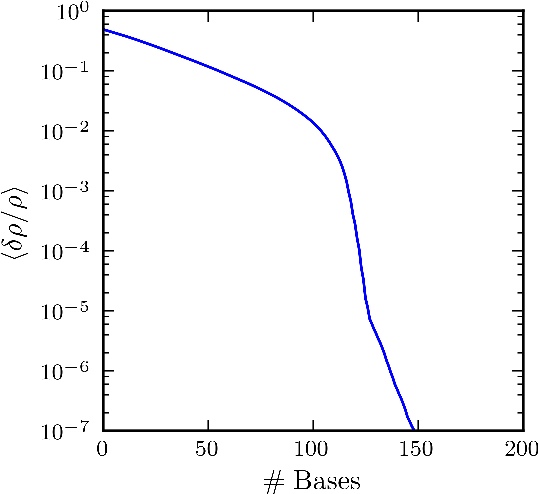}
\caption{SNR (Eq.~(\ref{eq:SNR})) for a CBC-BNS gravitational waves as a function of the number of SVD-basis elements -- valid for $SNR<10\%$ --. Figure taken from \cite{PhysRevD.82.044025}.}
\label{fig:bns-svd}
\end{center} 
\end{figure}

\end{example}

\subsection{PCA and POD} \label{sec:pca-pod}
Principal Component Analysis and Proper Orthogonal Decomposition are closely related from a mathematical point of view, though usually applied in different contexts. PCA is in general used in statistical analysis. It represents a solution to an optimization problem; namely, that one of  finding the uncorrelated directions of maximum statistical variability in a stochastic data set. POD is usually applied in dimension reduction modeling. As in PCA, POD also represents a solution to an optimization problem: finding the best low-rank approximation to a data set matrix. This is achieved by minimizing the Frobenius norm of the error matrix (Eq.~(\ref{eq:errorF})).

Suppose we have a data set matrix of size $L\times N$, $N$ being the number of variables and $L$, the number of samples corresponding to each variable. In PCA one is not interested in the ``length" of each variable and their individual components, but rather on their correlations. Put differently, if there are $N$ stochastic variables, the covariance matrix, regardless of the number of components in each stochastic variable, is in all cases of size $N \times N$. Compression can be achieved by ignoring the directions in which the stochastic data have little variance. 
\blank

\noindent{\bf Mathematical relationship between PCA and POD}

Although conceptually different, both methods are mathematically related. To see this, we make a simple observation concerning the snapshot matrix and its covariance matrix. Let ${\bf X}:=[X_1,\ldots,X_N]\in\complex^{L\times N}$ be the snapshot matrix with each column $i$ storing $L$ independent observations of the random variable $X_i$. We assume that each $X_i$ is identically distributed. Subtracting to each column $X_i$ the corresponding mean value $\langle X_i \rangle$, $X_{s,i}\rightarrow X_{s,i}-\langle X_i \rangle\,,\,s=1,\ldots,L$, and replacing $\bf{X}$ by this ``centered" matrix version, the associated covariance matrix $\bf{C}$ can be written as
\be\label{eq:cov}
{\bf C}=\frac{1}{L-1}{\bf X}^\dagger {\bf X}\,.
\ee
Without loss of generality, we will suppose $\bf X$ has full rank. Therefore, being an hermitian, non-negative definite matrix, $\bf C$ can be diagonalized as
$$
{\bf C}={\bf V}{\bf\Lambda}{\bf V}^\dagger\,,
$$
where ${\bf\Lambda}=\text{diag}(\lambda_1, ..., \lambda_N)$ is the matrix of non-zero descending-ordered real eigenvalues and $\bf V$ is the unitary matrix storing eigenvectors of $\bf C$. The columns of $\bf V$ are called {\it principal axes} or {\it principal directions} in the PCA framework. PCs correspond to the projection of the data onto these principal directions, the columns of the matrix product ${\bf X}{\bf V}$.

Now, perform a POD/SVD decomposition of the snapshot matrix {\bf X}
$$
{\bf X}=\mathbf{U}\mathbf{\Sigma}\mathbf{\tilde V}^{\dagger}
$$
and rewrite Eq.~(\ref{eq:cov}) as
\be\label{eq:cov2}
{\bf C}=\frac{1}{L-1}\, \,(\mathbf{U}\mathbf{\Sigma}\mathbf{\tilde V}^{\dagger})^\dagger \mathbf{U}\mathbf{\Sigma}\mathbf{\tilde V}^{\dagger}=\mathbf{\tilde V}\,\frac{\mathbf{\Sigma}^\top\mathbf{\Sigma}}{L-1}\,\mathbf{\tilde V}^\dagger\,.
\ee
The similarity relation between $\bf C$ and $\frac{\mathbf{\Sigma}^\top\mathbf{\Sigma}}{L-1}$ implies that both have the same eigenvalues:

$$
\lam_i = \frac{\sigma^2_i}{L-1}\,,\quad i=1,\ldots, N\,.
$$

Another benefit of this mathematical correspondence is that PC's are straigthforward to compute as ${\bf X}{\bf V}={\bf U}\mathbf{\Sigma}$.

\subsection*{Further reading}

The reader may consult \cite{JolliffePCA,trefethenLinearAlgebra} for clear expositions about the classical matrix factorization methods presented in this section. QR decompositions form another class of matrix factorizations which can be used for low-rank approximations. A theoretical framework for rank revealing QR decompositions of a matrix was developed in the the late 1980's and 1990's~\cite{chan1987, hong1992}, motivated partly by the SVD's high computational cost. QR decompositions form the basis for many modern fast algorithms~\cite{Harbrecht2012428,guruswami2012optimal,Deshpande2010,Civril2008d}. Actually, motivated by GW research an important resuIt has been shown: the  RB-greedy algorithm is completely equivalent to a certain type of QR decomposition  \cite{Antil2018}.

\section{Spectral expansions} \label{sec:spec}
In this section we discuss fast converging classical linear approximations, since they naturally lead to Reduced Basis when considering parametrized problems. 

Consider a set of complex functions,  
$$
\cF:= \{ h: \mathbb{C} \rightarrow \mathbb{C} \} \, .
$$ 
A standard linear approximation consists of the following: a basis with $n$ (non necessary orthogonal) functions is {\em somehow} chosen, 
$$
\{ e_i \}_{i=1}^n \, . 
$$
 Next, a function $h$ in $\cF$ is approximated by a linear combination of the chosen basis elements, 
\be
h(z) \approx  h^{(n)}(z) := \sum_{i=1}^n c_i e_i(z) \, , \quad c_i \in \mathbb{C}  .   \label{eq:expansion}
\ee

There are several criteria to choose the expansion coefficients $\{ c_i \}$. We recap least-squares (LS), in Section~\ref{sec:eim} we will discuss interpolation. In the LS approach the coefficients $\{ c_i \}$ are chosen such that the representation error is minimized with respect to some chosen norm, 
\be
\left \| h - \sum_{i=1}^n c_i e_i \right \|^2\, .   \label{eq:LS}
\ee
As was stated in Section \ref{sec:Projection}, the solution to the LS problem (Eq.~(\ref{eq:orthopt}), Section~\ref{sec:Projection}) is the orthogonal projection onto the span of the basis, 
\be
h^{(n)} = {\cal P}_n h \, . 
\ee 
That is, the approximation (\ref{eq:expansion}) minimizes the error (\ref{eq:LS}) when the coefficients $\{ c_i \}$ are chosen such that the residual 
$$
\delta^{(n)} h  = h - h^{(n)}
$$
satisfies $\langle \delta h , h^{(n)}  \rangle  = 0$, which implies
\be
\left \langle e_j (\cdot ) , h (\cdot) -  \sum_{i=1}^n c_i  e_i (\cdot) \right \rangle = 0 \quad \quad \text{ for } j=1,\ldots, n \, . \label{eq:LS_ort} 
\ee
The solution to (\ref{eq:LS_ort}) is 
\be
c_i = \sum_{j=1}^n({\bf G}^{-1} )_{ij} \langle e_j , h \rangle \, , \label{eq:optLS}
\ee
where ${\bf G}^{-1}$ is the inverse of the {\em Gram matrix} or {\it Gramian} ${\bf G}$, with entries
$$
{\bf G}_{ij} := \langle e_i  , e_j \rangle \, .  
$$
If the basis is orthonormal, this matrix is the identity and one recovers the familiar expression 
\be
{\cal P}_n h =  \sum_{i=1}^n \langle e_i , h \rangle e_i \, .  \label{eq:proj_GS}
\ee
The Gramian matrix can be very ill-conditioned in general, meaning that the calculation of its inverse, needed in (\ref{eq:optLS}), can have large numerical errors (see, for example \cite{taylor_1978}). Therefore, from a numerical conditioning point of view, it is convenient to work with an orthonormal (or orthogonal) basis since the Gram matrix is the identity. 

In iterative approximations one defines the {\it convergence rate} as the rate at which the representation error
\be
\| \delta^{(n)}h \| = \left \| h - {\cal P}_n h \right \| \label{eq:proj_error}
\ee
decreases as $n$ increases for any given $h$. In the context of RB, when one is dealing with parametrized systems, this error will depend also on the parameters of the system.

\subsection{Spectral methods}
\label{sec:spectral}
In terms of accuracy and optimal convergence rates for approximation of a space $\cF$, we have not yet discussed two related aspects:
\begin{enumerate}
\item The choice of a ``good'' basis or, more precisely, the approximation space $W_n$.  
\item The optimal choice of how many basis elements to use. One would think that the larger, the better. This is related to the {\it regularity} of the functions in $\cF$. That is, how smooth or differentiable they are. 
\end{enumerate}

The first comment might be puzzling, after all we have emphasized that given an approximation space $W_n$, the LS approximation is uniquely defined in a geometric way. The question really is what the approximation space $W_n$ should be to minimize the representation error. \\

\subsubsection{Fourier Expansions}
\label{sec:fourier}

We start with the simplest and most familiar case: that one of periodic functions in $[0,2\pi]$, unit weight, $\omega(x)=1$ \footnote{The choice of norm is crucial, see the next Section about Jacobi polynomials for more insights about this.}, and Fourier modes as (orthonormal) basis, 
$$
e_j (x) = \frac{1}{\sqrt{2\pi}} e^{ijx} \, , \quad j\in \integer
$$
Assuming for simplicity that $n$ is even, we have 
$$
{\cal P}_{n+1} h (x) = \sum_{-n/2}^{n/2} \hat{h}_j e_j(x)\, , 
$$
where $\hat{h}_j$ are the Fourier coefficients of $h$. 

One can show that if $h$ has $s$ derivatives, then there exists a constant $C>0$ independent of $n$ such that
\begin{equation}
\| h - {\cal P}_{n+1} h \| \leq C(n+1)^{-s} \left \| \frac{d^s h }{dx^s} \right \|
\label{eq:conv_fourier}
\end{equation}
for all $n\geq 0$. In particular, if $h$ is smooth ($h\in C^\infty$), then the representation error decays faster than any power law with $n\rightarrow \infty$, which is referred to as {\it spectral convergence}. Under further conditions, in particular the case in which $h$ is analytic, the error decay is actually exponential.  The Fourier case gives a very intuitive way of how this happens just by using integration by parts, for more details see Chapter 9 of \cite{Sarbach2012}. 

The summary here is that Fourier modes are not a good basis just because they are periodic functions, but they provide fast convergence as a representation space, as fast as the regularity of the function(s) being represented has. This is the core idea of spectral methods and, for parametrized systems, reduced basis. A brief summary of the non-periodic case follows. 

\subsubsection{Jacobi Polynomials}
\label{sec:jacobi}

In the case of non periodic functions a similar result to that one discussed for periodic functions holds, whether the domain is bounded or infinite. Again, the choice of weight is crucial. Any interval can be mapped into $[-1,1]$ or $(-1,1)$, that is why we usually refer to those intervals. The open interval case is because in some cases (such as Chebyshev) the weights actually diverge at the $\pm 1$ end points -- there is a reason for this but we shall skip it. 

Polynomials are a natural basis to use. Why? Just because after centuries we understand them well; going beyond that is one way of looking at Reduced Basis. Given a maximum degree, the span of polynomials is the same, so the question is only what kind of weights guarantee fast convergence upon regularity of the function to be represented. The following family of weights is a sufficient class, thought not a necessary one, 
\be
\omega(x) = (1-x)^{\alpha }(1+x)^{\beta} \, \quad x\in (-1,1) \label{eq:jacobi_weight}
\ee
with $\alpha, \beta > -1$.  

Under these assumptions, the following holds in a Jacobi Polynomial approximation:  
\begin{equation}
\| h - {\cal P}_{n+1} h\|  \leq C (n+1)^{-s} \left \|  (1-x^2)^{s/2} \frac{d^s h}{dx^s} \right\| 
\label{eq:conv_spectral}
\end{equation}
for all $n > (s-1)$ and $C$ independent of $n$. If $h$ is smooth there is {\em asymptotic} spectral convergence. Under additional conditions on the smoothness (or, but not necessarily, analyticity), the convergence is in fact exponential. In this review we will loosely associate smoothness with asymptotic exponential convergence of application-specific spectral expansions, without discussing these additional assumptions. \\

Most of spectral methods literature is based on solving numerical problems, prominently differential equations. In this review we shift this focus to that one of a more fundamental {\em representation problem}, after which all the calculus for approximating quadratures, taking derivatives, and solving differential equations follows in a rather straightforward way. This is not just a matter of taste, but will make the introduction of Reduced Basis and all its associated calculus and applications very natural and almost trivial.

\subsection*{Further reading} 
The weights (\ref{eq:jacobi_weight}) lead to {\it Jacobi polynomials}, which are solutions to a singular Sturm-Liouville problem, the properties of which guarantee the above mentioned spectral convergence. Standard examples are Legendre and Chebyshev polynomials. Their span is the same, it is with respect to which scalar product they guarantee fast convergence and their discrete version and relation to interpolation, discussed in Section~\ref{sec:eim}. They are also orthonormalized with respect to their own scalar products, which helps avoid the typical conditioning problem of inverting the associated Gramian matrix. For more details on and a quick glimpse at spectral methods, from a physicist or practitioner perspective, see Chapter 9 of Ref.~\cite{Sarbach2012}. Reference~\cite{trefethen2000spectral} is a surprisingly compact and efficient book to get started with spectral methods (despite its title, MATLAB is not necessary at all to digest the content). A long classic reference, especially for practitioners, is \cite{boyd2001chebyshev}. The book is legally available for free from the author's webpage, though with some typos. A great book, in some sense targeted at ordinary differential equations is \cite{funaro1992polynomial}.   For a modern presentation and the latest results, from theory to current approaches to beat Gibb's phenomena, see \cite{Jan-RB} and references therein. 

\section{Parametrized problems and optimal approximations} \label{sec:optimal}
We are interested in approximating some abstract space of parametrized functions $\cF$, which in general is not linear (the sum of two waveforms does not need to be a waveform) but we assume that it can be embedded in a Hilbert one $\cH$ (for example, the one of integrable functions in the $L_2$ sense).  We  denote the underlying space of parameters as $\Phi$, which we assume to be compact. For example, $\cF$ can be the space of $\lam$-parametrized solutions $u_{\lam}(x)$ of a partial differential equation representing the dynamics of a physical system. In this review we place emphasis in gravitational waves, parametrized for example by the mass and spin of each black hole in a binary collision.

The question discussed next is how well one can {\em theoretically} approximate {\em all of} $\cF$ by a set of $n$ basis elements of  $\cH$ in a linear and most compact way. This leads to the Kolmogorov n-width \footnote{There are other widths, see \cite{Pinkus}.} $d_n$ of $\cF$ with respect to $\cH$, 
\begin{align} \label{eq:width}
d_n = d_n(\cF, \cH):= \min_{\{ e_i \}_{i=1}^n \in \cH} \max_{\lam\in\Phi } \min_{c_i \in \mathbb{C}} \bigg\| h(\cdot ;\lam) - \sum_{i=1}^n c_i(\lam) e_i(\cdot) \bigg\|^2   \, .
\end{align}
\begin{comment}
We explain the meaning of (\ref{eq:width}), from right to left, with respect to the $\min, \max, \min$ properties.  
\begin{itemize}
\item The first $\min$ implies that the optimal representation with respect to the (so far arbitrary) basis $\{ e_i \}_{i=1}^n$ is used. That is, given a basis, the best representation (in the induced norm $\|.\|$) is considered. We already discussed that this is the orthogonal projection ${\cal P}_n$ onto the span of the basis $\{e_i \}_{i=1}^n$, so we can replace (\ref{eq:width}) by 
\begin{align} \label{eq:nwidth2}
d_n := \min_{\{ e_i \}_{i=1}^n \in \cH} \max_{\lam\in\Phi } \bigg\| h(\cdot ;\lam) - {\cal P}_n h(\cdot; \lam)  \bigg\|^2   \, .
\end{align}
\item Next,  the largest error in parameter space for such a best approximation is picked. That is the ``worst best'' approximation given a choice of basis.
\item Finally, a choice of basis (more precisely, the subspace $W_n$) which minimizes this worst best error is chosen and the associated error is by definition the n-width.  The compactness of $\Phi$ is important to guarantee that these minimum and maximum exist in the search through the parameter space.
\end{itemize}
\end{comment}
In other words, the $n$-width is an upper bound of  ``what is the best that one can do'' if one could optimally choose an optimal basis. 

This is mostly a theoretical problem, since solving for such a basis is impractical (or, actually, in most cases, intractable) from a computational point of view because it carries combinatorial complexity (all elements of the basis have to be simultaneously chosen). It is more of a theoretical upper bound against which to benchmark the quality of any computable approximation which seeks for an approximate solution to the $n$-width problem.

\noindent{\bf Parameter regularity and fast convergence} \label{sec:optimal}

Exact expressions for the $n$-width in general can only be achieved in a few cases (see, for example, \cite{MAGARILILYAEV200197}). In general, the best one can do is to set up upper bounds under specific assumptions and infer the dependence of the $n$-width with respect to $n$. Indeed, Kolmogorov calculated this distance for a special class of functions \cite{MAGARILILYAEV200197} with its first $(r-1)$ derivatives with respect to parameter variation {\it absolutely continuous} and obtained a power law dependency for the $n$-width, 
$$
d_n \sim n^{-r} \, . 
$$
If the set of functions considered were $C^{\infty}$ {\em with respect to parameter variation} -- the functions themselves can be discontinuous -- the rate of convergence of its optimal basis representation would be better than any power law. This is called {\it spectral convergence} and will be relevant in next chapters at the moment of quantifying the fast convergence of a Reduced Basis approach. This is exactly the case in many scenarios of GWs, since they depend smoothly on the parameters of their sources and one therefore expects an optimal approach to have very fast (in fact, exponential) convergence, as opposed to random sampling, for example, for which the convergence is sublinear. This is the main reason underlying the extreme high accuracy of very compact surrogate models based on reduced bases. That is, one expects in most GW scenarios
$$
d_n \sim e^{-an^b} \, , 
$$
and an optimal basis should be extremely accurate and compact; in fact in some cases super-exponential convergence ($b>1$) has been found in the GW context \cite{Field:2011mf}. The question then turns into how to find an approximate basis which is not only computable but is also nearly optimal with respect to the n-width. 

It is ``common knowledge'' that parametrized problems with regularity with respect to parameter variation show spectral convergence of the n-width in practice. One might ask how this observation can be possible at all when we have mentioned that the n-width is not computable in practice: we will return to this point when we discuss the greedy algorithm. But in fact, up to our knowledge there is no rigorous proof of this expectation for general parametrized systems. But there is a very compelling argument. Asymptotic exponential convergence is also observed in practice in GWs and, being physicists, that is good enough for us. The argument is as follow (courtesy of Albert Cohen): 

The convergence rate of the n-width depends on the smoothness/regularity of the functions with respect to the parameters of the problem. We now {\em argue} that if there is regularity with respect to parameter variation, as in many cases of interest, then one can expect fast (in fact, up to exponential or even super-exponential) decay of the approximation error. The argument is a spectral standard one (spectral methods were discussed in the previous section) applied to parameter variation and is as follows: 

If the parametric map 
$$
\lam \rightarrow h_{\lam}(\cdot )
$$
is smooth enough with respect to $\lam$, then it can be very well approximated
in some appropriate basis in the $\lam$ parameter variable,
for instance by Fourier or Jacobi polynomials. This means that there is an expansion of the form
\be
c_1(\lam)h_1 (\cdot) + \ldots + c_n(\lam) h_n(\cdot )+ \ldots \, , \label{eq:conv_argument}
\ee
with $h_i \in \cF$, that converges fastly towards $h_{\lam}(\cdot)$ in $\cF$ uniformly in $\lam$.

Now, a partial sum of the form (\ref{eq:conv_argument}) is a member of the span of $\{ h_1, \ldots ,h_n \}$, which means that this linear
space approximates well all functions in $\cF$.

What ``well'' means depends on the approximation result used. If the number of parameters 
is finite and the dependence is analytic, then
exponential rates of the form $exp(-n^{1/d})$
can easily be proved. If the dependence is only $C^s$,
then a rate $n^{-s/d}$ can also easily be proved. This argument establishes, then, that there are bases in which expansions of the form (\ref{eq:conv_argument}) converge very fast, and as a consequence the n-width can only decay faster. 

To summarize, for problems with smooth parametric dependence, fast convergence in terms of greedy reduced bases can be expected and it is not surprising that such global methods outperform local ones.

We mentioned that the functions themselves can be discontinuous, which might be confusing so it is worthwhile explaining it in more detail, even if qualitatively. Consider a problem of fluid dynamics, where the solutions might develop shocks at a finite time, and imagine that one is solving the partial differential equations for a parametrized family of initial data. This is completely fine in terms of the fast convergence of the n-width, since the location, shape, etc., of the shock depend smoothly on the initial data; the time series (in this case) itself does not need to have regularity with respect to the physical variable (time in this example) but with respect to parameter variation. 

\break

\section{Reduced Basis} \label{sec:rb}

In this section we present the RB-greedy framework for parametrized systems in order to address the resolution of the Kolmogorov problem in a quasi-optimal way. First mathematical conventions related to the RB-scheme are presented, followed by a discussion of its near-optimality with respect to different $n$-width behaviors.

\subsection{Introduction}

In previous sections we have discussed linear approximation of functions by orthogonal projection onto the span of a basis. The latter was taken to be a generic, problem-independent one, such as Fourier modes or polynomials. The Reduced Basis (RB) approach is a framework for efficiently solving parametrized problems, representing the solutions in a compact way, and predicting new ones based on an offline-online decomposition.  

In parametrized problems one is interested in functions of the form 
$$
h_{\lam} = h_{\lam }( \cdot ) = h(\cdot ; \lam) \, . 
$$
where is (in general a multidimensional) parameter $\lam$, the discretization of which will define the training space. 

In fields related to scientific computing and data science one is usually interested in multiple evaluations and analyses of functions in real time. The approach of RB is especially tailored to that one in which some numerical problem has to be solved in order to obtain each function, and obtaining such numerical solutions is very expensive. The approach is also very powerful when large existing data sets are known and a sparse representation is needed and/or multiple, fast operations on them. 

Then a minimal, nearly optimal set of representative such solutions (waveforms, for the purpose of this review) is sought for in order to construct a reduced basis ({\tt rb}) for the whole solution set. These {\tt rb} solutions constitute an application-specific basis. That is, a set 
\be
\{h_{\Lam_i} \}_{i=1}^n \label{eq:rb-basis}
\ee
of functions in the space of interest for carefully chosen parameter values $\Lam_i$  is used as a basis itself, as opposed to generic basis such as polynomials or Fourier modes.

The scalar product in this parametrized space is taken at fixed parameter values; that is, with respect to the physical dimension(s), 
$$
\langle h_i , h_j  \rangle = \langle h_{\lam_i} , h_{\lam_j}  \rangle = \langle h({\cdot ; \lam_i}) , h({\cdot ; \lam_j})  \rangle \;  .
$$

The same results of Section~\ref{sec:spec} follow through for the parametrized and application-specific case. Namely, the optimal solution to the least-squares approximation is 
\be
h (\cdot; \lam) \approx {\cal P}_n h (\cdot ; \lam) = \sum_{i=1}^n c_i (\lam ) e_i (\cdot) \, ,  \label{eq:Pm}
\ee
with the coefficients $c_i$ given by the equivalent of Eq.~(\ref{eq:optLS}), in the physical dimension(s) and the reduced basis \ref{eq:rb-basis} relabeled as $\{e_1, \ldots, e_n\}$. Namely, 
\be
c_i(\lam) = \sum_{j=1}^n({\bf G}^{-1} )_{ij} \langle  e_j (\cdot) , h (\cdot; \lam) \rangle  \, , \label{eq:optLSlam}
\ee
where the entries of the Gramian matrix are 
$$
{\bf G}_{ij}:= \langle e_i  , e_j \rangle \, .  
$$
Clearly, the Gramian coefficients depend on the special parameters $\Lam_i$ associated to each basis element.

So far, the span of the reduced basis uniquely determines the {\tt rb} representation. This reduced space depends primarily on the choice of the selected parameter points in order to define a starting  basis. As of a convenient basis itself, for the same reasons discussed in Section~\ref{sec:spec}, from a conditioning numerical perspective, it is convenient to work with an orthonormal basis built out of the {\tt rb} solutions through a simple orthonormalization procedure. This does not change the span of the {\tt rb} and is numerically convenient. 

\subsection{The Training set}
\label{sec:training}

A commonly used approach to construct a basis, described below in Section~\ref{sec:greedy}, is through a greedy algorithm. In its simplest version, the algorithm identifies a set of $n$ points in parameter space out of a representative enough set of functions of interest that are actually known. We call this set of functions the {\it training set}:
\be
{\cK} := \{ h_{\lam_i} \}_{i=1}^N\,.
\ee
The subset $\{ h_{\Lam_i} \}_{i=1}^n$ constitutes a nearly optimal basis for application-specific spectral expansions of any function in $\cK$ in a precise sense discussed in Section~\ref{sec:near-opt}. If there is partial redundancy/similarity in the latter, then $n<N$ or even $n \ll N$. Note here that we have used the symbol $\cK$ to represent a discretization of the space of interest $\cF$ in order to perform actual computer calculations. 

The training set can be constructed by any means, including simple random or uniform sampling, more sophisticated stochastic methods \cite{Messenger:2008ta}, the metric approach \cite{Owen:1995tm}, or those of Ref.~\cite{Manca:2009xw}, for example. For large problems, resampling it while constructing the basis can be critical; for an application in the case of GWs see for example \cite{Blackman:2014maa}. Regardless of the method used to populate the training set, the RB-greedy formalism produces a compact and highly accurate representation of the training space catalog. The reduced basis is used to approximate other functions in the space of interest, whether they were in the training set catalog or not, through linear combinations that represent an orthogonal projection onto its span, 
\be
h_{\lam} = {\cal P}_n h_{\lam} + \delta h_{\lam } \, , \label{eq:approx}
\ee
where the RB approximation is ${\cal P}_n h_{\lam} $ and satisfies, by construction, $\langle {\cal P}_n h_{\lam}, \delta h_{\lam } \rangle=0$.

\subsection{Greedy algorithms}
\label{sec:greedy}

We mentioned that a commonly used way to generate a reduced basis is through a greedy algorithm. In its simplest form, such as when the waveforms are inexpensive to compute or the data is already somehow known, the greedy algorithm, outlined in Algorithm \ref{alg:Greedy}, has as input a discretization of the parameter and solution space,  
\be
{\cal T}:= \{\lam_i \}_{i=1}^N \quad {\cK} = \{ h_{\lam_i} \}_{i=1}^N  \label{eq:training}
\ee
with the elements of $\cT$ usually called {\it training points}. To put emphasis on structure instead of form, all functions in the training set in this review are  normalized. How to recover the norm, particularly for parameter estimation of any detected signal, is discussed in Section~\ref{sec:roq-construction}.

The scheme needs an arbitrary seed $\Lam_1 \in {\cal T}$ to initialize it, and a threshold error $\epsilon$ for a target representation accuracy (or {\it greedy error}). Part of the output of the algorithm is a sequential selection of $n$ parameter {\it greedy} points 
$$
\{\Lam_1,\Lam_2,\ldots, \Lam_n  \} \subset {\cal T}
$$  
and their associated waveforms 
$$
\{h_{\Lam_1},h_{\Lam_2},\ldots, h_{\Lam_n} \} \subset \{ h_{\lam_i} \}_{i=1}^N\,.
$$
The set of waveforms $\{h_{\Lam _i}\}_{i=1}^n$ constitutes the {\em reduced basis}. As was already discussed, for numerical conditioning purpose it is sometimes convenient (see Eq.~(\ref{eq:projection-LS})) to work with an orthonormalized set $\{ e_i \}_{i=1}^n$ instead of directly the $\{ h_{\Lam_i} \}_{i=1}^n$. 

Another output of the algorithm is precisely the set of projection coefficients for functions in the training space catalog, whereas coefficients for any other known function can be computed through projection onto the basis. In Section~\ref{sec:eim} we discuss how to approximate these coefficients through interpolation in the physical dimension (time in the case of gravitational waveforms), and in Section~\ref{sec:surrogates} a predictive approach for accurate and fast evaluation of new (unknown) functions in $\cF$ through {\it surrogate models}. These are predictive models: as opposed to a known function being projected into a compact basis, they {\em predict} new solutions (waveforms). 

\hspace{0.5cm}

{\scriptsize
\begin{algorithm}[H]
\caption{Greedy algorithm for reduced basis}
\label{alg:Greedy}
\begin{algorithmic}[1]
\State {\bf Input:} $ \{ \lam_i \, , h(\cdot;\lam_i) \}_{i=1}^N$,  $\epsilon$ 
\vskip 10pt
\State Initialize $i=0$ and define $\sigma_0 = 1$
\State {\bf Seed choice} (arbitrary):  $\Lam_1 \in {\cal T}$, $e_1=h(\cdot; \Lam_1)$
\State {\tt rb} = $\{e_1\}$
\While{$\sigma_i \geq \epsilon$}
\State $i=i+1$
\State $\sigma_i = \max_{\lam\in\cT}\| h(\cdot;\lam) - {\cal P}_{i} h(\cdot;\lam) \|^2$
\State $\Lam_{i+1} = \text{argmax}_{\lam\in\cT}\| h(\cdot;\lam) - {\cal P}_{i} h(\cdot;\lam) \|^2$ 
\State $e_{i} = h(\cdot;\Lam_{i+1}) - {\cal P}_{i} h(\cdot;\Lam_{i+1})$ ~(Gram-Schmidt)
\State $e_{i+1} = e_{i+1} / \| e_{i+1} \|$ ~(normalization)
\State {\tt rb} = {\tt rb} $\cup \, e_{i+1}$
\EndWhile
\vskip 10pt

\State {\bf Output:} {\tt rb} $\{ e_i \}_{i=1}^n$ and greedy points $\{ \Lam_i \}_{i=1}^n$
\end{algorithmic}
\end{algorithm}}

\begin{comment}
\begin{itemize}

\item Greedy-type algorithms are global optimization procedures used in contexts outside reduced basis or even dimensional reduction.
\item Being a global optimization algorithm, the choice of the seed $\Lam_1$ is largely irrelevant. This was explicitly discussed as a sidenote in Ref.~\cite{Caudill:2011kv}, see Fig.~\ref{fig:seed-choice} below. We also refer to Section~\ref{sec:ringdown} for a discussion of Ref.~\cite{Caudill:2011kv} in the context of RB for multi-mode black hole ringdown.
\item What the greedy algorithm does in Step 8 is to select the waveform for which its representation error onto the existing basis with $i-1$ elements is  {\em worst}, and in Step 11 adding it to the enrichment of the  {\tt rb} representation. 
\item In steps 9 and 10 the {\tt rb} waveforms are orthonormalized to avoid ill-conditioning of the computation of the projection (see the discussion in Section~\ref{sec:spec}, before \ref{sec:spectral}). 
\item Given an arbitrary user-defined tolerance error $\epsilon$, the algorithm stops when the approximation (\ref{eq:Pm}) meets the error tolerance -- introduced as an input in the greedy algorithm --,
$$
\| h_{\lam} - {\cal P}_n {h_{\lam}} \|^2 \leq \epsilon \,\,\,\, \forall \,\, \lam \in {\cal T}.  
$$
\item The expected exponential convergence of the method for the problems of interest implies that $\epsilon$ can be made arbitrarily small with a relatively small number $n$ of basis element, with $n < N$ and, in many cases, $n \ll N$.
\end{itemize} 
\end{comment}

\begin{figure}[h]
\begin{center}
\includegraphics[width=0.6\linewidth]{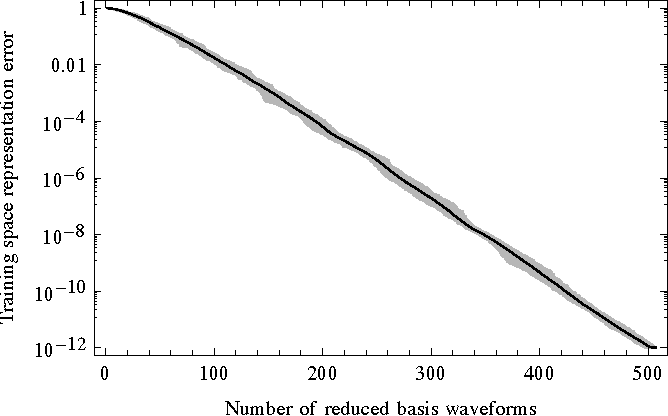}
\caption{Taken from \cite{Caudill:2011kv}. This figure shows the representation error as a function of the number of reduced basis waveforms for a single QNM  catalog. The authors iterate over all possible seed waveforms in the training set. The dark line represents the average, and the shaded area the maximum dispersion around it. This numerical experiment confirms that the seed choice becomes nearly irrelevant due to the global nature of the the greedy algorithm at each step, as intuitively expected being a global optimization.}.
\label{fig:seed-choice}
\end{center} 
\end{figure}

In practice, one uses a better conditioned orthogonalization algorithm in step $9$ than the standard (i.e., ``classical'') Gram-Schmidt one.  
The naive implementation of the classical Gram-Schmidt procedure is actually ill-conditioned. This is related to the fact that the Gramian matrix, which would have to be inverted, can become  nearly singular \cite{taylor_1978}. To overcome this one can use an iterated Gram-Schmidt algorithm or a QR decomposition. A popular alternative to the classical Gram-Schmidt, largely used, and called the {\it modified} Gram-Schmidt, is also ill-conditioned, contrary to common perception since its ill conditioning only becomes apparent for large large data sets. See \cite{Ruhe1983591,Giraud,stewart_gram_schmidt} for discussions about the conditioning and numerical stability of different orthonormalization procedures.

\blank

\noindent{\bf Error definitions}. We define the {\it greedy error} of the {\tt rb} $\{{e_i}\}_{i=1}^n$ as
\begin{align} \label{eq:greedyErr}
\sigma_n := \max_{\lam} \bigg\| h_{\lam} - {\cal P}_n h_{\lam} \bigg\|^2 \leq \epsilon  \, ,
\end{align}
where $\epsilon$ is the user-defined tolerance error, which depends on the number of basis elements.

The quantity $\sigma_n$ represents the largest error in the parameter space of the best approximation by the greedy-reduced basis. 

As discussed in \cite{Field:2011mf}, in the limit of sufficiently dense training spaces the greedy error is comparable to the minimal match ($\MMm$) through 
\be
\sigma_n \sim 1 - \MMm {\rm ~~as~~} N \rightarrow \infty. 
\ee
Since the RB framework allows us to compress the information presented in the training set, it is useful to introduce the quantity 
\be
C_r:= N/n \,,\label{eq:compression}
\ee
called the {\it compression ratio} \cite{Salomon2010}, to measure it.\\

\subsubsection{Convergence rates and near-optimality} \label{sec:near-opt}
The greedy algorithm chooses, in a precise sense, a nearly-optimal basis for the function spaces $\cK$ or $\cF$, depending on whether one is discussing the discrete or continuum cases. In order to quantify this near-optimality, recall the definition of the $n$-width from Section \ref{sec:optimal}, Eq.~(\ref{eq:nwidth2}), but write it in a more geometric way:
\begin{align} \label{eq:nwidth}
d_n = d_n(\cF, \cH) = \min_{W_n \in \cH} \max_{\lam } \bigg\| h_\lam - {\cal P}_n h_\lam  \bigg\|^2   \,.
\end{align}

The search space here is the whole Hilbert space and one looks for an optimal $n$-dimensional subspace $W_n$ for approximating the function manifold $\cF$. As was stated in Section~\ref{sec:optimal}, $d_n$ is a theoretical upper bound to any practical algorithm to perform a linear approximation. In practice, solving such optimization problem becomes unfeasible due to its intrinsic combinatorial complexity. The nested nature of the greedy algorithm becomes crucial for reducing the complexity of the Kolmogorov problem. This means that each $H_m={\tt Span}\{e_i\}_{i=1}^m$  satisfies $H_1\subset H_2\subset\ldots\subset H_n$ and this feature dramatically reduces the search space in~(\ref{eq:nwidth}). One expects that this reduction of the search space is at the expense of losing completely -- if not a modicum -- the $n$-width optimality but, as we discuss in the next paragraph, this expectation is the opposite from being true.

In which sense does the reduced basis-greedy procedure degrade the optimality in the Kolmogorov sense? To answer this, lets summarize two important results in relation with the greedy algorithm: if $d_n$ decays exponentially with $n$ then so does $\sigma_n$ \cite{DeVore2012},
\be
d_n \leq D \mathrm{e}^{-a n^\alpha}  \implies \sigma_n \leq \sqrt{2D}\gamma^{-1}\mathrm{e}^{ -a'_{\alpha} n^{\alpha}} \, , \label{eq:exp}
\ee
where $D$, $a$, $\alpha$ are positive constants.

Similarly, if the n-width has polynomial decay then so does the greedy error,
\be
d_n \leq D n^{-\alpha} \implies \sigma_n \leq D'_{\alpha}n^{-\alpha} \,,\label{eq:poly}
\ee
where $D,\alpha>0$.
More generally, for any decay rate of the n-width, 
\be
\sigma_n \leq 2\gamma^{-1}d_{n/2} \,.\label{eq:genwidth}
\ee
The factor $\gamma$ in Eqs.~(\ref{eq:exp}) and~(\ref{eq:genwidth}) is a constant in $(0,1]$ \cite{DeVore2012}. In recent years there have been efforts to improve these bounds. For details see \cite{WOJTASZCZYK2015685, NGUYEN2020105344}.

In light of these results, we see that the reduced basis-greedy procedure inherites the optimality of the $n$-width. If the latter has exponential convergence, so it does the greedy error. In this precise sense the reduced basis-greedy approach is nearly-optimal: we cut down the complexity of the Kolmogorov problem at the very low expense of losing quality which, in most practical applications, becomes insignificant.  

The convergence rate of the $n$-width (and, in consequence, of the greedy error) depends on the parametric smoothness/regularity of the functions/waveforms. Indeed, if the functions are analytic and can be extended to a complex region, exponential decay for the greedy error can be proven (see \cite{quarteroni2015reduced}, section 5.5, and citations therein). Therefore, the {\tt rb} expansion is expected to have very fast convergence to the original waveform as the number of basis elements is increased. This is similar in spirit to the standard fast convergence of spectral methods, but here smoothness in the parameters of the problem is exploited, and the basis are elements themselves of the space of functions of interest. For this reason RB is sometimes referred to as an {\it application-specific spectral expansion}. 

To summarize, for problems with smooth parametric dependence, fast convergence in terms of greedy bases can be expected and it is not surprising that such global methods outperform local ones. 
\begin{comment}
Not only the choice of basis is important, but a also global expansion. To illustrate this consider again the example of the space of smooth periodic functions from section \ref{sec:fourier}. If one is interested in a family of them, and is able to choose $n$ Fourier modes to represent them as best as possible, one could:
\begin{itemize}
\item build a discretization of the space of periodic functions and choose $n$ Fourier modes which when compared against the whole training set gives the best global approximation error; this approach is a global variation of the well known {\it best m-term approximation} problem, in which one is interested in approximating some function and wants the best set of $m$ elements chosen from a dictionary of functions to do it;
\item follow the standard approach, in which the functions of interest are projected onto the span of the first $n$ Fourier modes. If the functions of interest are smooth, this approach gives exponentially decaying representation errors with the number of Fourier modes. This is, of course, not a predictive model, since the function to be projected has to be known.
\end{itemize}
The first path would return a global and problem-dependent basis by construction. The second one, on the contrary, will return a standarized basis with no further feeling for the problem than the minimization of the maximum error due to the projection. 
\end{comment}

\subsubsection{Complexity, scaling, computational aspects}
Besides its near-optimality the greedy algorithm has several implementation advantages:
\begin{enumerate}
\item It is simple to implement (see Algorithm~\ref{alg:Greedy}).
\item It follows a stricter aproximation criteria (see section \ref{sec:best_space}) than POD/SVD, namely, the minimization of the worst error, as opposed to the average one,  in parameter space due to the projection onto the {\tt rb} basis. It provides a stricter error control. 
\item It is embarrassingly parallel. Each {\it greedy sweep}, defined by Step 7 of Alg.~\ref{alg:Greedy}, can be carried out simultaneously and independently for different values of $\lam \in \cT$.
\item As we remarked in Sec.~\ref{sec:near-opt}, the greedy points and associated basis are hierarchical (nested). This is particularly important if the algorithm is to be used to guide which numerical relativity simulations should be carried out. Since they are so expensive, it is desirable that if further accuracy is needed, then it should be possible to achieve it by adding more points (as opposed to starting the process from scratch).
\item Low complexity. That is, the cost of all steps 6-12 in Alg.~\ref{alg:Greedy} are independent of $i$, leading to a total cost linear in $n$, the total number of basis elements. This observation follows from the fact that 
\be
{\cal P}_{i+1}h_{\lam} = {\cal P}_{i}h_{\lam } + \langle e_{i+1} , h_{\lam }\rangle e_{i+1}\,. \label{eq:projec_reuse}
\ee
So, if the projection coefficients are stored while building the basis, at each sweep only the dot product between the elements of the training set and the {\em last} basis element has to be computed. 
\end{enumerate}

\subsection*{Further reading}
A comprehensive review of reduced basis is \cite{quarteroni2011}. For textbooks see \cite{Jan-RB} and \cite{quarteroni2015reduced}. 

\part{Predictive Models}

\newpage

\section{Polynomial interpolation} \label{sec:interp}

There are several reasons why we find it useful to discuss classical polynomial interpolation:
\begin{itemize}
\item It is used in ROM for gravitational wave physics and other fields; we discuss some advantages and disadvantages. 
\item There is a close relationship between interpolation, quadrature rules, spectral methods and Reduced Basis. It is easier to introduce the standard theory first, followed by the application specific version at the heart of ROM for parametrized systems.
\end{itemize}
The emphasis of this section is on approximation quality quantified through convergence rates and error bounds. Except for some side comments pointing out implementations known to be poorly conditioned, we shall not discuss numerical algorithms or their implementation which can be found in  standard references.

\subsection{Representation versus prediction}
\label{sec:prov_vs_pred}
In standard approximation theory, discussed in Sections~\ref{sec:low_rankd} and \ref{sec:spec}, one expands a function in some basis. Assuming that the approximation quality is measured by some weighted $L_2$ norm, the optimal approximation is the orthogonal projection onto the representation space. This projection is independent of the basis, in the sense that it only depends on its span. Assuming these are orthonormal for simplicity and conditioning purposes, the orthogonal projection is given by 
\be
h(x) \approx {\cal P}_n h (x) = \sum_{i=1}^n \langle e_i, h \rangle e_i(x)\, . \label{eq:l2_proj}
\ee 

Approximating $h$ by its projection as in (\ref{eq:l2_proj}), while perhaps useful by itself, is not a {\em predictive} tool: we need to know the values of $h(x)$ for all $x$ in order to calculate the projection coefficients in Eq.~(\ref{eq:l2_proj}). Here, by prediction it is meant a new evaluation within the range used to build the approximating model. This means, for example, within the same parameter and physical (time/frequency for example) ranges in the case of GWs. We are not dealing with a much more difficult problem: namely, that one of prediction of time series, for example, based on previous history. Then, to get a predictive power we need to first turn to an interpolation scheme. 

Building an interpolant for $h(x)$ requires only the knowledge of $n$ values of $h(x)$ whereas the unknown values of the function are predicted from this interpolant. As previously mentioned, the applications are many, including numerical schemes to solve partial differential equations, the numerical computation of integrals and building predictive surrogate models.  As we will see in Section~\ref{sec:project_or_interpolate} these apparently radically different approaches for approximation, such as interpolation and projection, are closely related, and in some cases both are equivalent within certain frameworks.

\subsection{Prediction through polynomial interpolation}
\label{sec:pol_interpolation}

Suppose we have a collection of $(n+1)$ different points $\{ x_i \}_{i=0}^n$\ft{Note that we count from $i=0$, only in this Section, in order to match our notation with the common definition of Lagrange polynomials.}, and a set of associated function values $\{ f_i := f(x_i)\}_{i=0}^n$. We assume that $f$ is a real valued, one-dimensional function. We will discuss the more challenging multivariate case in Section \ref{sec:multivar}. 

The interpolation problem consists of using the partial function sampling $\{ f_i \}_{i=0}^n$ to approximate $f(x)$ by another function which agrees with $f$ at the interpolating {\it nodes} $\{ x_i \}_{i=0}^n$. To be concrete, let's focus on linear interpolation: we require the interpolant of $f$ to be expressible as a linear combination of the $(n+1)$ independent basis functions $\{ e_i(x) \}_{i=0}^n$,
\be
\cI_n[f](x):= \sum_{i=0}^nC_i e_i(x) \, .  \label{eq:interp1}
\ee
We already discussed in section \ref{sec:spec} that the optimal least-squares-sense approximation of the above form is provided by orthogonal projection onto the basis  $\{ e_i(x) \}_{i=0}^n$. In the interpolation scheme one trades the least-squares criteria for one which requires less information about $f$. Let's define this criteria by making the interpolant being equal to the function at the nodes, 
\be
\cI_n[f](x_j) = f(x_j) \, , \quad j=0,\ldots, n \, . \label{eq:interp2}
\ee
When the basis are linearly independent polynomials of degree $\leq n$ we call this special case {\it Polynomial interpolation}. One can show by explicit construction that there is a unique polynomial which satisfies the interpolation problem (\ref{eq:interp1},\ref{eq:interp2}). Before doing so we give an example.

\begin{example} \label{ex:interp1}

Suppose we have two points, $\{ x_0,x_1\}$, sampling a function $f(x)$. Let the sample values be $f_0 = f(x_0)$ and $f_1 = f(x_1)$. The functions $\{1, x\}$ comprise a suitable basis for our $n=1$ polynomial interpolant. Then the interpolating polynomial is of degree $\leq 1$, i.e.~a straight line joining the two points:
$$
\cI_1[f](x) = ax + b
$$
We explicitly solve for $(a,b)$ by writing the interpolation problem ~\eqref{eq:interp2}
\begin{eqnarray}
\cI_1[f](x_0) &=&  ax_0 + b = f_0 \\
\cI_1[f](x_1) &=&  ax_1 + b = f_1 
\end{eqnarray}
which has as solutions
\be
a = \frac{f_0 - f_1}{x_0-x_1} \,\, , \,\, b= \frac{x_0f_1 - x_1f_0}{x_0-x_1} \,.
\ee
Thus, 
\be
\cI_1[f](x) = \left( \frac{f_0 - f_1}{x_0-x_1} \right)  x + \left(  \frac{x_0f_1 - x_1f_0}{x_0-x_1} \right)  \label{Ex1} \, . 
\ee
\end{example}

\noindent {\bf General solution: Lagrange polynomials}

\blank
\noindent{\bf Existence}. The existence of a polynomial which solves the interpolation problem~\eqref{eq:interp2} can be shown by explicit construction. Suppose we have access to a basis $e_i(x) = \ell_i^{(n)} (x)$ which satisfies 
\be
\cI_n[f](x) = \sum_{j=0}^n f_j l_j^{(n)}(x) \label{Lag_int}
\ee
where, for each $j=0\ldots n$, $l_j^{(n)}(x)$ is a polynomial of degree $\leq n$ such that
\be
l_j^{(n)}(x_i) = \delta_{ij} \,\,\, \mbox{for } i=0\ldots n \label{Lag_pol_def}
\ee
and $\delta_{ij}$ is the Kronecker delta ($\delta_{ij} = 1$ if $i=j$ and zero otherwise). Then $\cI_n[f](x)$ as given 
by Eq.~(\ref{Lag_int}) would be a solution to the interpolation problem. 

The {\it Lagrange polynomials} are precisely those with the property \eqref{Lag_pol_def}. They are given by 

\be
l_j^{(n)}(x) = \left( \prod_{k=0,k \neq j}^n (x-x_k) \right)/ \left(  \prod_{k=0,k \neq j}^n (x_j-x_k) \right) \, . \label{eq:lag_polynom}
\ee

\begin{example}

We look at the same case ($n=1$) of Example~\ref{ex:interp1}, but now we construct an interpolant through Lagrange polynomials. By Eq.~(\ref{eq:lag_polynom}) these are
\be
\l_0^{(1)}(x) = \frac{x-x_1}{x_0-x_1} \;\;\; , \;\;\; \l_1^{(1)}(x) = \frac{x-x_0}{x_1-x_0} \, , 
\ee
and using (\ref{Lag_int}) the Lagrange form of the interpolant becomes
\bea
\cI_1[f](x) &=& f_0 l_0^{(1)}(x) + f_1 l_1^{(1)}(x) \nonumber \\
& = & \left( \frac{f_0-f_1}{x_0-x_1} \right) x + \left( \frac{x_0f_1 - x_1f_0}{x_0-x_1} \right) \label{Ex2}
\eea
The result (\ref{Ex2}) is the same as (\ref{Ex1}) due to the uniqueness of the polynomial interpolant, which we now prove.
\end{example}

\noindent{\bf Uniqueness}. Suppose there are two polynomials $P_n$ and $R_n$ of degree $\leq n$ satisfying the interpolation condition $P_n(x_j)=R_n(x_j)=f_j,\, j=0, \ldots, n$. The polynomial $Q(x):=P_n(x)-R_n(x)$ then satisfies $Q(x_j)=0,\, j=0, \ldots, n$. This means that $Q(x)$ has $(n+1)$ zeros, in contradiction with the hypothesis that it is at most of degree $n$. Ergo, $Q(x)=0$ and $P_n(x)=R_n(x)$. 

\blank
\noindent{\bf The Vandermonde form}

Yet another way of showing existence and uniqueness for the polynomial interpolant in a constructive way is the following. We write down the polynomial interpolant using monomial basis functions, 
$$
\cI_n[f](x) = \sum_{j=0}^n a_j x^j\, . 
$$
The interpolation conditions (\ref{eq:interp2}) become a set of linear equations for the coefficients $a_j$
\be
{\bf V} a = f \label{van} \, , 
\ee
where ${\bf V}_{ij} = x_i^{j-1}$, $a=[a_0,\ldots, a_n]^\top$ and $f=[f(x_0),\ldots, f(x_n)]^\top$. The matrix ${\bf V}$ is called the {\it Vandermonde} matrix and its determinant, also called the {\it Vandermonde polynomial}, takes the form 
$$
\det({\bf V}) = \prod_{0\leq i< j  \leq n}(x_j-x_i) \, .
$$
In order to have one and only one solution to Eq.~(\ref{van}) this determinant has to be non-zero. In one spatial dimension this condition is automatically satisfied whenever the interpolation points are unique. There are ways of inverting ${\bf V}$ in order to solve Eq.~(\ref{van}) using Lagrange polynomials, which we will not pursue here. The main reason for briefly mentioning the Vandermonde matrix is because a generalization of it will play a prominent role in the Empirical Interpolation Method, described in Section~\ref{sec:eim}.

\subsection{Convergence rates, collocation points and Runge's phenomenon} \label{sec:runge}
Here we discuss the behavior of the error of the polynomial interpolant $\cI_n[f](x)$ of some function $f(x)$ at $(n+1)$ nodes, to illustrate the difficulties in obtaining an accurate interpolation scheme on a unstructured set of nodes which, from the point of view of ROM, one wishes to be sparse. 

Leaving aside both sparsity and unstructured meshes, one might imagine that for ``nice enough'' functions (for example, infinite differentiable), this error decreases as the number of nodes and, therefore, degree of the polynomial basis $n$ increase. This is not the case, the classical counterexample is Runge's one, here described, and the overall problem is referred to as {\it Runge's phenomenon}. 

The distinction is between {\em local} approximations of fixed degree and {\em global} ones. Because these two concepts play a key role in reduced order modeling we discuss them in this context with some detail.  Suppose that one has $\{ x_i \}_{i=0}^n$ nodes at which to interpolate a function. 

\noindent{\bf Local interpolation}

We consider local interpolants of degree one because we have already explicitly discussed them in Example~\ref{Ex1}. One would partition the interval of interest, where on each subinterval 
\be
I_j:= [x_j, x_{j+1}]\,, \quad  j=0, \ldots , n-1 \label{eq:subinterval}
\ee
one builds a local interpolant of degree one using the end points as interpolating nodes, 
\be \label{eq:subinterval_interp}
\cI[f](x) = \left( \frac{f_j - f_{j+1}}{x_j-x_{j+1}} \right)  x + \left(  \frac{x_jf_{j+1} - x_{j+1}f_j}{x_j-x_{j+1}} \right)  \,, \quad \text{for } x\in I_j\, ,  \quad j=0, \ldots, n-1 \, . 
\ee
Such fixed order interpolants, where their degrees are independent of the number of nodes $n$, is -- as explained later below -- guaranteed to converge to the interpolated function as the number of nodes in a given fixed physical interval increases (the gridspacing decreases). Furthermore, such convergence is guaranteed regardless of the structure/location of the interpolating nodes, they can even be randomly located. Even more, to some degree, convergence is also guaranteed regardless of the smoothness of the data; this is why fixed order methods, especially very low order ones, are in general used for noisy data. 

This class of methods is usually very robust, but at the expense of accuracy. The latter is exacerbated in the absence of a dense amount of data. On the contrary, in Reduced Order Modeling one seeks very sparse representations of very high accuracy. This combination of, and even trade-off between, robustness (convergence) and accuracy is a delicate and challenging issue in ROM.

\noindent{\bf Global interpolation}

\begin{figure}
\begin{center}
\includegraphics[width=0.7\linewidth]{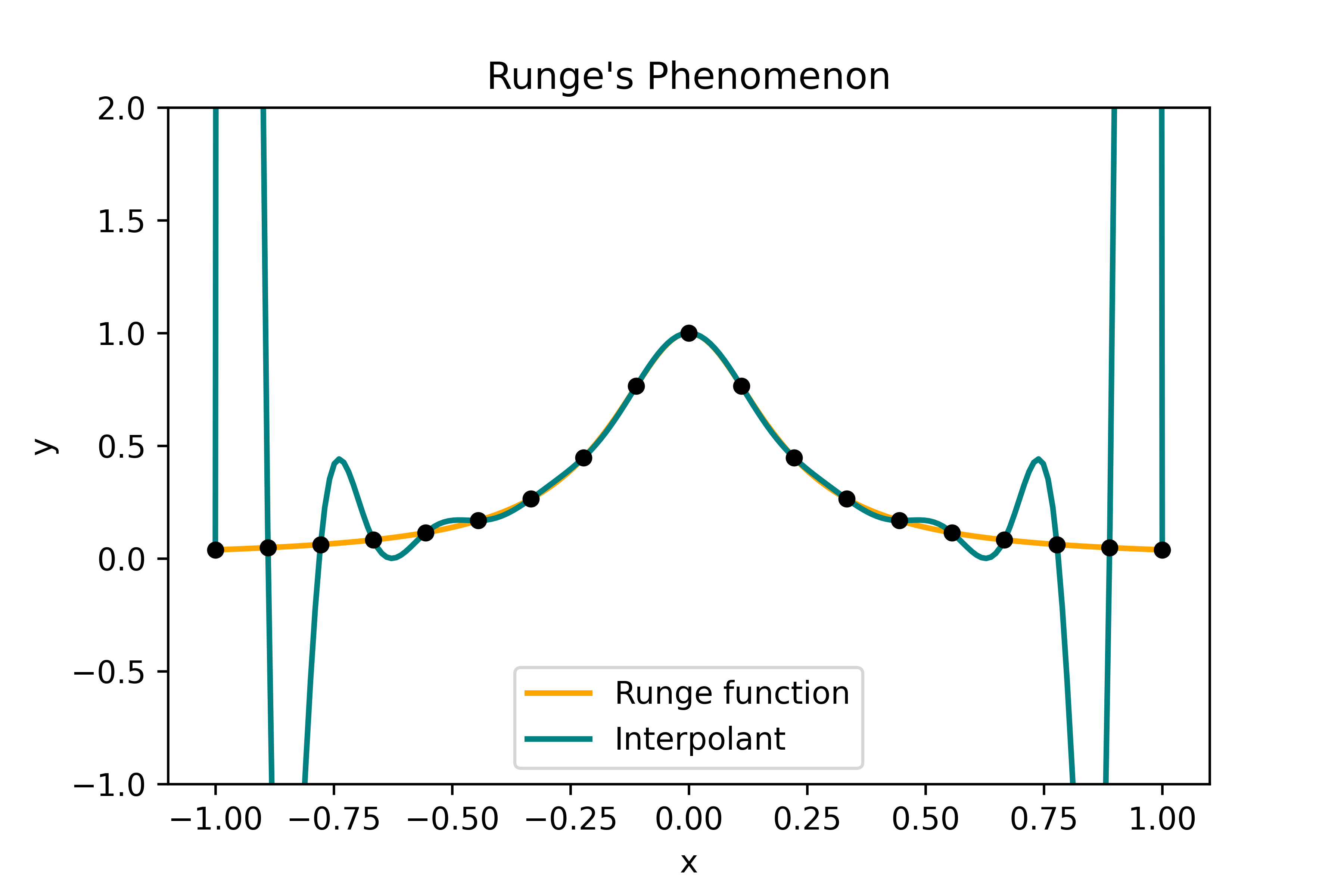}
\caption{Runge's phenomenon in polynomial interpolation. In this example the approximation uses 19 equispaced nodes over the interval $[-1, 1]$ and the interpolant polynomial is of degree 18. The interpolation tends to diverge near the boundaries. This issue cannot be solved by increasing the resolution of the sampling but the opposite: divergencies get worse for equispaced nodes.}
\label{fig:runge}
\end{center} 
\end{figure}

Instead, one could use a global interpolant, where all $(n+1)$ nodes and function values are used to build a single interpolant, 
$$
\cI_n[f](x) = \sum_{i=0}^n \ell_i ^{(n)} (x) f_i \, . \quad \forall x \in [x_0, x_n]\, .  
$$
As the density of points and $n$ increases, one would keep using all of them, thereby increasing the degree of the interpolant. This might seem desirable and more accurate than a local approach of fixed degree, since more information is used. The classical counterexample is due to Runge~\cite{Runge1895}, and it consists of interpolating the function 
\be
f(x) = \frac{1}{1+(5x)^2} \; \, , \quad x\in[-1,1] \,  . \label{eq:runge}
\ee
This function is $C^{\infty}$ (has infinite derivatives) yet the error of a global polynomial interpolant at equally spaced points {\em diverges} near the boundaries~\cite{10.2307/2323093} at higher orders (see Fig.~\ref{fig:runge}). For unstructured points, such as those usually appearing in reduced models, the lack of convergence of global approximations is in general worse. 
As discussed below, this can be resolved and fast convergence achieved if the interpolation nodes can be appropriately chosen. However, standard choices for such special nodes are in general not suitable for reduced order modeling. In particular, they are neither necessarily sparse nor adapted to the problem of interest. They are generic and in general (and most important) not hierarchical, requiring recomputing from scratch the sought approximation if higher accuracy is required. 
\blank

\noindent{\bf Error analysis and convergence I}

In what follows we assume that $f$ is smooth enough as needed from the context. Denoting the interpolation error by 
$$
E_n(x) := \left | f(x)- \cI_n[f](x) \right |
$$
the following result holds: 
\be
E_n (x) =  \left | \frac{1}{(n+1)!} f^{(n+1)}(\xi) \omega _{n+1}(x) \right | \label{eq:error_interp}
\ee
where 
$$
\omega _{n+1}(x) := \prod_{i=0}^n (x-x_i)
$$ 
is called the nodal polynomial of order $(n+1)$,  and $\xi$ 
is in the {\em smallest} interval $I_x$ containing $x_0 \ldots x_n$ and $x$. 
Equation~(\ref{eq:error_interp}) can be proven as follows. Let's define the function
$$
F(t)=f(t)-\cI_n[f](t)-\alpha(x) \omega_{n+1}(t)
$$
where
$$
    \alpha(x)=\frac{f(x)-\cI_n[f](x)}{\omega_{n+1}(x)}
$$
This function is well defined since $x\neq x_i$, and it is as smooth as $f(t)$. Observe that $F(x)=0$ and, in particular, $F(x_i)=0$ ($i=1,...,n$). This means that $F(t)$ has $(n+2)$ distinct zeros in the interval in which 
$f(t)$ is smooth. 
Now, the Mean Value Theorem tells us that, between any couple of zeros, there is one zero for the derivative $F'(t)$. Then the function $F'(t)$ has $(n+1)$ zeros. Repeating the previous argument, 
we conclude that $F^{(n+1)}(t)$ has exactly one zero. Let's call it $\xi$. Thus $F^{(n+1)}(\xi)=f^{n+1}(\xi)-\alpha(\xi)(n+1)!=0$. The statement 
in (\ref{eq:error_interp}) follows directly from this relation.

Let's make a comment here. The point $x$ at which the interpolant is evaluated when computing the error (\ref{eq:error_interp}) does {\em not} need to be in the smallest interval containing $x_0\ldots x_n$. Assuming an ordering $x_0<\ldots<x_n$, sometimes the process of approximating $f(x)$ by $\cI_n[f](x)$ is called interpolation only if $x\in [x_0,x_n]$ and {\it extrapolation} otherwise.

We get back to the error formula (\ref{eq:error_interp}) and analyze its behavior as $n$ increases. In  particular, we ask ourselves under what conditions does 
$$
E_n(x) \rightarrow 0 \;\;\; \mbox {as} \;\;\; n \rightarrow \infty \, . 
$$
The error (\ref{eq:error_interp}) can be decomposed in two terms, one related to the behavior of the derivatives of $f$,
$$
\left | \frac{f^{(n+1)}(\xi)}{(n+1)!} \right |
$$
and another one related to the distribution of nodes through $\omega _{n+1}(x)$. 

We first discuss why fixed order interpolants are guaranteed to converge, regardless of the distribution of nodes (assuming the interpolated function is smooth enough), since the argument is simple. 
\begin{discussion}[Local interpolants]

For definiteness consider the case of equally spaced points (the generalization to an arbitrary set is straightforward), 
$$
x_j = x_0 + j \Delta x\, , \quad j=0, \ldots n\, .
$$
Then the error of the interpolant (\ref{eq:subinterval_interp}) on each subinterval (\ref{eq:subinterval}) is bounded by 
\be
E \leq  \frac{1}{(m+1)!} \max_{x\in I_j} \left |  f^{(m+1)} (x) \right | (\Delta x)^m   \, ,  \label{eq:bound_interp_error}
\ee
where $m$ is the (fixed) number of interpolation nodes and degree of the interpolants (clearly, $m\leq n$). Since $m$ is fixed, the convergence rate is determined by $\Delta x \rightarrow 0$, and is of order $m$. 
\end{discussion}

\begin{discussion}[Global interpolants] \label{disc:global_interp}

In this case $m=n$. One can see from the interpolation error (\ref{eq:error_interp}) that convergence is guaranteed if the derivatives of $f$ decay with $n$ sufficiently fast. In fact, the interpolant converges in the infinity norm if all derivatives $f^{(s)}$ are bounded by the same constant in the interval of interest. However, it is not obvious from (\ref{eq:error_interp}) to find out in a straightforward way the allowed growth rate of the derivatives of $f$ such that convergence/divergence of the error takes place. 

In fact,  it is not easy to see how such error formula explains Runge's phenomenon without resorting to complex calculus~\cite{10.2307/2323093}. However, we can ask ourselves the next related questions:

({\bf Q1}) What can be said about the maximum of $|\omega _{n+1}(x)|$ in the interval of interest?

({\bf Q2}) If the physical problem allows such freedom, how can one choose the nodes so as to minimize such maximum?

For simplicity and without loss of generality we focus on the interval $x\in [-1,1]$. We can answer these questions as:

Answer to ({\bf Q1}): For all choices of nodes
\be
\max _{[-1,1]} | \omega _{n+1} (x) | \geq 2^{-n}  \label{monic}
\ee	

Answer to ({\bf Q2}):
If 
$$
\omega _{n+1}(x) = 2^{-n}T_{n+1}(x)\,,
$$
where $T_{j}$ denotes the {\it Chebyshev polynomial} of degree $n$, then 
$$
\max _{[-1,1]} | \omega _{n+1} (x) | = 2^{-n} \, . 
$$
In other words, using the roots of the Chebyshev polynomial as interpolating nodes minimizes the error 
associated with the location of such nodes. It is in this sense that Chebyshev nodes are many times referred to as optimal points for interpolation. They have the disadvantage, though, that they are note hierarchical/nested, or adapted to the problem of interest.  The former failure means, from a ROM perspective, that any model relying on them, including perhaps expensive numerical simulations, have to be completely built from scratch if higher accuracy is required. This is solved by a nearly-optimal set of nested nodes for application-specific interpolants designed with ROM problems in mind, discussed in Section~\ref{sec:eim}. 
\end{discussion} 
\blank

\noindent{\bf Error analysis and convergence II}

In principle, direct approximation through projection requires the knowledge of the continuum waveform, while interpolation only needs information at a (specified by the user) set of nodes. We next address what is the relationship between these two strategies. In terms of representation accuracy, we focus on the $L_2$ norm, since it is directly related to the overlap error commonly used in data analysis. Since, as discussed, given a basis and linear approximation, projection is optimal in the least squares sense, one knows that 
\be
 \| h - {\cP}_n h \|^2\leq \| h - \cI_n [h] \|^2  \, .  \label{eq:proj_vs_interp}
\ee
The question is how ``sub-optimal'' in the above sense interpolation is. The answer obviously depends on the basis and interpolation nodes. A general framework to study this is through the definition of the Lebesgue constant. 

For any arbitrary basis and interpolation scheme one can derive the more precise bound related to (\ref{eq:proj_vs_interp}): 
\be
\| h  - {\cal I}_n [h] \|^2 \leq \Lam_n  \| h - {\cP}_n h \| ^2\, ,  \label{eq:bound352}
\ee
where 
\be \label{eq:leb_AppB} 
\Lambda_n  := \| {\cal I}_n \| ^2 \, . 
\ee

\begin{comment}
\begin{enumerate}
\item The $\Lam_n$'s are referred as the {\it Lebesgue constants} and depend, given a vector norm, purely on the interpolation operator ${\cal I}_n$.
\item In terms of accuracy, the strategy is to build an interpolant (by choosing both the basis and nodes) such that the Lebesgue constant grows as slow as possible with $n$. It is a computable quantity, so one can actually quantify how much is ``lost'' with respect to projection through Eq.~(\ref{eq:bound352}). 
\item Strictly speaking, the Lebesgue constant is usually referred to in the context of the infinity norm, 
$$
\| u \|_{\infty}:= \max_i | u_i | \, , 
$$
but throughout this review, unless otherwise stated, we refer to it in the $L_2$ norm. 
\end{enumerate} 
\end{comment}

As an example, and going back to Runge's phenomenon at equally spaced versus Chebyshev nodes for global polynomial interpolation, the following illustrates the advantages of the latter and explains why it is such a popular choice -- when it is a feasible approach at all within the problem of interest, which is not always the case in practice.

\begin{example} \label{ex:leb}

Here  the $\| . \|_\infty$ norm is used. For large $n$ one has the following behavior (see, for example, \cite{hesthaven2007spectral}).
\begin{enumerate}
\item For equally spaced nodes, the Lebesgue constant grows exponentially, as 
\be
\Lam_n \sim \frac{ 2^{n+1} }{n \log n } \label{eq:leb_equally}\,.
\ee
\item For Chebyshev nodes, it only grows logarithmically, 
\be
\Lam_n \sim \log{(n)} \label{eq:leb_cheb}\,.
\ee
\end{enumerate}
\end{example}
In the context of polynomial interpolation, it can be proved that the Lebesgue constant grows at least logarithmically in the infinity norm for any selection of nodes. In particular, Chebyshev nodes belong to a family that induce a nearly-optimal behavior on the L{ebesgue constant, since the growth (\ref{eq:leb_cheb}) is only logarithmic. For thorough discussions around this topic see \cite{hesthaven2007spectral}.

\subsection{Discrete expansions and interpolation} \label{sec:project_or_interpolate}

In approximation through projection one in principle needs full knowledge of the function $f$ to be approximated through the projection coefficients $\langle e_i , h \rangle$ in Eq.~(\ref{eq:proj_GS}). In interpolation, instead, one only needs knowledge of the function at the interpolating nodes. 

In practice, the integrals to compute $\langle e_i , h \rangle$ are replaced by a {\it quadrature} rule, i.e. a numerical approximation (we discuss quadrature rules and specifically those built using ROM in Section~\ref{sec:roq}). These are called then {\it discrete} projection approximations. Here it suffices to say that, if the quadrature nodes can be chosen, one can find an optimal family of quadrature rules that maximizes the degree for which the quadrature is exact for polynomials. These are called {\it  Gaussian quadratures} and the nodes {\it Gaussian nodes}; examples of the latter are Legendre and Chebyshev nodes. The relevant result here relating approximation by discrete projection and interpolation is the following:
\begin{result} [Discrete projection and interpolation] \label{res:proj_and_interp}

If the projection coefficients are approximated by Gaussian quadratures and are used Gaussian nodes for interpolation, then the resulting approximation ${\cal P}_n h$ {\em exactly} equals the polynomial interpolant $\cI[h]$~\cite{hesthaven2007spectral}.
\end{result}

Through this result one can guarantee fast convergence of interpolation using Gauss nodes. The problem is that in many parametrized problems of interest, one cannot choose Gaussian nodes in parameter space as representative solutions. Either because data is not available (for example, if taken from experiments), or because it is not efficient to do so from a modeling perspective. In Section~\ref{sec:eim} we discuss a generalization for ROM and parametrized systems that attempts to generalize Result~\ref{res:proj_and_interp}.

\subsection{Multiple dimensions}\label{sec:multivar}

\noindent{\bf Evaluation cost of multivariate polynomials}

Even when counting with enough data for multi-dimensional fits, enough for accurate local approximations which avoid Runge's phenomenom, a perhaps not so well known fact is that the evaluation cost of multivariate polynomials in general grows exponentially with the dimensionality of the problem. 

Naive evaluation of a $1$-dimensional polynomial of degree $n$, that is, evaluating all the monomials in the standard format
\be
p_n(x) = a_0 + a_1 x + \ldots + a_n x^n \label{eq:pol_monomials}
\ee
carries an operation count of $\cO(n^2)$. Through a simple factorization, Horner's rule
 (see \cite{Stoer-Bulirsch-1980}), the operation count can be reduced to $2n$. Suppose we want to evaluate $p_n(x)$. First note that $p_n(x)$ can be decomposed in a nested way
$$
p_n(x)=a_0 + x(a_1 + x(a_2+\cdots x(a_{n-1}+a_n x)\cdots)\,.
$$  
Horner's rule consists in computing iteratively the recursion
\begin{equation*}
\begin{split}
b_0=&a_n\\
b_ k =&a_{n-k} + b_{k-1}x\quad k=1,\cdots, n
\end{split}
\end{equation*}
reaching $p_n(x)=b_n$ in $2n$ operations (multiplication and addition, $n$ times). It can be shown that this is the minimum number of operations to evaluate a one-dimensional problem unless some offline factorization is carried out for multiple online evaluations, in which case the computational cost can be slightly reduced.

Next, consider multivariate polynomial evaluation. Suppose one attempts to evaluate the generalization of (\ref{eq:pol_monomials}) to the $d$-dimensional case. For definiteness, the form in the ${\tt dim}=2$ case would take the form 
$$
p_n(x,y) = \sum_{i=0}^{n_x}\sum_{j=0}^{n_y} a_{ij} x^i y^j \, . 
$$
The evaluation cost of such an approach in the $d$-dimensional case, assuming for simplicity the same degree $n$ in each dimension, is of order $\cO(n^{2d})$. Ideally, one would wish a generalization of Horner's rule with, say, evaluation cost of order $\cO(d \times n)$. Unfortunately not such algorithm is known, neither a proof of which could be the optimal number of operations needed to perform a multivariate polynomial evaluation. In recent years there have been several proposals to face this problem, ranging from rigorous mathematical attempts (see e.g.~\cite{Sauer1998, Moroz2013FastPE, Umans2007FastPF, 10.1007/978-3-642-15582-6_55, Ballico2011, Ballico2011, Kedlaya2008FastPF, VANDERHOEVEN2020101405}) to the proposal of heuristic greedy algorithms~\cite{Ceberio2003} aiming to reduce the number of calculations.

Multivariate polynomial evaluations scale exponentially with the dimensionality of the problem. This is a severe issue in most cases of interest which cannot be underestimated. In the context of ROM, the limitations of multivariate polynomial approximations appear in almost every possible context. This point is discussed in more detail and emphasized in Section~\ref{sec:chall}. 

\subsection*{Further reading}

Reference~\cite{polynom_multivariate} reviews multivariate polynomial interpolation, and~\cite{NarayanX12} presents a proposal for polynomial interpolation on unstructured grids in multiple dimensions. 

There are schemes for fast {\em multiple} evaluations of multi-variate polynomials though. They are based on an offline factorization of the given polynomial (which can be expensive, but done only once), with a total cost of ${\cO}(n^{{\tt dim} +1})$ for $\cO (n^{\tt dim})$ evaluations, leading to an average of $\cO(n)$ for each evaluation~\cite{Lodha97aunified}. To our knowledge this problem of multi-dimensionality has not been yet tackled in ROM for gravitational waves and is still largely an open problem in approximation theory. Instead, in GW science local interpolants in the form of splines are being used, which results in degradation of the accuracy of the surrogate models being built. In terms of higher (than one) dimensions, sparse grids are being used. One of their problems is that they are not necessarily hierarchical, leading to a problem when building training sets from numerical relativity to later apply the RB framework. 

\break

\section{The Empirical Interpolation Method} \label{sec:eim}
The Empirical Interpolation Method (EIM) was proposed in 2004~\cite{Barrault2004667} as a way of identifying a good set of interpolation nodes on multi-dimensional unstructured meshes and has since found numerous applications (for a {\em very} short list, see \cite{Aanonsen2009,Chaturantabut5400045,sorensen2010,Eftang:2011,Maday_2009}). When the basis is application-specific, so is the EIM. 

Here, perhaps contrary to what might be suspected, interpolation is in the physical dimension(s), not on the parametric ones. For definiteness, in the case of gravitational waves this is either the time or frequency domain. 

Before getting into technical details, we highlight some of the properties of the EIM:
\begin{itemize}
\item It is hierarchical (nested). The ``most relevant'' points in the physical dimension(s) and associated interpolants are selected and built, respectively. This follows a process that is dual, in a precise sense, to the construction of a reduced basis using a greedy algorithm. 
\item It is designed to overcome the difficulties of stability, accuracy and evaluation cost on sparse and scattered grids, especially in multiple dimensions. 
\item It is highly accurate. By design the algorithm {\em attempts} to control the behavior of the Lebesgue constant (see Section~\ref{sec:cond}) at each greedy EIM sweep.
\item It provides an {\it affine parametrization} which has multiple consequences. In particular, it allows for the design of {\it Reduced Order Quadratures}, described in Section~\ref{sec:roq}, which lead in particular to fast likelihood evaluations, as discussed in Section \ref{sec:PE}. 
\end{itemize}

The material below relies on standard interpolation ideas which we have summarized in Section~\ref{sec:interp}. We will refer to the physical variable(s) as $x$, the nodes selected by the EIM as $X_i\,(i=1\ldots n)$ \footnote{Except for the GW case, in which we will explicitly use $t$ and $T$ -- or $f$ and $F$, in the case of frequency.}, the (in general multidimensional) parameter as $\lam$ and its corresponding greedy points as $\Lam_i\,( i=1\ldots n)$. It is not a coincidence that there is the same number of EIM nodes and greedy points, in fact they go hand by hand.  
\subsection*{EIM: algorithm and properties}
Within the EIM method one seeks to find an empirical (that is, problem-dependent) {\em global} interpolant, which also provides an affine parametrization of functions. That is, the EIM approximation of a function $h(x;\lam)$ is of the form 
\be
h(x; \lam) \approx {\cal I}_n [h](x; \lam)= \sum_{i=1}^n B_i(x) h(X_i; \lam) \label{eq:affine_eim}\, . 
\ee
The affine parametrization of ${\cal I}_n [h](x; \lam)$ refers to the fact that the r.h.s. of (\ref{eq:affine_eim}) is a sum of terms which depend only on $x$ (the $B_i$  coefficients) multiplied by other ones which depend only on $\lambda$ (the $h(X_i; \lam)$); one might also want to refer to this as ``separability''.  
  
Below in Section~\ref{sec:proj_to_interp} we discuss how one arrives at (\ref{eq:affine_eim}), how to compute in the offline stage the $B_i$ coefficients, and the quality of this approximation. 

In addition, even though it is perhaps not usually viewed this way, or not emphasized enough compared to online evaluations to solutions of parametrized PDEs, the EIM also provides an application-specific down- and upsampling that, for practical purposes, beats Nyquist downsampling, with implications for signal or more generally data processing. See for example Section~\ref{sec:surrogate_eim} and discussion around Figures~\ref{fig:SurrogateSchematic} and~\ref{fig:eim_gw}. \\

\subsection{From projection to interpolation}\label{sec:proj_to_interp}

In Section \ref{sec:Projection} we discussed that the optimal linear representation in any weighted $L_2$ norm, given a basis (assuming for simplicity and conditioning that it is orthonormal) of cardinality $n$ is through an expression of the form 
\be
h(x ; \lam) \approx \cP_n h(x; \lam):= \sum_{i=1}^n c_i (\lam) e_i (x) \, ,  \label{eq:proj_rb}
\ee
where 
\be
c_i (\lam) = \langle  e_i (\cdot ), h(\cdot;\lam)  \rangle\, . \label{eq:proj_c}
\ee
This is the standard approximation by projection, the most common one being using polynomials or Fourier modes as basis. Now, approximation through projection requires knowledge of the function to be represented at sufficient values of $x$ so as to accurately compute the coefficients (\ref{eq:proj_c}). In Section~\ref{sec:interp} we discussed how this is usually replaced by interpolation, and the strong relationship between projection and interpolation in standard spectral theory.  Next we describe the EIM strategy, which mimics the spectral approach. 

The approximation through projection onto a reduced basis, Eq.~(\ref{eq:proj_rb}), is replaced by interpolation (in the physical dimension(s), $x$) as follows. First a {\tt rb} $\{ e_i (x) \}$ is chosen, for example through a POD or greedy approach.  The interpolant then is sought to have the form 
\be
{\cal I}_n [h](x; \lam) := \sum_{i=1}^n C_i (\lam) e_i(x)\, ,  \label{eq:interpdef}
\ee
where on purpose the {\em expansion, which are not projection,} coefficients $C_i$ are denoted by capital letters to distinguish them from the projection ones (\ref{eq:proj_c}). Instead, they are defined to be solutions of the interpolation problem; namely that the interpolant exactly agrees with the function at the EIM nodes (the construction of which we discuss below) 
\begin{align} \label{eq:intproblem}
{\cal I}_n [h](X_i; \lam) = h(X_i;\lam) , \qquad \forall \, i=1,\dots,n .
\end{align}
For the moment, we shall assume that the EIM nodes $X_i$ are known and proceed to describe how to use them to find the EIM interpolant. This can be somewhat misleading, since it is not the way it works in practice, which is: the first EIM node is found, its associated interpolant built, the second EIM node found, the interpolant enriched to take it into account, and so on. They are not disjoint processes, unlike (again) standard spectral methods, where all the Gaussian nodes are found and afterwards the interpolant built. This is not only a procedural difference, but highlights a big difference of the EIM: namely that the approach (nodes and interpolant) is hierarchical. Hopefully this gradual presentation is intuitive and pedagogical, later in this section we will present the full, coupled algorithm together with its numerical intricacies.

Equation~(\ref{eq:intproblem}) is equivalent to solving the $n$-by-$n$ system 
\be
\sum_{i=1}^n {\bf V}_{ji}C_i(\lam) = h (X_j; \lam )\,, \quad j=1, \ldots, n
\ee
for the coefficients $\{ C_i \}_{i=1}^n$,  where the interpolation matrix
\begin{equation} \label{eq:InterpMatrix}
  \bf{V} := \left(  \begin{array}{cccc}   
              e_1(X_1)  &  e_2(X_1)            & \cdots & e_{n}(X_1)      \\
              e_1(X_2)  &  e_2(X_2)            & \cdots & e_{n}(X_2)       \\
              e_1(X_3)  &  e_2(X_3)          & \cdots & e_{n}(X_3)   \\              
              \vdots    & \vdots             & \ddots & \vdots                       \\
              e_1(X_{n})  & e_2(X_{n})    & \cdots & e_{n}(X_{n})  \\               
             \end{array}
   \right) 
\end{equation}
is a generalization of the Vandermonde matrix -- see Eq.~(\ref{van}) -- when polynomial basis are used. We have already discussed that in that case the Vandermonde matrix can easily be very ill-conditioned if the nodes are chosen, for example, equally spaced, not to mention if they are scattered. So one can anticipate that the solution to the problem (\ref{eq:interpdef}) is already non-trivial and one of the goals of the EIM is to make sure that it does not lead to an ill conditioned problem, as well as providing a high accuracy interpolant. 

The choice of empirical nodes given by the EIM together with the linear independence of the reduced basis ensure that $\bf{V}$ (Eq.~(\ref{eq:InterpMatrix})) is invertible so that 
\be
C_i = \sum_{j=1}^n \left( {\bf V}^{-1}\right)_{ij}h (X_j; \lam)   \label{eq:CEIM}
\ee
is the unique solution to (\ref{eq:intproblem}).  
It then follows upon substituting (\ref{eq:CEIM}) into (\ref{eq:interpdef}) that the empirical interpolant is
\be \label{eq:EIM_with_B}
{\cal I}_n [h](x;\lam) = \sum_{j=1}^n B_j (x) h(X_j; \lam )
\ee 
where 
\be
B_j(x) :=  \sum_{i=1}^n e_i (x)  \left( {\bf V}^{-1} \right)_{ij}    \label{eq:BEIM}
\ee
is independent of $\lam$.

Note that (\ref{eq:EIM_with_B}) is  a linear combination of the waveform itself evaluated at the empirical nodes.
The coefficients $\{ B_i \}_{i=1}^n$ satisfy $B_i(X_j)=\delta_{ij}$ and are built directly from the reduced basis. They provide a clean offline/online separation. Because of this the $\{B_i\}_{i=1}^n$ functions can be pre-computed offline once the reduced basis is generated while the (fast) interpolation is evaluated during the online stage from (\ref{eq:EIM_with_B}) when the parameter $\lam$ is specified by the user. Evaluations of the waveform at the EIM nodes are still needed at the arbitrarily chosen parameter $\lam$ in order to construct the interpolant in (\ref{eq:EIM_with_B}). One can wonder how this can be of any use. In Section~\ref{sec:surrogates} we explain how to build surrogate predictive models for the evaluations of $ h(X_j; \lam )$ for any $\lam$ (that is, not present in the original training space). 

Next we discuss how to compute the EIM nodes in a rather qualitative way,  before presenting the general algorithm later on. 
\begin{example} Finding the EIM nodes and building the interpolant. 

Consider for definiteness a time series. The algorithm takes as input the basis set $\{ e_i \}_{i=1}^n$ and an arbitrary number and choice of time samples $\{ t_i\}_{i=1}^{L}$ from which the empirical interpolation nodes $\{ T_i \}_{i=1}^n$ are to be selected. The EIM algorithm proceeds as follows
\begin{enumerate}
\item The first time node is chosen to maximize the value of $| e_1(t_i) |$; that is, $\left| e_1(T_1) \right| \geq \left| e_1(t_i) \right|$ for all time samples $t_i$. 
\item Next,  an empirical interpolant for the second basis function is built using only the first basis function: From Eqs.~(\ref{eq:interpdef},\ref{eq:intproblem})  we have ${\cal I}_1 [e_2](t)= C_1 e_1(t)$ where $C_1 = e_2(T_1) / e_1(T_1)$ has been found from Eq.~(\ref{eq:intproblem}) with $i=1$. 
\item The second empirical interpolation node is chosen to maximize the value of the pointwise interpolation error of ${\cal I}_1 [e_2](t) - e_2(t)$; that is,  
$\left| {\cal I}_1 [e_2] (T_2) - e_2(T_2) \right| \geq \left| {\cal I}_1 [e_2](t_i) - e_2(t_i) \right|$ for all data samples. 
\item Steps $2$ and $3$ are then repeated to select the remaining $(n-2)$ nodes. 
The full algorithm for generating the EIM nodes is shown in Algorithm~\ref{alg:EIM}.
\end{enumerate}
\end{example}

\begin{comment}
As described, the EIM follows a greedy approach, albeit somewhat different from that one usually used to build a reduced basis. While a greedy algorithm to build a RB selects the most relevant points in parameter space, the EIM selects the most relevant points in the physical dimension(s). In addition, the former uses a (possibly weighted) $L_2$ norm while the EIM uses the infinity one. 
\end{comment}

\begin{comment}
Notice also that even though constructing the EIM is also done offline, it has a much smaller computational cost than constructing the basis, since it operates only on the latter, the training set {\it is not} involved anymore. 
\end{comment}

The following example graphically shows the first three iterations of the EIM algorithm applied to Legendre polynomials.

\begin{example} \label{ex:leg-eim}

Consider the set $\{P_i(x)\}_{i=0}^{n}$ of $(n+1)$ normalized Legendre polynomials defined on $[-1,1]$. These form an orthonormal basis for the space of degree $n$ polynomials wrt the weight function $\omega (x) \equiv 1$. Their ordering is important for approximations: expanding, by orthogonal projection, a smooth function using the first $n$ Legendre polynomials typically results in exponential convergence \cite{hesthaven2007spectral, Wang2011OnTC}. Said another way, $P_0$ is the most important basis function, $P_1$ the next most important, and so on.

Given a convergence rate for the aforementioned Legendre projection-based approximation we might wonder how much accuracy is lost by trading it for the interpolation (\ref{eq:interpdef}) and how to optimally choose the nodes $\{X_i\}$. When the relevant error measurement is the maximum pointwise error, Chebyshev nodes are known to be well suited for interpolation, bringing an additional error which is bounded by $\log(n)$~\cite{Press92,Quarteroni2010}, as discussed in Section~\ref{sec:runge} (see Example \ref{ex:leb}). For this reason, Chebyshev nodes are specially tailored to benchmark the nodes selected by the EIM in this example. 

We select the EIM nodes according to the same preferential ordering as the Legendre polynomials themselves; the first six normalized polynomials are shown in Fig.~\ref{fig:LegPolys}. 

\begin{figure}[H]
\begin{center}
\includegraphics[width=0.6\linewidth]{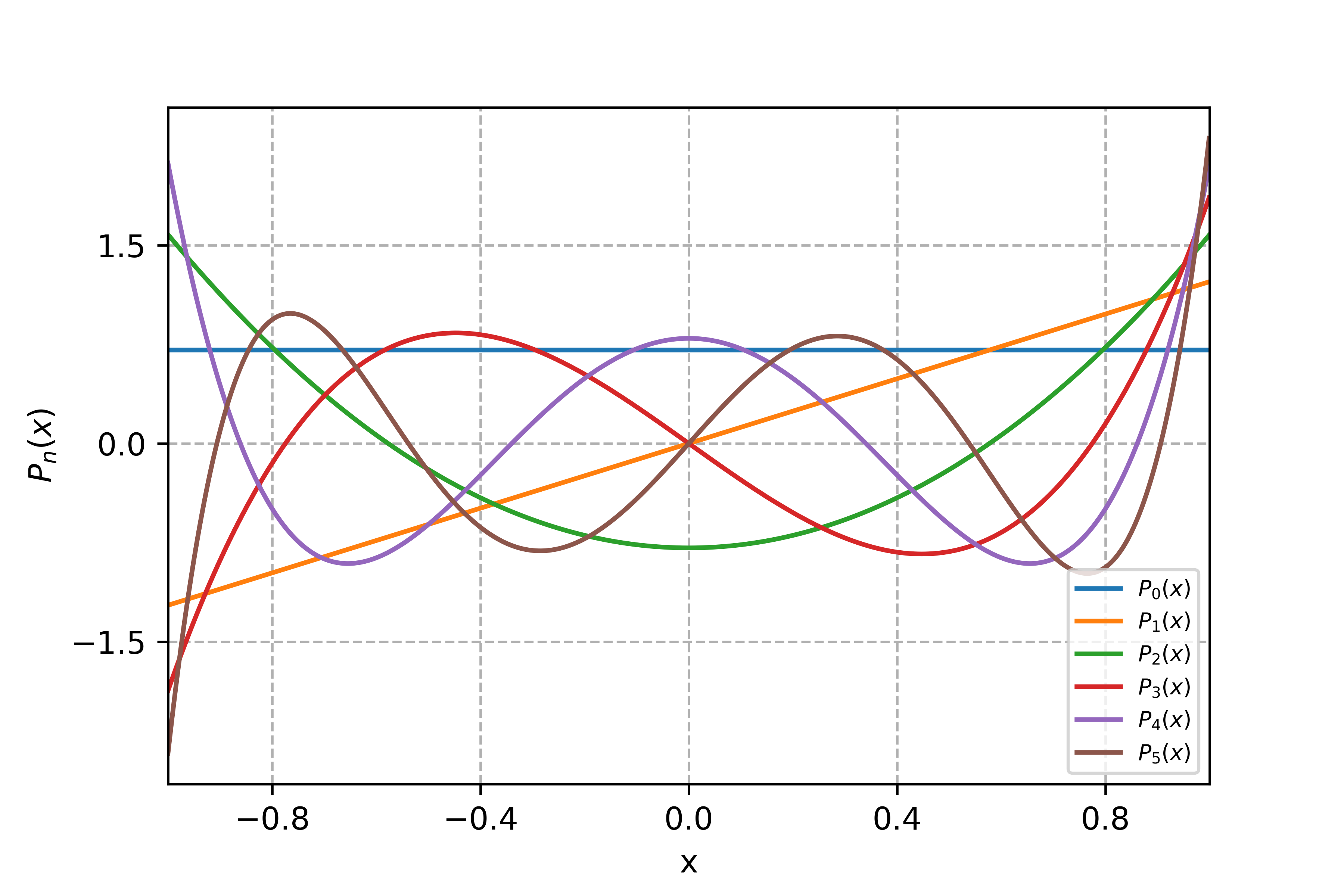} 
\end{center}
\caption{The first six normalized Legendre polynomials.}
\label{fig:LegPolys}
\end{figure}

For this example we chose the first 24 Legendre polynomials as the initial basis for feeding the EIM algorithm. Figure~\ref{fig:LegEIM} shows the process of the EIM algorithm to choose the interpolation nodes for the Legendre basis. The first EIM node is defined by the location of max$(|P_0|)$. Since $P_0$ is a constant function there is no preference in the search, so we simply select the middle point $x=0$.  To identify the second node we
\begin{enumerate}
\item build the empirical interpolant ${\cal I}_0[P_1]$ of $P_1$ using $P_0$ as the basis and  $x=0$ as interpolation node (the first EIM node),
\item compute the pointwise error $r_1 = P_1 - {\cal I}_0[P_1] $, and
\item select the second EIM node by the location of max$(\left|r_1\right|)$. In this case we find $r_1 = P_1$ since ${\cal I}_0[P_1]$ is zero due to the fact that $P_1(0) = 0$. 
\end{enumerate}
The process continues until the number of EIM nodes equals the number of basis elements. 

\begin{figure}[H]
\begin{center}
\includegraphics[width=0.45\linewidth]{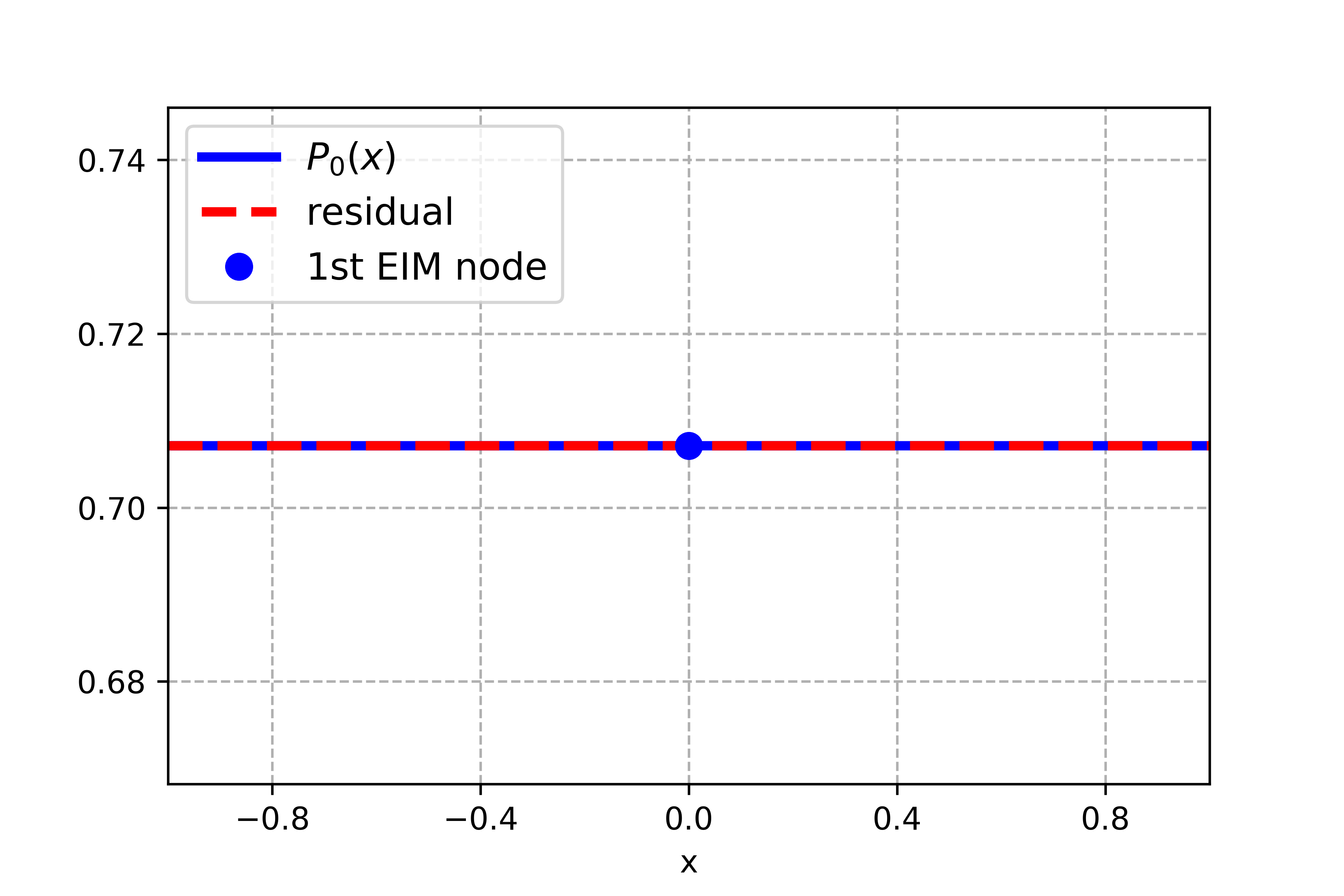} 
\includegraphics[width=0.45\linewidth]{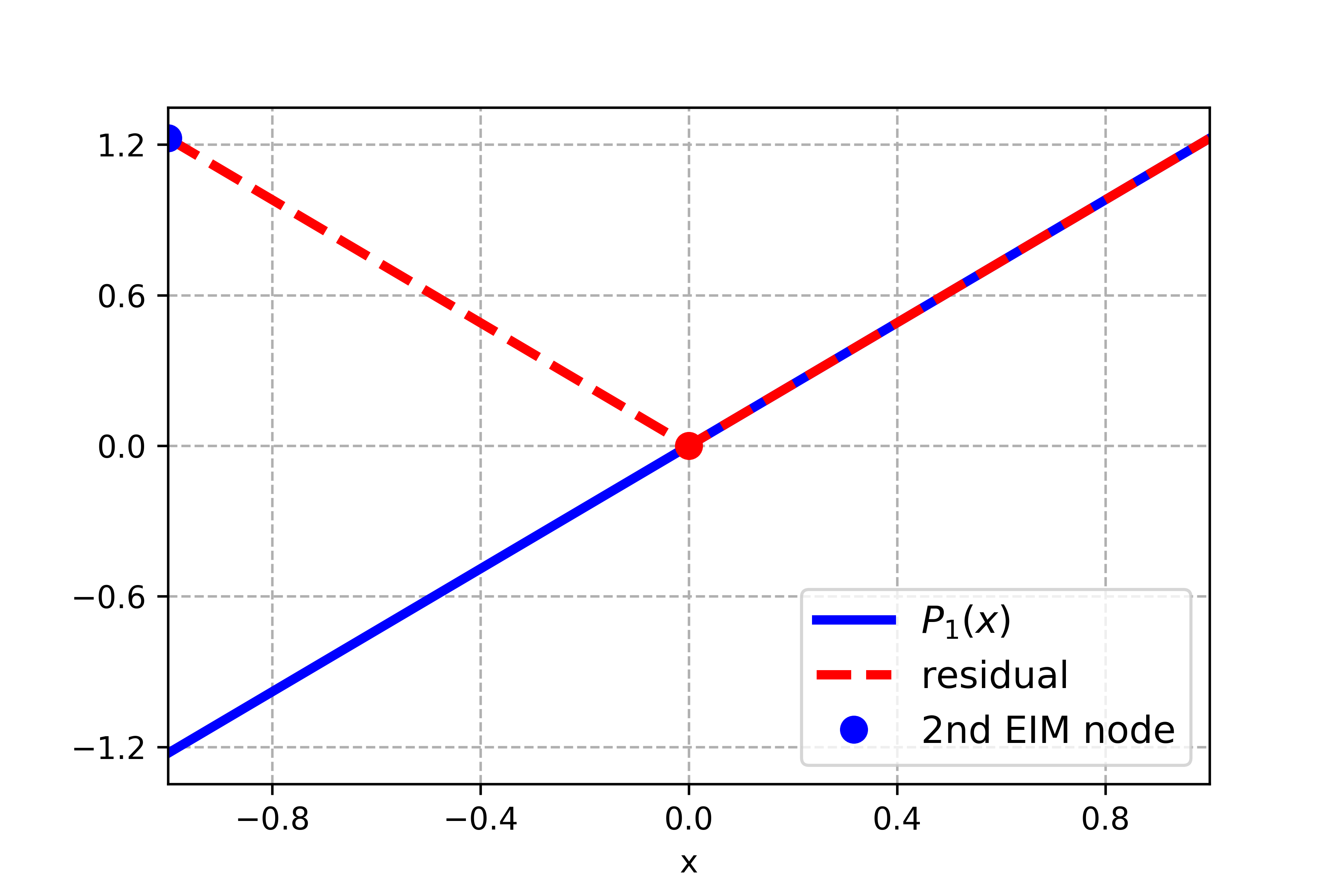}
\includegraphics[width=0.45\linewidth]{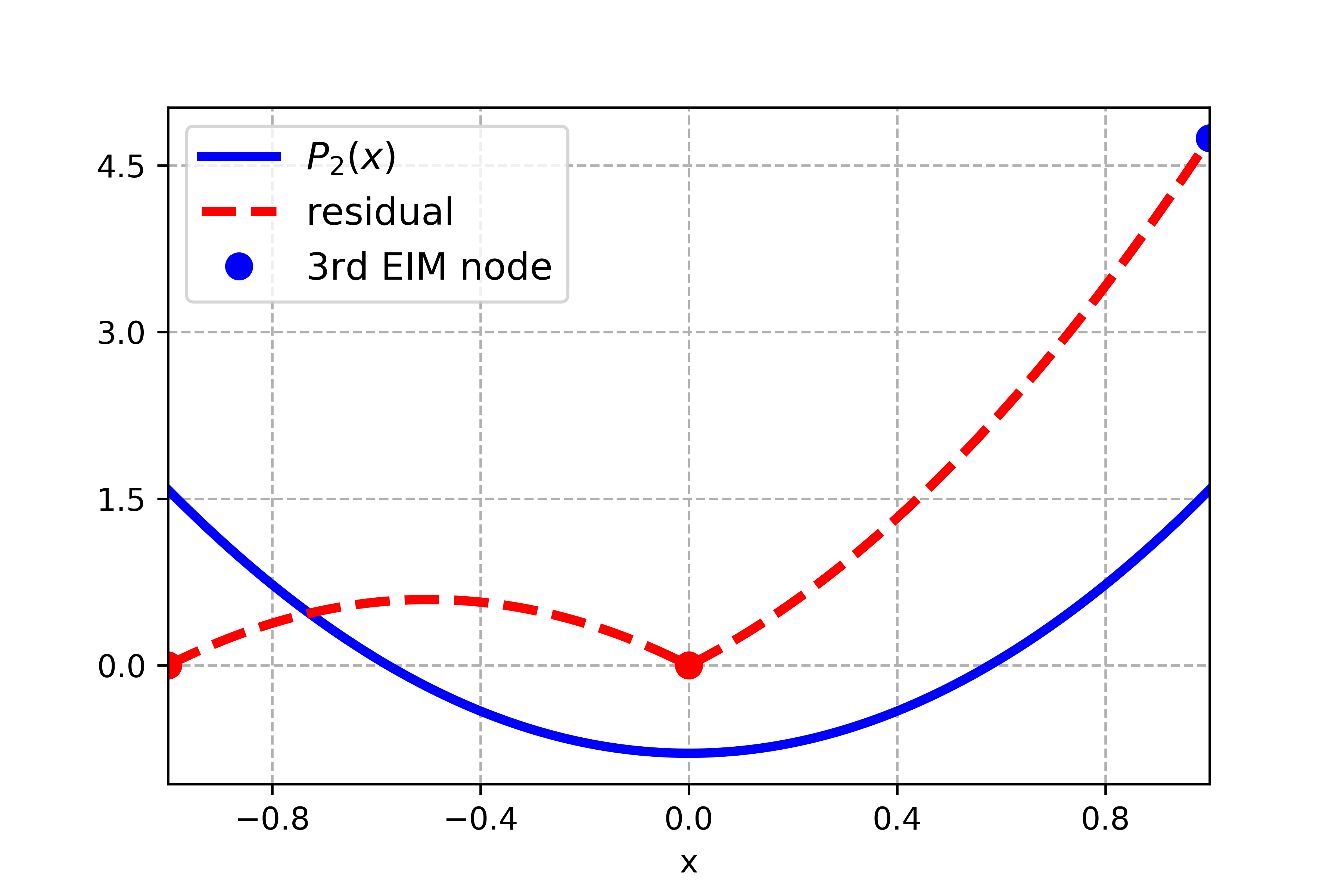}
\includegraphics[width=0.45\linewidth]{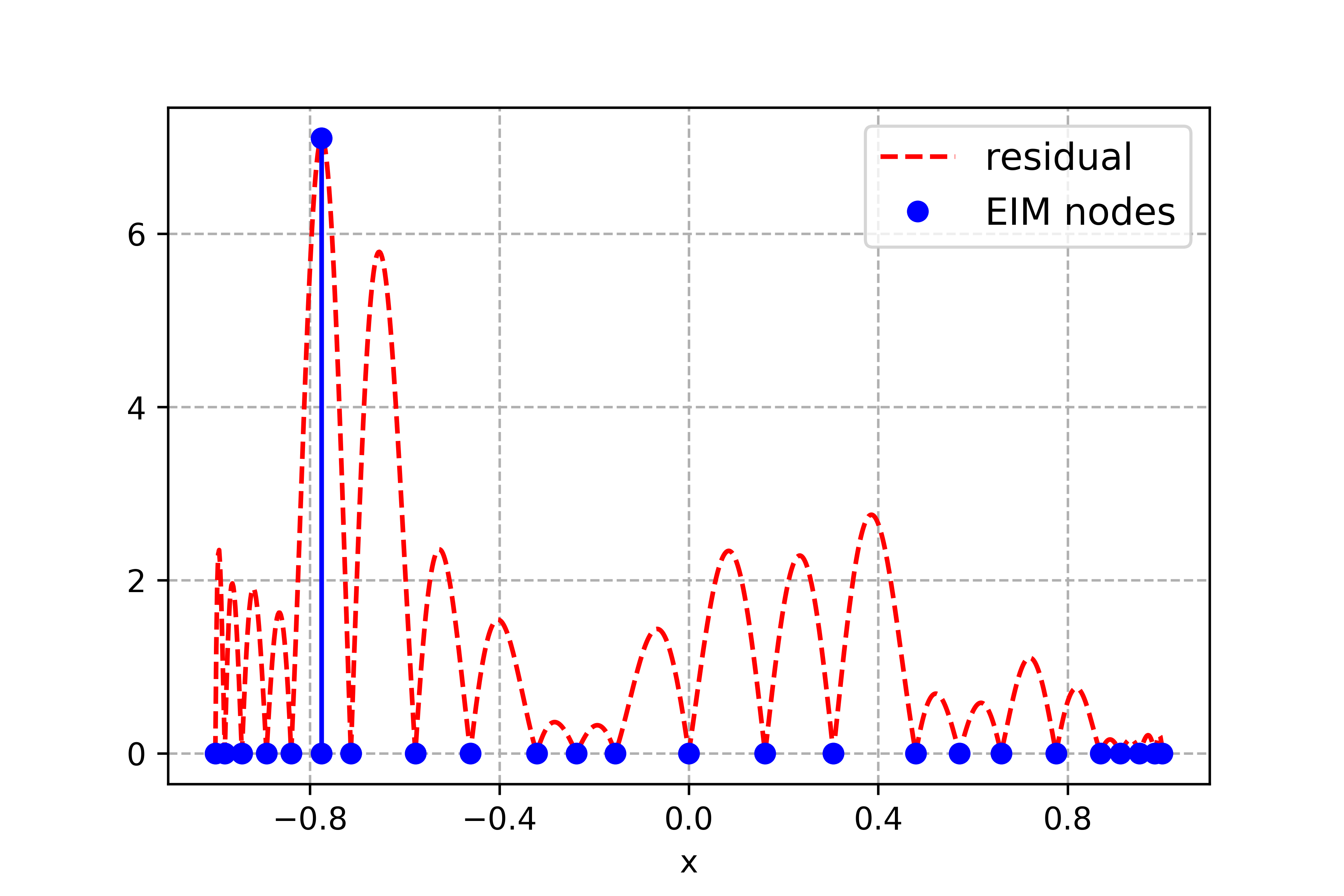}
\caption{Representation of the EIM algorithm acting on the first 24 normalized Legendre basis. We show the bases, residuals and EIM nodes involved in the iterations. The first three figures (upper-left, upper-right and bottom-left) illustrate the two steps of the algorithm that maximize residuals and select the EIM nodes. The last panel (bottom-right) shows the latest iteration of the algorithm, which corresponds to the $24$th basis. The vertical line corresponds to the residual's maximum.
\label{fig:LegEIM}}
\end{center}
\end{figure}

In Fig.~\ref{fig:cheb} we show the $24$ EIM nodes found by the algorithm and compare them to the first 24 Chebyshev nodes in the interval $[-1, 1]$. Both node distributions are qualitatively similar. Note that the EIM distribution emulates the clustering of the Chebyshev nodes at the edges $x\pm 1$. Reference~\cite{Maday_2009}, which has inspired our numerical experiment with Legendre polynomials, also compares the Lebesgue constant(s) for the EIM nodes of this example and the and Chebyshev ones. 

\begin{figure}[H]
\begin{center}
\includegraphics[width=0.6\linewidth]{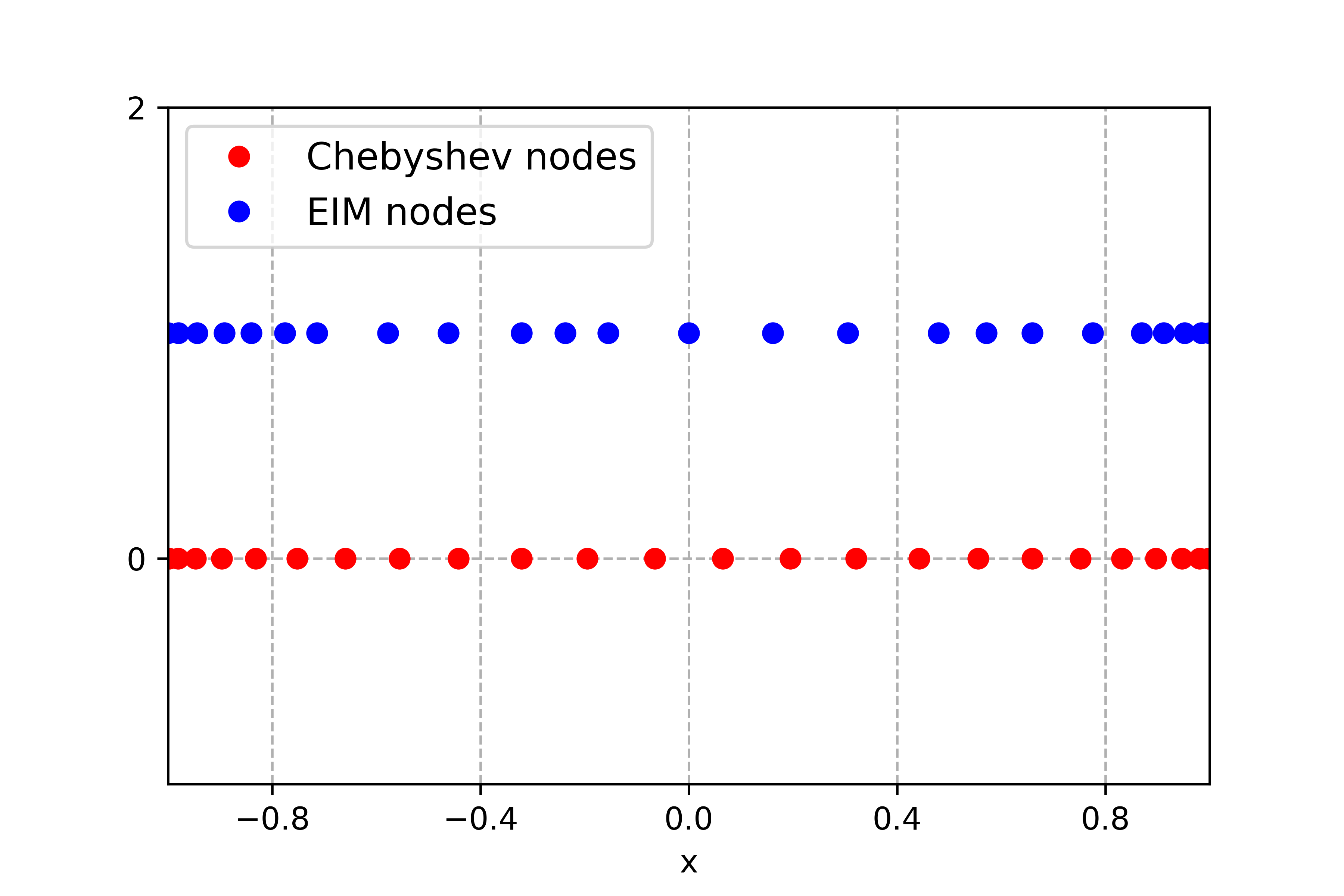}
\caption{A direct comparison between the distribution of the 24 EIM nodes for the Example~\ref{ex:leg-eim} and the first 24 Chebyshev nodes.} 
\label{fig:cheb}
\end{center}
\end{figure}

\end{example}

\begin{example} \label{ex:burst}

This example, taken from Ref.~\cite{Canizares:2013ywa}, graphically shows the EIM algorithm's application to basis functions for which a good set of interpolation nodes is unknown. This is a sine-Gaussian family of the form 
\be
h(t; \lam ) := A e^{-(t-t_c)^2/(2\alpha^2)}\sin(2\pi f_0 (t-t_c)) \, ,
\label{eq:h_t}
\ee
which models gravitational waves from generic burst processes. Here $A$, $f_0$ and $\alpha$ are the amplitude, frequency and width of the waveform $h$ respectively, $t_c$ is the arrival time of the GW-burst signal, and $t\in[-\infty,\infty]$. The Fourier transform (FT) of this waveform is given by
\begin{align}
{\tilde h} (f,t_c; \lam )= e^{i2\pi f t_c}{\tilde h} (f; \lam )\,,\label{h_f1}
\end{align}
where ${\tilde h} (f; \lam )$ is the FT of the GW-burst at $t_c = 0$: 
 \begin{align}
{\tilde h} (f; \lam ) = i2A\alpha \sqrt{2\pi} \sinh(4\pi^2\alpha^2f_0f) e^{-2\pi^2\alpha^2(f_0^2+f^2)} .
\label{h_f2}
\end{align}
This waveform family is, then, described by four free parameters $\lambda = (\alpha,f_0, t_c, A)$. We focus on the two parameters $(\alpha,f_0)$, since the others are extrinsic and can be handled differently. The parameter space considered is 
\be 
\alpha = [.02,2]\sec \quad \, , \quad f_0 = [.01,1]{\rm Hz} \, , \label{eq:range_pars}
\ee
Figure~\ref{fig:deimalg} provides a graphical illustration of the EIM first iterations. 
 
\begin{figure}[H]
\begin{center}
\includegraphics[width=0.4\linewidth]{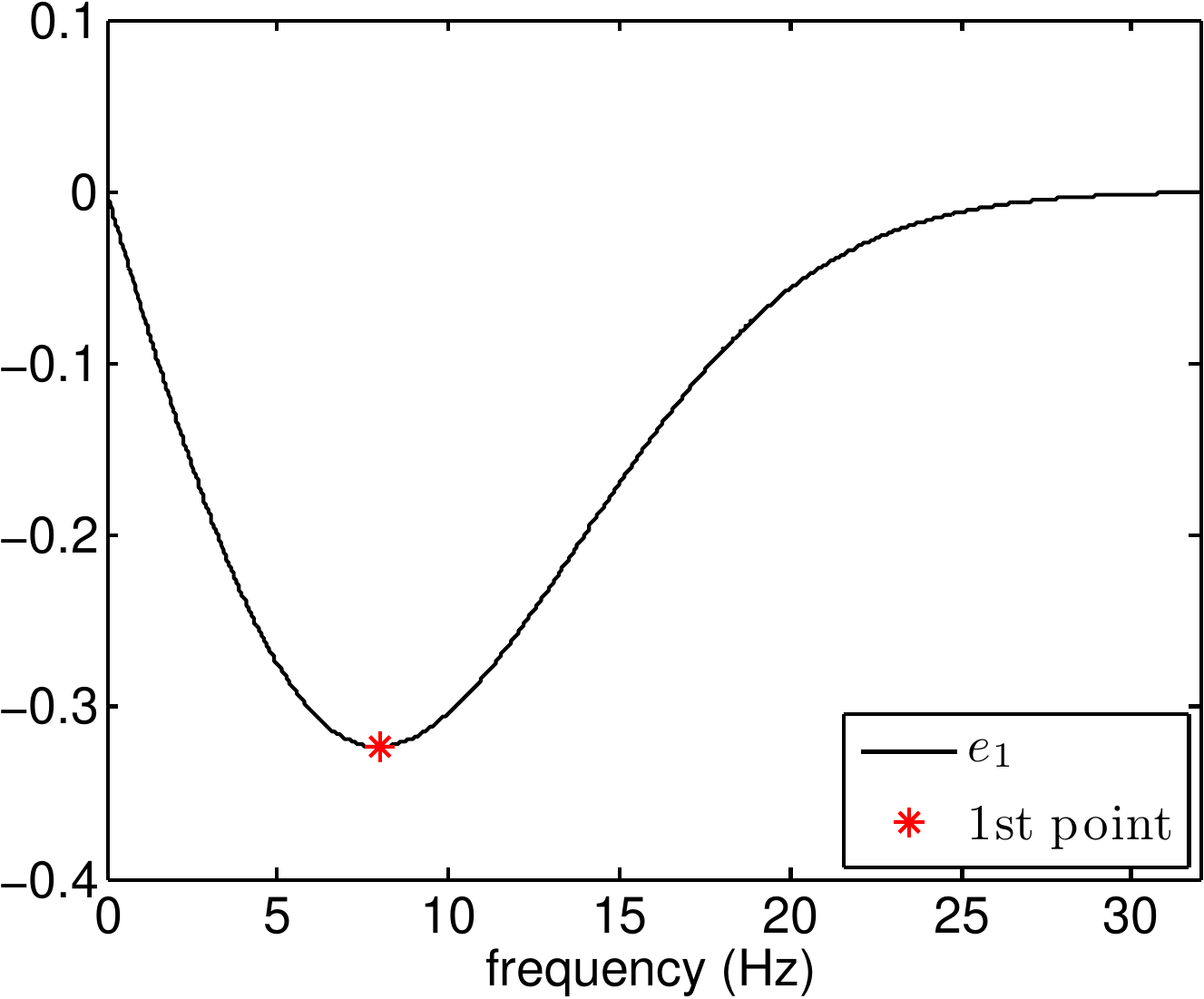} 
\includegraphics[width=0.4\linewidth]{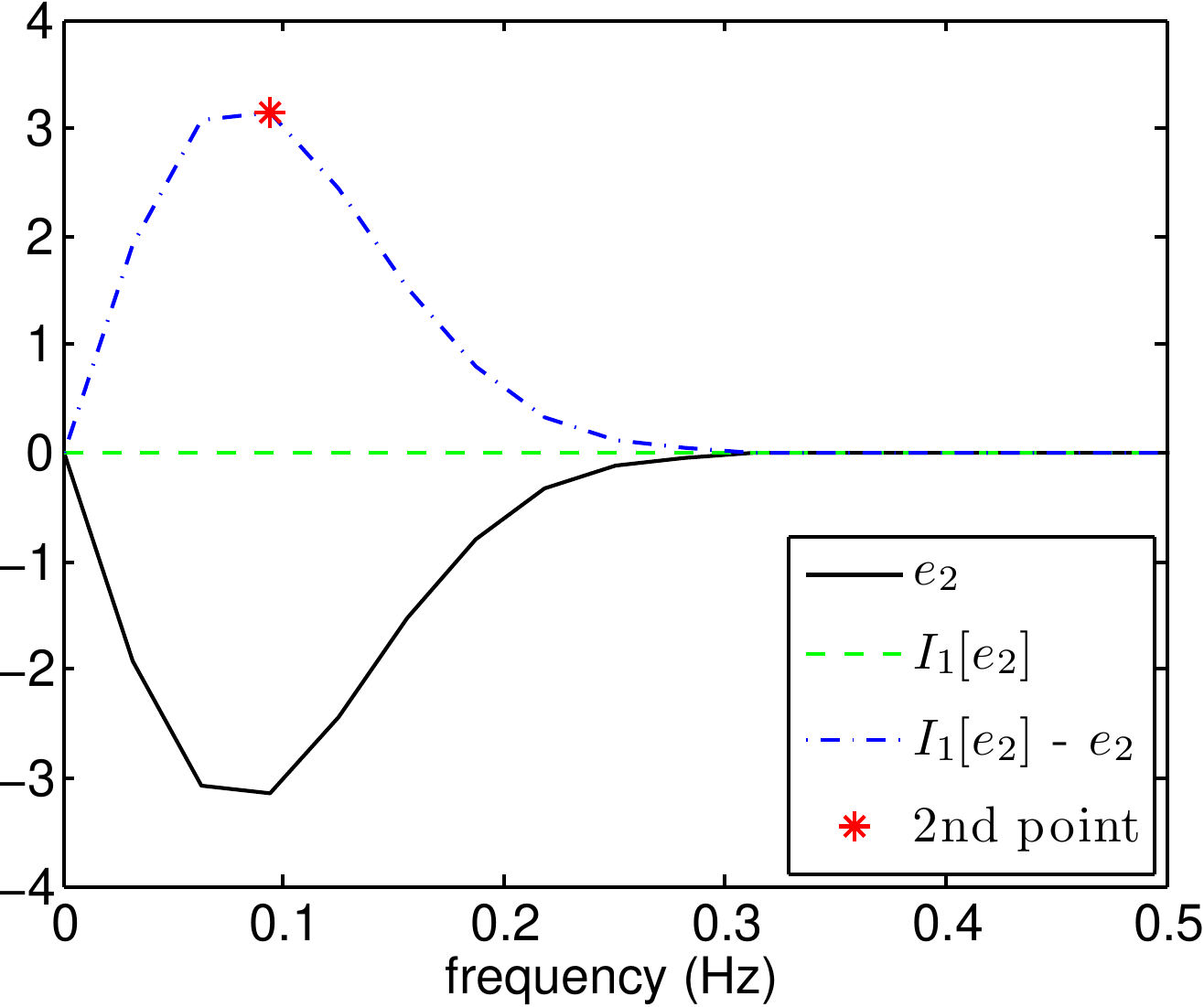}
\caption{Iterations 1 (left) and 2 (right) of the EIM algorithm for Example ~\ref{ex:leg-eim}. The first EIM node is defined by the location of max$(|e_1|)$. To identify the second node one: i) builds the empirical interpolant ${\cal I}_1[e_2]$ of $e_2$ using $e_1$ and the EIM node $F_1$, ii) computes the pointwise error ${\cal I}_1[e_2] - e_2$; iii) the second EIM node is then defined by the location of max$(\left|{\cal I}_1[e_2] - e_2\right|)$. The process continues until all $n$ empirical interpolation nodes are found. Figures taken from~\cite{Canizares:2013ywa}.
\label{fig:deimalg}}
\end{center}
\end{figure}

\end{example}

\subsection{EIM algorithm}

We present the EIM algorithm as used/introduced in Ref.~\cite{PhysRevX.4.031006}.

{\scriptsize
\begin{algorithm}[H]
\caption{The Empirical Interpolation Method}
\label{alg:EIM}
\begin{algorithmic}[1]
\State {\bf Input:} {\tt rb} = $\{ e_i \}_{i=1}^n$
\vskip 10pt
\State $X_1 = \text{argmax}_x | e_1|$ 
\For{$j = 2 \to n$} 
\State Build ${\cal I}_{j-1} [e_j](x)$
\State $r_j(x) = e_j(x)-{\cal I}_{j-1} [e_j](x)$  ($r$ stands for residual)
\State $X_j = \text{argmax}_x |r_j|$
 \EndFor
\vskip 10pt
\State {\bf Output:} EIM nodes $\{ X_i \}_{i=1}^n$ and interpolant $\cI_n$
\end{algorithmic}
\end{algorithm}
}
\begin{comment}
We make some remarks on the algorithm:
\begin{itemize}
\item It uses a greedy algorithm in the infinity norm. 

\item The iteration is performed only on the reduced basis, and there is no need of a training set as it was the case in the construction of the basis itself.

\item At each greedy sweep it chooses the node that differs most from the interpolant built so far.

\item The selection of EIM nodes is basis-dependent: different choices of the basis should lead to different nodes. However, once the nodes are computed, the interpolant becomes basis independent~\cite{sorensen2010}. From a geometric viewpoint, one realizes that the interpolant is a projector (though not orthogonal) onto the span of the basis. Once the reduced space and the nodes are chosen, the interpolant becomes uniquely defined.
\end{itemize}
\end{comment}

\subsection{Accuracy and conditioning of the EIM} \label{sec:cond}
In Section~\ref{sec:runge} we discussed how the quality of an interpolant in a weighted $L_2$ norm is related to the optimal projection representation. Interpolation in general loses accuracy but at the advantage of needing less information for computing the approximation. One can bound the interpolation error defining a {\it Lebesgue constant} (see Eq.~(\ref{eq:bound352})):
\be\label{eq:ineq}
\| h_\lam  - {\cal I}_n [h_\lam] \|^2 \leq \Lambda_n  \| h_\lam - {\cP}_n h_\lam \| ^2\,,\quad\Lambda_n  := \| {\cal I}_n \| ^2\,.
\ee

Taking the maximum in both sides of the inequality, and recalling the definition of the greedy projection error in Eq.~(\ref{eq:greedyErr}), we obtain

\be
\max_{\lam} \| h{_\lam} - {\cal I}_n [h_{\lam}] \|^2 \leq \Lambda_n \sigma_n \, . \label{eq:errorEIM}
\ee

\begin{comment}
The maxima in (\ref{eq:errorEIM}) involve some ambiguities, as they might refer to those with respect to the training set or those of the underlying space of interest. It is problem and user-dependent to clarify which is the case, but the above results apply to both of them.
\end{comment}

\begin{comment}
As discussed below, the Lebesgue constant is an a posteriori computable number, so if one has an estimate of the projection error $\sigma_n$, also has an estimation of the interpolation one. Assuming there is enough training data, an estimate for $\sigma_n$ can be accomplished by standard validation tests. Otherwise, if the data is sparse, one may apply other standard methods from machine learning such as k-fold or leave-one-out cross-validations. 
\end{comment}

In the context of the EIM, some obvious questions arise from these observations: how is the  EIM related to the Lebesgue constant? Does it optimize for accuracy in some way? These questions were addressed in Ref.~\cite{Villanueva2020}. 

First notice that, at each step of the EIM, when building the interpolant, a Vandermonde matrix ${\bf V}_n$ (Eq.~(\ref{eq:InterpMatrix})) has to be inverted for computing the interpolation coefficients. The nature of the algorithm already ensures the invertibility of ${\bf V}_n$ at all steps. It turns out that the EIM in fact  optimizes with respect to this invertibility.
\blank

\noindent{\bf Theorem}. Define $\textup{det}({\bf V}_j):=V_j(X_1,\ldots,X_j)$. Then, the residual $r_j(x)$ computed in the $(j-1)$-iteration of the EIM-loop satisfies 

\be\label{eq:prop}
r_j(x)=\frac{V_j(X_1,\ldots, X_{j-1},x)}{V_{j-1}(X_1,\ldots,X_{j-1})}\quad j=2,3, \ldots n \, . 
\ee
\blank
In consequence, once $X_j$ is chosen in Step 6, the residual at this node becomes
\be
r_j(X_j)=\frac{V_j(X_1,\ldots, X_{j-1},X_j)}{V_{j-1}(X_1,\ldots,X_{j-1})}=\frac{\text{det}({\bf V}_j)}{\text{det}({\bf V}_{j-1})}\,. \label{eq:residual}
\ee

The proof can be found in \cite{Villanueva2020}. This result says that, at each $j$-step, the EIM algorithm selects a new node $X_j$ in order to maximize the module of the determinant of ${\bf V}_j$, making the Vandermonde matrix as invertible as possible at each iteration. 

The following is a useful identity to compute Lebesgue constants in practice~\cite{antil2012two}: if the basis vectors are orthonormal in the 2--norm $\|\cdot\|_2$, then

\be
\Lambda_n=\|{\bf V}_n^{-1}\|^2_2\,.
\ee

We use this relation and the formulation of the inverse of a matrix in terms of its adjoint to rewrite the Lebesgue constant as

\be\label{eq:lambda_adj}
\Lambda_n=\frac{\|\text{adj}({\bf V}_n)\|^2_2}{|\text{det}({\bf V}_n)|^2}\,.
\ee

In doing this, we see that when maximizing the determinant of the Vandermonde matrix at each step the EIM {\em attempts} to {\em partially} control the growth of $\Lambda_n$, in the sense of making the denominator of (\ref{eq:lambda_adj}) as large as possible, but without controlling its numerator.  

So, the algorithm does not solve for the optimization problem of finding {\em global} (since it is a nested approach) nodes that maximize the determinant of the Vandermonde matrix, but neither the partial problem of minimizing the whole Lebesgue constant at each step, as is usually thought. Analogous observations can be made for the conditioning of ${\bf V}_n$, since both the Lebesgue constant and the condition number of ${\bf V}_n$ are directly related by
$$
\kappa_n = \|{\bf V}_n\|_2\Lambda_n\,.
$$
For further discussions around these topics in the context of GWs, see \cite{Villanueva2020}.

Finally, we point out that, in the context of the EIM, a rather pessimistic bound for the Lebesgue constant can be derived~\cite{sorensen2010}:
$$
\Lambda_n\leq (1+\sqrt{2L})^{n-1} \|e_1\|_{\infty}^{-1}\,,
$$
where $L$ stands for the size of the sampling of the $x$ variable. As an a priori estimate, this bound is not useful in practice since it does not reflect the observed growth of $\Lambda_n$, which is, in most cases of interest, much slower. Sharper a priori estimates are difficult to prove. In practice, one computes the Lebesgue constant a posteriori.

\subsection*{Further reading}

The original papers introducing the EIM are \cite{Barrault2004667} and \cite{Maday_2009}. Reference \cite{sorensen2010} presents a very clearly explained discrete version  with emphasis on solutions to non-linear time-dependent parametrized equations. In the latter case one is interested in approximating non-linear terms, in similarity with Galerkin versus collocation methods when solving partial differential equations. An hp-refinement of the standard EIM approach is presented in \cite{Eftang:2011}; this in general should be of very practical importance for several scenarios, such as when there are discontinuities with respect to parameter variation, but it has so far not been used in GW science. The study of conditioning of Vandermonde matrices can be tracked to \cite{GAUTSCHI1983293}, where orthogonal polynomials are used to improve the condition number. See also \cite{Gautschi2011OptimallySA,Gautschi2012HowU}, and, for a short review, \cite{Higham2013NumericalC}. For different studies around Vandermonde-like matrices see \cite{Kuian2019OptimallyCV, Demmel2006AccurateSO,Pan2015HowBA,Reichel1991ChebyshevVandermondeS}.

\newpage
\section{Surrogate models}
\label{sec:surrogates}
The first kind of surrogate models for gravitational waves followed the lines of fitting for the projection coefficients of a reduced basis, be it obtained through a POD or greedy approach. That is, given a basis $\{ e_i \}_{i=1}^n$ the RB approximation for a waveform $h$ is 
\be
h(t ; \lam) := \sum_{i=1}^n c(\lam_i) e_i(t) \, , \label{eq:rep}
\ee
where the basis elements $\{ e_i\}$ can be waveforms themselves as in the RB-greedy approach, or linear combinations of them as in the POD approach. 

This representation requires knowledge of the projection coefficients $\{ c_i\}$, which are only known if the waveform itself is known. That is, as it stands, Eq.~(\ref{eq:rep}) is not a {\it predictive} tool but a {\it representation} one. A natural step to predict new waveforms is to build a surrogate by simply fitting in some way for new values of the parameter $\lam$: 
\be
h_{\tt s}(t, \lam) := \sum_{i=1}^n c_{{\tt s}\,, i}(\lam) e_i(t) \, ,
\ee
where $\{ c_{{\tt s}\, , i} \}$ are $n$ approximations to the true projection coefficients $\{ c_i \}$. There is a large variety of ways to fit for these coefficients, such as polynomial interpolation, splines, radial basis, or ML regression approaches. A number of them (without much success) were studied in Ref.~\cite{kaye2012}.

One of the problems found when the basis elements are not gravitational waves themselves, as in the POD case, or if using the auxiliary orthonormal basis in the RB-greedy approach, is that they have non-physical complex structure in their parameter dependence that is difficult to represent, this is also discussed in Appendix F of~\cite{PhysRevX.4.031006}. One solution is to use as basis elements the waveforms themselves, something that is not possible within the POD approach, since it does not provide a set of the most representative waveforms to use. Another approach, if a dense training space is available, is to use local fits (such as splines) to find these unknown projection coefficients. The limitation of needing a dense training set is precisely being able to generate it. 

The main source of errors in these approaches is the fitting procedure, which in general has slow convergence. Therefore, fitting only at the EIM nodes, as described below, serves at least two goals: i) to minimize the source of fitting errors by doing so only at the EIM nodes, ii) to provide very fast to evaluate surrogate models. Impressive results have been achieved, which we discuss in Section~\ref{sec:binary-surrogates}, using this approach. Other approaches follow non-algorithmic schemes and the literature in this field is rather new and in constant evolution. To mention only a few, see for example~\cite{Hesthaven2018NonintrusiveRO, WANG2019289} in which neural networks are combined with reduced order modeling or~\cite{PhysRevLett.122.211101} for the same approach in the context of GWs. We leave for future updates of this review a thorough discussion of these topics.

\subsection{Surrogate models for components} \label{sec:comps}
Even though we have emphasized data-driven approaches, {\em some} domain knowledge {\em is} still (very) useful. For example, the GWs emitted by two black holes have an apparent complexity, but can be decomposed into components with simple structure and easier to build models for. For example, in the non-precessing case the GWs are essentially oscillatory functions of time with an increasing amplitude, until around the time of merger, followed by the ringdown regime. Thus, it is advantageously to decompose them into phase and amplitude:
\be
h(t ; \lam) = A(t, \lam) e^{i \Phi (t, \lam) } \, . \label{eq:comps}
\ee
The structure of the waveforms themselves as well as phase and amplitude for the case of two black holes initially in quasi-circular orbit and without spin) are shown in Figures~\ref{fig:22mode} and~\ref{fig:amp-phase}. One can notice the simplification in structure when decomposing into phase and amplitude. 

\begin{figure}[ht]
\begin{center}
\includegraphics[width=0.8\linewidth]{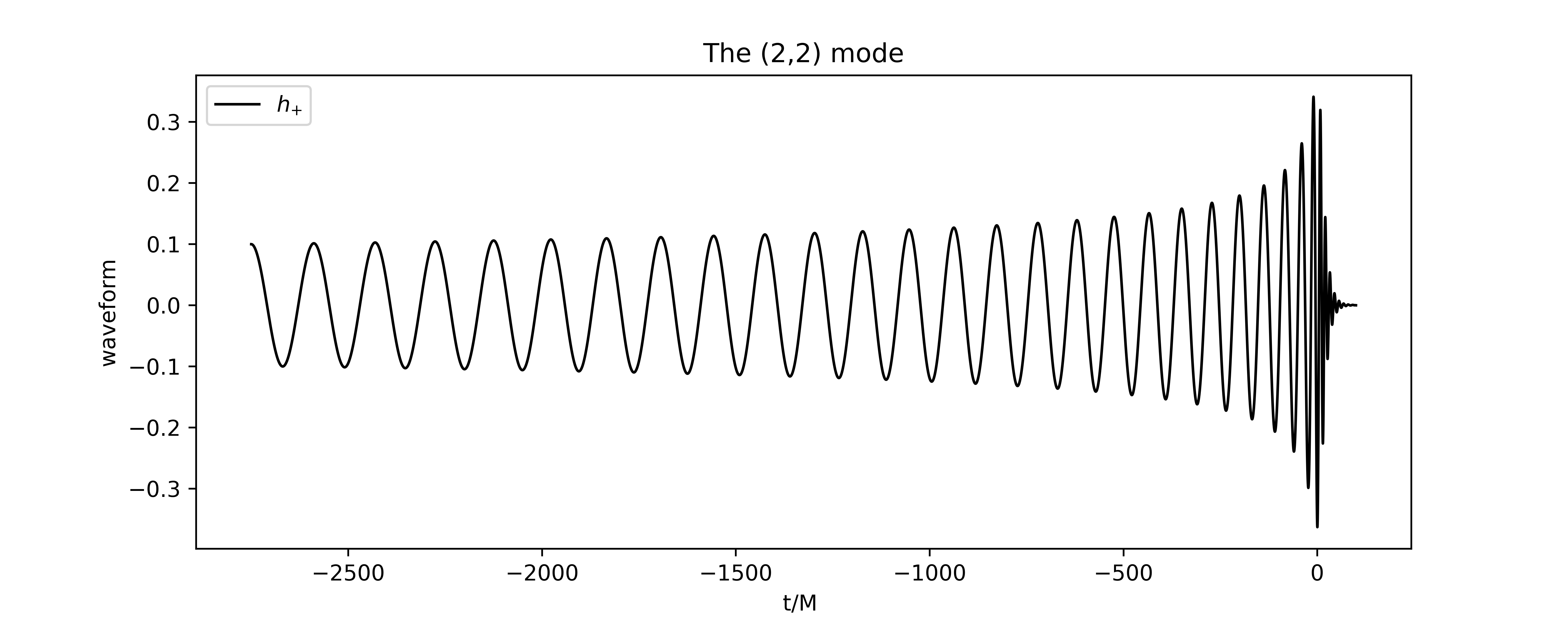}
\caption{Representation of the $h_+$ polaritazion of a waveform corresponding to the mode $(2,2)$ for a BBH coalescence process initially in quasi-circular orbit without spin and mass ratio $q=m_1/m_2=1.7$. This picture corresponds to the surrogate model {\tt SpEC\_q1\_10\_NoSpin}~\cite{PhysRevLett.115.121102}, and was generated using the gwsurrogate Python package~\cite{gwsurr}.}
\label{fig:22mode}
\end{center}
\end{figure}

\begin{figure}[ht]
\begin{center}
\includegraphics[width=0.48\linewidth]{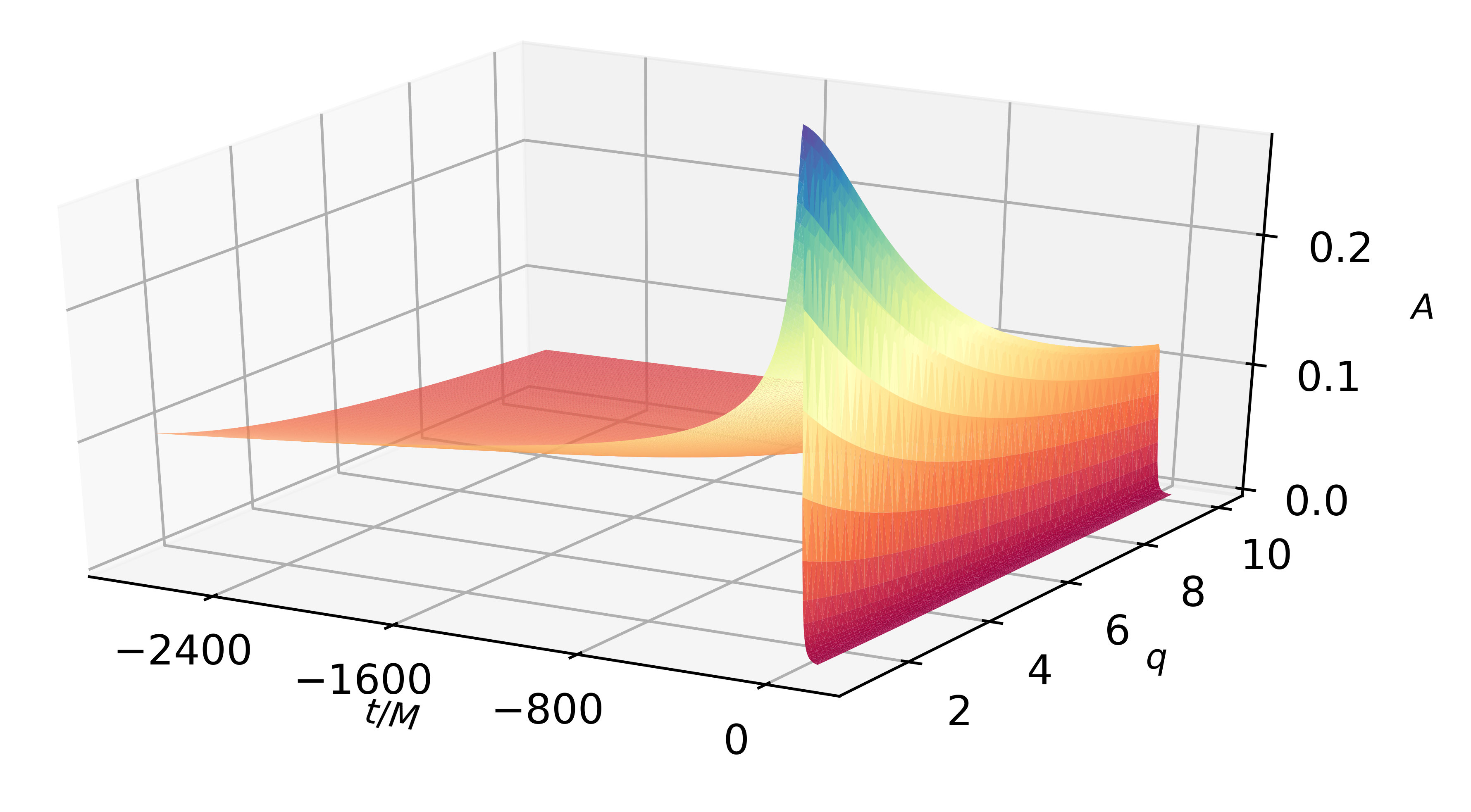} 
\includegraphics[width=0.48\linewidth]{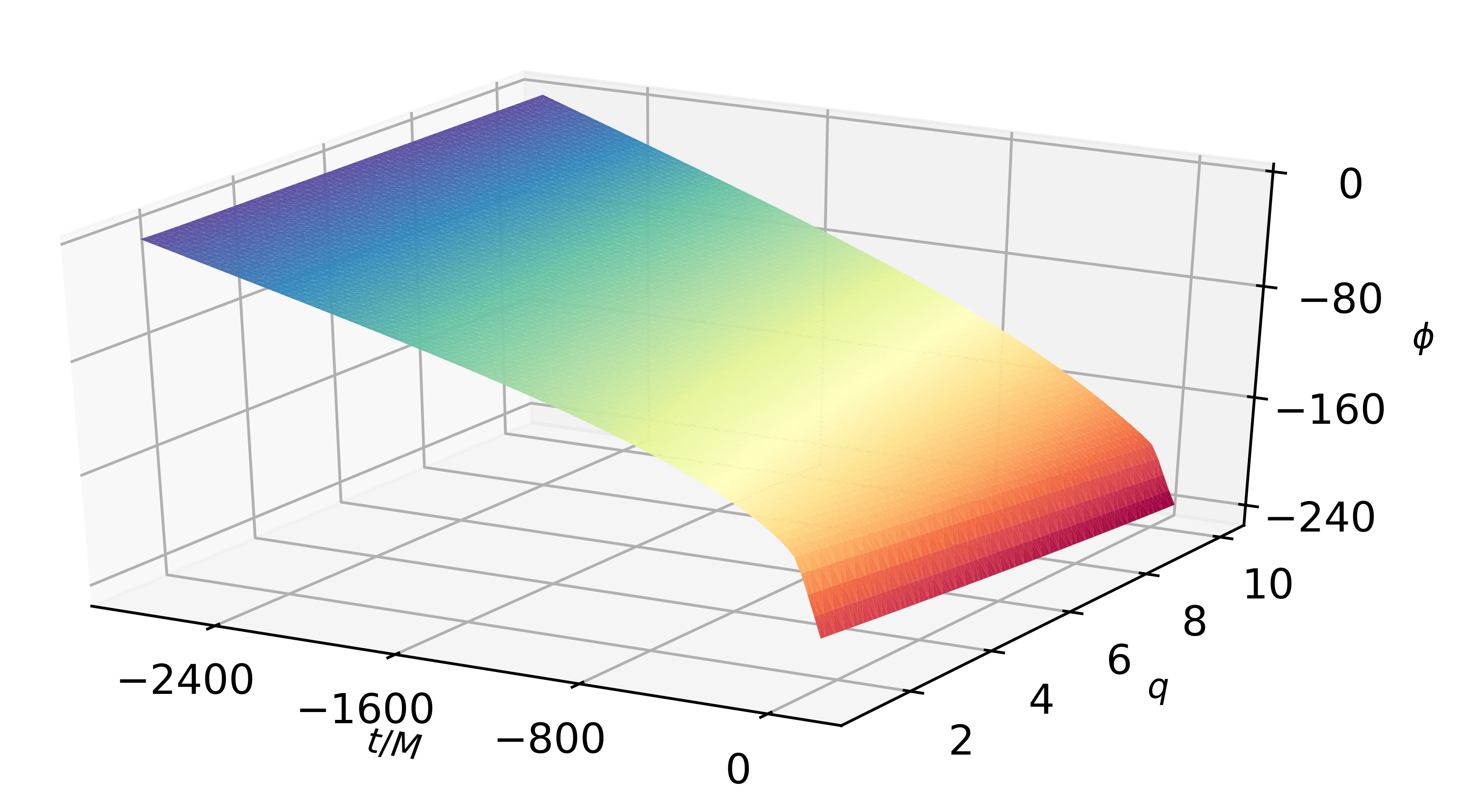}
\caption{Representation of amplitude $A(t, \lam=q)$ and phase $\phi(t, \lam=q)$ surfaces for the waveform mode $(2,2)$ corresponding to a symbolic model~\cite{tiglio2019ab} for the surrogate plotted in Fig.~\ref{fig:22mode} above.}
\label{fig:amp-phase}
\end{center}
\end{figure}

Similarly, in the precessing case using a co-precessing frame simplifies the structure of the waveforms, and it is advantageous to include the coordinate transformations themselves as components to model for. A question arises of whether to build a reduced basis for the waveforms and later build surrogate models for the system components, or to start by building bases for the latter and then reconstruct the waveform from these surrogate components or pieces. In general, the latter approach should be more accurate, even if at the expense of higher offline cost. 

\subsection{Empirical interpolant based surrogates} \label{sec:surrogate_eim}

The surrogate approach here described was introduced in Ref.~\cite{PhysRevX.4.031006}, the presentation of which we follow closely. It represents, with some conceptually relatively minor variations, the state-of-the art as of this writing. The approach has three offline stages, which are described below in decreasing order of computational cost, and a fourth, very fast to evaluate online surrogate model. \\

\noindent{\bf Offline stages}
\begin{enumerate}
\item Select the most relevant $n$ points in parameter space (shown as red dots in Fig.~\ref{fig:SurrogateSchematic}) using, for example, a greedy algorithm as described in Section~\ref{sec:greedy}. The waveforms/functions associated with these selections (shown as red lines) provide a nearly optimal reduced basis for the space of interest $\cF$.
\item Identify $n$ time (or frequency) samples of the full time series using the EIM, to build an interpolant that accurately reconstructs any fiducial waveform for all times if it is known at the EIM nodes. These nodes are shown as blue dots on the vertical axis in Fig.~\ref{fig:SurrogateSchematic}.
\item  At each EIM node perform a fit with the method of choice in the parameter dimension for the amplitude and phase of the waveform using the greedy data from Step 1. The fits are indicated by blue horizontal lines in Fig.~\ref{fig:SurrogateSchematic}.  
\end{enumerate}

\begin{figure}[H]
\begin{center}
\includegraphics[width=0.4\linewidth]{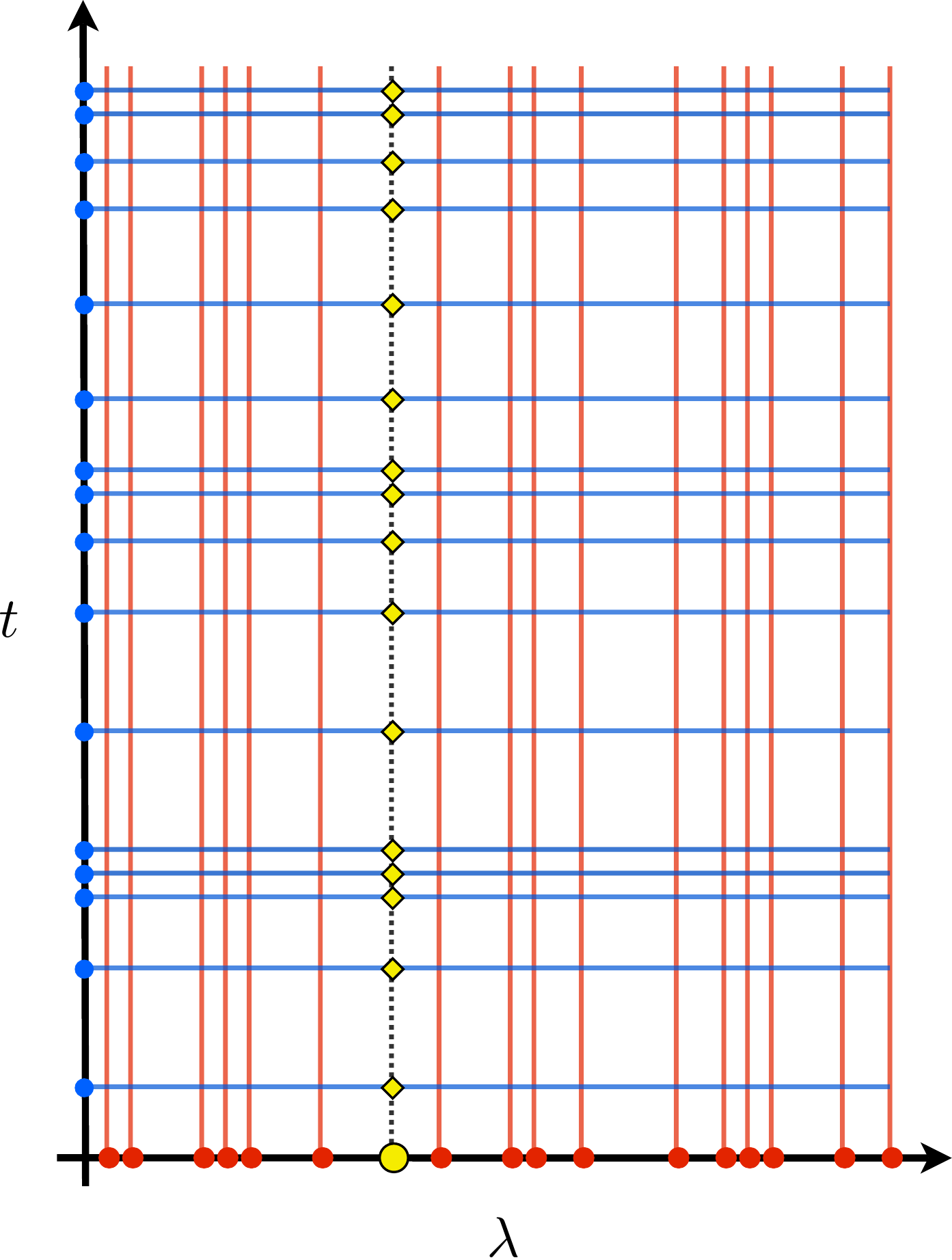}
\caption{A schematic of the method for building and evaluating the surrogate model, taken from \cite{PhysRevX.4.031006}. The red dots show the greedy selection of parameter points for building the reduced basis (Step 1, offline), the blue dots (Step 2, offline) show the associated empirical nodes in time from which a waveform can be reconstructed by interpolation with high accuracy, and the blue lines (Step 3, offline) indicate a 
fit for the waveform's parametric dependence at each EIM node. The yellow dot shows a generic parameter, which is predicted at the yellow diamonds and filled in between for arbitrary times using the EIM, represented as a dotted black line (Step 4, online).} 
\label{fig:SurrogateSchematic}
\end{center} 
\end{figure}

For further illustration, we show in Fig.~\ref{fig:eim_gw} the distribution of EIM nodes for gravitational waves corresponding to an EOB model, taken from~\cite{PhysRevX.4.031006}. Notice the sparsity of the distribution of EIM subsamples (nodes): only 19 empirical nodes are needed to reconstruct the {\it whole} time series consisting of roughly $65-70$ waveform cycles for the {\it entire} parameter space. From a signal processing perspective, this may sound quite strange, since the EIM sampling does not seem to meet the minimal Nyquist sampling needed to reconstruct the time series. This apparent contradiction is resolved by realizing that the reduced basis already encodes the physical dependency of the model, so, there is indeed information in the background working for a faithful reconstruction of the waveforms.

\begin{figure}[H]
\begin{center}
\includegraphics[width=0.7\linewidth]{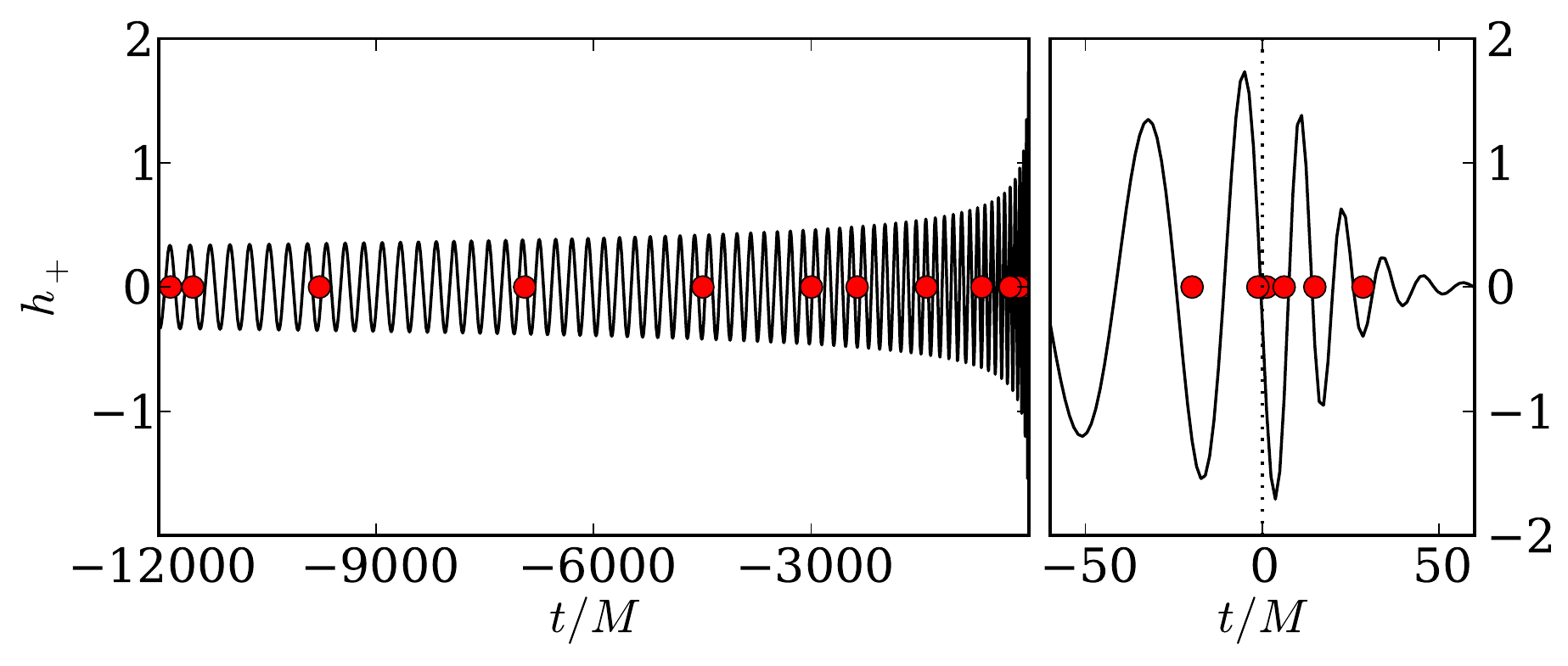}
\caption{Example of a distribution of the EIM nodes corresponding to a fiducial model composed by EOB waveforms. Only $19$ nodes are needed to represent $65-70$ cycles within about machine accuracy.  Figure taken from~\cite{PhysRevX.4.031006}.} 
\label{fig:eim_gw}
\end{center} 
\end{figure}

\noindent{\bf Step by step offline construction}\\

Put in a more specific way, the surrogate model is built as follows. Given $\{ \Lam_i \}_{i=1}^n$ greedy parameters and functions (waveforms) the empirical interpolant for an arbitrary parameter value $\lam$ has the form (Section~\ref{sec:eim})
\be
\cI[h](t, \lam) = \sum_{i=1}^nB_i(t) h(T_i, \lam) \approx h(t, \lam) \, , \label{eq:eim2}
\ee
where the $\{ B_i \}$ functions are computed offline from the basis, as well as the EIM nodes $\{ T_i \}$. 

In the case of non-spinning binaries, it is convenient to decompose all waveforms $h$ into their phase and amplitudes as in Eq.~(\ref{eq:comps}) and,  in particular at the EIM nodes, 
\be
h(T_i, \lam) = A(T_i, \lam) e^{i \Phi (T_i, \lam) } \, .  \label{eq:surrogate1}
\ee
For each $i=1\ldots n$, build fits for amplitude and phase, 
$$
A(T_i, \lam) \approx A^{\tt fit}(T_i, \lam) \quad , \quad \Phi(T_i, \lam) \approx \Phi^{\tt fit}(T_i, \lam)\quad \forall \lam \, . 
$$
Replacing these approximants, which are so far only valid for the EIM nodes $\{T_i\}$, as opposed to arbitrary values of $t$, into (\ref{eq:surrogate1}) yields
\be
h_{\tt s}(T_i, \lam) := A^{\tt fit} (T_i, \lam) e^{i \Phi^{\tt fit} (T_i, \lam) } \, .  \label{eq:surrogate1}
\ee
Finally replacing $h$ by $h_{\tt s}$ in Eq.~(\ref{eq:eim2}) leads to the final surrogate model for all values of $t$ and $\lam$, 
\be
h_{\tt s}(t, \lam):= \sum_{i=1}^n B_i(t) A^{\tt fit} (T_i, \lam) e^{i \Phi^{\tt fit}} (T_i, \lam) \quad \forall t,\lam\, . \label{eq:surrogate_final}
\ee
Expressions similar to Eq.~(\ref{eq:surrogate_final}) are the final expressions of the surrogates used so far within numerical relativity, while decomposing the waveforms into more components in the case of spin and precession. Importantly, once the training set has been processed to build the bases, the surrogate model only relies on the knowledge of the bases, the training set is no longer involved. Its accuracy (error estimates) and evaluation costs are discussed next. \\

\noindent{\bf Online stage, evaluation cost}\\

We now discuss the cost, in terms of operation counts, to evaluate a surrogate model. For simplicity we count every arithmetic operation as a single one regardless of their actual complexity. 
 
Also for simplicity and definiteness, still restricting the discussion to the non-spinning case, the complete surrogate model is given by Eq.~(\ref{eq:surrogate_final}), where the $n$ coefficients $B_i(t)$ and the $2n$ fitting functions $\{ A_i (\lam) \}_{i=1}^n$ and $\{ \phi_i (\lam) \}_{i=1}^n$ are assembled offline as described above.
 
In order to evaluate the surrogate model for some parameter $\lam_0$ we only need to evaluate each of those $2n$ fitting functions at $\lam_0$, recover the $n$ complex values $\{ A_i(\lam_0) e^{-i \phi_i(\lam_0)} \}_{i=1}^n$, and finally perform the summation in \eqref{eq:surrogate_final}.  
Each $B_i(t)$ is a complex-valued time series with $L$ samples, where $L$ is the desired number of time or frequency outputs. 

Therefore, the overall operation count to evaluate (\ref{eq:surrogate_final}) at each $\lam_0$ is $(2n-1)L$ plus the cost to evaluate the fitting functions. At least in one-dimension, the evaluation cost of polynomials of order $n_{\tt fit}$ using Horner's method (which is known to be optimal in terms of operation counts~\cite{Pan1966}) is $2n_{\tt fit}$. We discussed the exponential cost of evaluating polynomials in higher (than one) dimensions in Section~\ref{sec:multivar} and the final comments of that whole Section. We present some further discussions on this and related topics related to the fitting stage in Section~\ref{sec:challenges}. To summarize, the cost of evaluating the surrogate (\ref{eq:surrogate_final}) is 
\be
{\tt cost_{s}} = \left[ (2n-1) + c_{\tt fit }\right ] L
\ee
where at least in $1$-dimension $c_{\tt fit} \leq 2n$ and therefore 
\be
{\tt cost_{s}} \leq \left( 4n-1 \right ) L \sim \cO(L) \label{eq:cost_surrogate}
\ee
\begin{comment}
\begin{itemize}
\item The evaluation of the surrogate model (\ref{eq:surrogate_final}) is embarrassingly parallel in the number of outputs $L$ (each output evaluation is completely independent of the other ones) and therefore the total computational time can be speed up as wished. In contrast, time dependent equations (ordinary or differential) are very difficult to parallelize in time. 
\item The cost of most ordinary and partial differential equations solvers is at best linear in $L$. The speedup of the described surrogate approach for evaluation, if parallelized, is actually independent of the number of desired outputs $L$. 
\item In Eq.~(\ref{eq:cost_surrogate}), the linearity in $L$ is simply due to the number of independent output sample evaluations. The optimal cost would be ${\tt cost_s} = L$, corresponding to one arithmetic operation to evaluate the surrogate at any output $x$ (time or frequency, in the context of this review) of interest. If $n$ is small enough, as it happens in practice for GWs, the operation count $(\ref{eq:cost_surrogate})$ is in a precise sense, very efficient.  \\
\end{itemize}
\end{comment}

\noindent{\bf Error estimates} \\

One of the errors of interest for the complete surrogate model is a discrete version of the $L_2$ normed difference between a fiducial waveform and its surrogate. For definiteness we consider equally spaced $L$ time output samples,   
\be
\| h(\cdot, \lam) - h_{\tt S} (\cdot, \lam) \|^2 :=  \Delta t \sum_{i=1}^L\left| h(t_i; \lam) - h_{\rm S}(t_i; \lam)\right|^2 \, , \label{eq:l2_surrogate_error_d}
\ee
where $\Delta t = (t_\mathrm{max} - t_\mathrm{min})/(L-1)$. Other errors of interest are the pointwise ones for the phase and amplitude, 
\be \label{eq:pointwise}
\left | \frac{A(t; \lam) - A_{\rm S}(t; \lam)}{A(t; \lam)} \right | \, , \quad | \phi(t; \lam) - \phi_{\rm S}(t; \lam) |  \,.
\ee
The following error bound for the discrete error (\ref{eq:l2_surrogate_error_d}) can be derived:  
\be
\| h(\cdot, \lam) - h_{\tt S} (\cdot, \lam) \|^2\leq\Lambda_n \sigma_n + \Lambda_n  \Delta t \sum_{i=1}^n \left[  h(T_i, \lam) - h_{\rm S}(T_i, \lam) \right]^2  \, . 
 \label{eq:sErrorb}
\ee
For a proof, see Ref.~\cite{PhysRevX.4.031006}. This bound identifies contributions from two sources. The first term in the r.h.s. of (\ref{eq:sErrorb}) describes how well the EIM interpolant (i.e., the basis and EIM nodes) represents $h(t;\lam)$. The expected exponential decay of the greedy error $\sigma_n$ with $n$ along with a slowly growing Lebesgue constant $\Lambda _n$ results in this term being very small. The second term in the r.h.s of (\ref{eq:sErrorb}) is related to the quality of the fit. Currently, the fitting step is the dominant source of error in the surrogate models constructed, compared to the first two steps of generating the reduced basis and building the empirical interpolant. Improving on this source of error is still a remaining goal. 

\subsection*{Further reading} \label{sec:challenges}
Certified approaches refer to those which can a priori guarantee error bounds for the reduced model compared to the ground truth. Up to our knowledge these have been so far restricted to elliptic (coercive) partial differential equations. We again refer to \cite{quarteroni2011}, \cite{Jan-RB} and \cite{quarteroni2015reduced}. For an implementation of the Reduced Basis and Empirical Interpolation Methods in a concise and user-friendly API, see the Arby~\cite{arby} Python package, a fully data-driven module for building surrogate models, reduced bases and empirical interpolants from training data.

\newpage
\section{Surrogates of compact binaries}
\label{sec:binary-surrogates}

To date the most accurate binary black hole (BBH) numerical relativity (NR) surrogates have used a reduced bases-greedy approach, the EIM, and surrogate construction as discussed in this review, with some variations, and the SpEC code \cite{spec} for numerical relativity training simulations. However, other public catalogs of NR waveforms such as that one of the RIT group could be equally used \cite{PhysRevD.100.024021,RIT}. The surrogate catalog can be found in \cite {catalog} and can be evaluated with the gwsurrogate~\cite{gwsurr} Python package. Another package for surrogate evaluations is surfinBH~\cite{surfinBH}, built on top of gwsurrogate. What is found in all these surrogates is that when compared to numerical relativity, they are at least an order of magnitude more accurate than other existing models such as hybrid, phenomenological or effective ones. 

In addition to NR surrogates, we also discuss other ones using as fiducial models post-Newtonian, Effective One Body, and Ringdown approximations.

\subsection{Numerical relativity binary black holes}
\label{sec:bbh-nr-surrogates}

\begin{enumerate}
\item {\bf Non spinning (1 dim):}

The first BBH NR surrogate was presented in~\cite{PhysRevLett.115.121102}, and referred to as {\tt SpEC\_q1\_10\_NoSpin}. The black holes are initially non-spinning with initial orbital eccentricity smaller than $10^{-3}$, in the time range $[-2750, 100]M$ where, as is common practice, the waveforms have been aligned so that $t=0$ stands for the peak of the amplitude (which is around the merger of the two black holes), corresponding to about $15$ orbits before merger. This is a one-dimensional parameter problem, with the mass ratio $q$ chosen in the range $q:= m_1/m_2 \in [1,10]$ .

The selected $17$ greedy values were taken from an EOB model, seeded with existing five parameters corresponding to numerical relativity simulations using the Einstein equations. Next,  numerical relativity was used to solve for the remaining $17$ parameters and later the reduced basis built, of $22$ elements total. Impressively, the resulting surrogate model includes all  spherical harmonic modes up to $\ell =8$. 

One could also ask for a symbolic model representing the numerical surrogate one. In this line, in \cite{tiglio2019ab} the authors constructed the first ab initio free-form symbolic model (that is, analytical expressions in terms of elementary functions) for gravitational waves using symbolic regression (SR) through genetic programming. The fiducial model corresponds to the principal $(2,2)$-mode of the surrogate {\tt SpEC\_q1\_10\_NoSpin} described in \cite{PhysRevLett.115.121102}, which is taken as ground truth solution of the Einstein field equations, since it is practically indistinguishable (meaning, within the numerical relativity errors themselves) from supercomputer numerical simulations. The approach is ab initio, meaning no approximations to the Einstein equations are taken, such as stitching PN waveforms with EOB, NR and ringdown ones. The search for closed expressions is completely free, meaning that no prior hypothesis related to the type of functions is made. This is at the heart of genetic programming: successive models evolve under evolutionary pressure until reaching a tolerance error (or another stopping condition) without incurring in any human bias. To generate the training set of waveforms for SR, instead of performing naive samplings such as equally spaced grids (which showed prohibitive in terms of convergence times), the key step was to sample the time domain with the $22$ empirical interpolation nodes (EIM nodes) used in the assembly of the fiducial model {\tt SpEC\_q1\_10\_NoSpin}. This sampling represented the minimal set of pieces of information to represent the whole fiducial model without loss of structure, then accomplishing fast convergence in the regression instance and ending with closed-form expressions of maximum overlap error of 1\% with respect to the NR surrogate model. One of the salient features of these closed expressions is that they are not divided into expressions for the inspiral regime, for the merger, and finally for the ringdown, but cover the whole inspiral-merger-ringdown regime.

\item {\bf Spinning, motivated by GW150914 (4-5 dim):}

Continuing the work of Refs.~\cite{PhysRevLett.115.121102}, Ref.~\cite{PhysRevD.95.104023} presented the first spinning BBH surrogate model, with the parameter region motivated by the first GW detection, GW150914.

Two surrogates were built: {\tt NRSur4d2s\_TDROM\_grid12} and {\tt NRSur4d2s\_FDROM\_grid12}. The physical range is $q \leq 2$, dimensionless spin magnitudes $\chi_{1,2}\leq 0.8$, and the initial spin of the smaller black hole along the axis of the orbital angular momentum. The parameter region includes LIGO's first detection, GW150914, though with less cycles (using for training NR simulations with $25 - 35$ cycles before merger, while GW150914 has around $55$ cycles). 

{\tt NRSur4d2s\_TDROM\_grid12}  was built using $276$ NR simulations as training set and PN to find the greedy points, for which the PN surrogate reaches a floor error of $10^{-3}$. The parametric fits, used in the surrogate assembly, are fixed to a particular ``order''  -- not necessarily polynomial order, since trigonometric functions were included as part of the dictionary of fitting functions . The authors attribute the reason for this floor to that fixing, the general problem of this issue is discussed in Section~\ref{sec:chall}. Next, NR simulations were performed for those PN greedy points using a minimally rotating and co-precessing frame and, together with the coordinate transformations to an inertial frame, reduced bases for the different ``pieces'' or components of the waveform were built. 

The initial spin of the smaller black hole by construction lays along the axis of the orbital angular momentum, reducing the parameter dimensionality to $5$. Next, the azimuthal component of the spin of the larger black hole at a reference time $t_0 = t_{\text{peak}}-4,500M$ is included through an analytical approximation, thus effectively reducing the parameter dimensionality to 4, while modeling $5$ dimensions. Thus, the model does include a physical approximation and is not purely based on NR simulations, while being able to include precession in the modeling.

Model {\tt NRSur4d2s\_FDROM\_grid12} is built not from the 276 NR training set simulations but from the time domain surrogate {\tt NRSur4d2s\_TDROM\_grid12}, which can be quickly evaluated at a much larger number of points to populate a new, more dense, training set.

\item{\bf Spinning (7d):} 

The first surrogate model, named {\tt NRSur7dq2}, for the full 7-D parameter space of GWs emitted for a non-eccentric BBH coalescence was presented in \cite{PhysRevD.96.024058}. It used $744$ NR simulations to construct the training set with parameter ranges $q\leq 2$, $\chi_1, \chi_2 \leq 0.8$, for about $20$ orbits prior to merger, and $\ell \leq 4$. The authors  ``recycled'' the $276$ NR simulations used in the spinning 4-D case described above and complete the total of  $744$ NR simulations with a metric-based population criteria to select the remaining parameter points. Then the authors extend the number of simulations by means of symmetry arguments to $886$. As in \cite{PhysRevD.95.104023}, but with certain improvements, they decompose waveforms in data pieces and proceed to construct the surrogate following as a guideline the 4-D case \cite{PhysRevD.95.104023}. 

The {\it in-sample} errors computed for the $886$ NR waveforms show that the largest surrogate errors are comparable to the largest NR resolution errors ($\sim 10^{-2}$). For estimating the {\it out-of-sample} errors, the authors performed cross-validation over the training set by randomly dividing it in 20 sets of $44-45$ waveforms. They left out $1$ set at each step and built a trial surrogate for the remaining $19$ sets to compare it against the one that left out. This resulted in mistmatches similar to those of the {\it in-sample} case. The authors also compare the mismatches for a fully precessing EOB model ({\tt SEOBNRv3} \cite{PhysRevD.89.084006}) and for a phenomenological waveform model which includes some effects of precession ({IMRPhenomPv2} \cite{PhysRevLett.113.151101}). Mismatches more than an order of magnitude larger than the {\tt NRSur7dq2} surrogate model are found.

In \cite{Varma:2019csw} the 7-D parameter space is covered, for about $20$ orbits before merger, mass ratios $q\leq 4$, arbitrary spin orientations with dimensionless magnitudes $\chi_1, \chi_2 \leq 0.8$, $\ell \leq 4$ multipole modes, and initial orbital eccentricity also less than $10^{-3}$. This model, {\tt NRSur7dq4}, extends the previous 7-D one and is to date the most exhaustive and general surrogate model for BBHs from NR. The length of these simulations is sufficient to represent some but not all the BBH signals measured by LIGO and Virgo in the first two observation runs. It is proposed in the reference to hybridize the NR waveforms with PN approximations for higher values of mass ratio and spin magnitudes (see Sec.~\ref{sec:hyb}). In contrast to the above Ref.~\cite{PhysRevD.96.024058}, here $1,528$ NR simulations were used for the training set. In addition, a surrogate model, named {\tt NRSur7dq4Remnant}, is built for the mass, spin, and recoil kick velocity of the remnant black hole. To test their models, the authors perform a 20-fold cross-validation study on the training simulations. First they randomly divide the $1,528$ training simulations into 20 groups of $\sim 76$ simulations each. For each one, they built a trial surrogate using the $\sim 1,452$ remaining training simulations and test against these $\sim 76$ validation ones. They show that the mismatches for {\tt NRSur7dq4} against NR, computed with the Advanced LIGO design sensitivity noise curve, are always $\lesssim 8\times10^{-3}$ at the 95 percentile level over the mass range $50-200M$. For {\tt NRSur7dq4Remnant} the 95th percentile errors are $\sim 5\times10^{-4}M$ for mass, $\sim 2\times10^{-3}$ for spin magnitude, and $\sim 4\times10^{-4}c$ for kick magnitude. Compared to the spinning EOB waveform model {\tt SEOBNRv3}\cite{PhysRevD.89.084006}, they found in the two models an improvement of errors of at least one order of magnitude.
\end{enumerate}

\begin{table}[H]
\begin{center}
\begin{tabular}{ ||c|c|c|c|c|c|c|| }
\hline
Name & mass ratio & $\chi_{1,2} $ & precession & {\tt dim} & $\ell $ & GW cyles \\ [0.5ex]
\hline \hline
{\tt SpEC\_q1\_10\_NoSpin} & $\leq 10$ & 0 & no & 1 & $\leq 8$ & $\sim 25-31$ \\
\hline
{\tt NRSur4d2s\_TDROM\_grid12} & $\leq 2$ & $\leq 0.8$ & yes & 4 & $\leq 3$ & $\sim 25-35$ \\ 
\hline 
{\tt NRSur7dq2} & $\leq 2$ & $\leq 0.8$ & yes & 7  & $\leq 4$ & $\sim 40$ \\
\hline
{\tt NRSur7dq4} & $\leq 4$ & $\leq 0.8$ & yes & 7 & $\leq 4$ & $\sim 40$ \\ [1ex]
\hline
\end{tabular}
\end{center}
\caption{Characterization of the waveform surrogates described so far corresponding to NR binary black holes.}
\label{tab:surr}
\end{table}
\subsubsection*{Kicks} \label{sec:kicks}

Before the release of accurate surrogates for GWs covering the full 7-D parameter space of BBH dynamics, black hole kicks were mostly modeled with fitting formulas based on PN theory with the subsequent calibration based on NR simulations. Reference \cite{Gerosa:2018qay} presented the first effort to determine remnant properties from BH mergers using a ROM-based surrogate model. More specifically, the authors used the {\tt NRSur7dq2} model \cite{PhysRevD.96.024058} to generate a waveform template and analyze the linear momentum dissipation due to the emission of GWs. Their procedure provides the velocity accumulation profile ${\bf v}(t)$ and the final kick speed $v_k$ of the remnant black hole. The comparison for recoil speeds between NR simulations and {\tt NRSur7dq2} shows well agreement within an order of $\sim 10^{-4}c$, with some outliers in the order of $\sim 10^{-3}$c. The authors suggest that, even in the case where the surrogate accurately models post-merger strains, small errors might propagate to the phase of the center-of-mass-oscillation causing a relatively large error on the final kick velocity. 

More recently, in \cite{Varma:2018aht} the first surrogate models for remnant oscillations were constructed using Gaussian process regression (a type of machine learning regression method, see, for example, \cite{Doctor:2017csx}) and NR simulations for training. These fits are able to provide remnant mass, spin vector and recoil kick vector with high accuracy for 1) precessing BHs with mass ratio $q\leq 2$ and spin magnitudes $\chi_1 , \chi_2\leq 0.8$; 2) non-precessing BHs with mass ratio $q\leq 8$ and anti-aligned spin magnitudes $\chi_1 , \chi_2\leq 0.8$.

\subsection{Numerical relativity hybrid binary black holes}\label{sec:hyb}

 The previous surrogate models do not cover the entire LIGO band. To remedy this, in \cite{PhysRevD.99.064045}  a hybridized non-precessing model named {\tt NRHybSur3dq8} was presented. In that work NR waveforms are ``stitched'' at early times with PN and EOB ones, thus being able to cover the entire band of advanced LIGO with a starting frequency of $20$Hz and for systems with mass as low as $2.25 M_{\odot}$. This model is based on $104$ NR simulations for the 3-D parameter region $q\leq8$, $|\chi_{z1}|, |\chi_{z2}|\leq 0.8$ for modes $\ell\leq 4$ and $(5,5)$, excluding $(4,1)$ and $(4,0)$. To populate the training space, the authors performed $91$ NR simulations and completed for a total of $104$ with $13$ waveforms added using BHs exchange symmetry (equal mass, unequal spin). The parameters for the $91$ NR simulations were selected by a greedy procedure, iteratively constructing a PN surrogate model, testing it with a dense validation set and selecting the next greedy-parameter for the largest model error. At the hybridization stage, the early-inspiral waveforms were stitched with NR ones minimizing a cost function by varying the time and frame shifts between waveforms in an appropiate matching region. The matching region was settled to start at $1,000M$ after the start of the NR waveform and end after $3$ orbits of the binary inspiral. At the end, the authors find that its hybridized surrogate model {\tt NRHybSur3dq8} performs well within NR truncation errors and outperforms the {\tt SEOBNRv4HM} spinning-EOB model \cite{PhysRevD.98.084028} by about two orders of magnitude.
 
As an application case, {\tt NRHybSur3dq8} was used in \cite{PhysRevD.93.044064} to generate GWs to study tidal effects by means of a PN tidal splicing method. The resulting model was named {\tt NRHybSur3dq8Tidal}. It was also added to the last update of the SXS Collaboration catalog \cite{Boyle_2019} of numerical simulations for BBH coalescences.

\subsection{Extreme mass ratio inspirals}

In \cite{rifat2019surrogate} the authors built a surrogate model, named {\tt EMRISur1dq1e4}, for extreme mass ratio inspirals using non-spinning point particle black hole perturbation theory (ppBHPT) through numerical solutions of the Teukolsky equation with a point particle as source. The trajectory of the particle was determined by an adiabatic inspiral at early times, a late-stage geodesic plunge, and a transition region. The mass ratio was taken to be in the range $[3,10^3]$.

After a mass rescaling the surrogate model agrees remarkably well with NR waveforms (solving the full Einstein equations), which are available for mass ratios $q\leq 10$. The mass rescaling was empirical, in the sense that it was chosen as a function of the mass ratio and numerically chosen to minimize the difference with NR waveforms for the $(2,2)$ mode. Even so, the degree of agreement after rescaling is surprisingly good and unexpected since a priori there is no reason why such a good agreement should be present at all. As the authors point out, however, their result seems to be in line with growing evidence that suggests perturbation theory with self-force corrections might be applicable to nearly equal mass systems.

\subsubsection{Eccentric inspirals}

In \cite{Barta:2018tyg} the authors introduce a SVD-based ROM technique to model waveforms emitted by the coalescence of compact binaries with any residual orbital eccentricity. They apply this framework to eccentric-PN waveforms generated with the {\it CBwaves} open-source software \cite{Csizmadia:2012wy} and build a reduced order model for a 3-D subset of waveforms of the full 8-D parameter space corresponding to total mass $M$, mass ratio $q$ and eccentricity $e_0$. The ranges covered by the template bank were $2.15M_\odot \leq M\leq 215M_\odot$, $0.01\leq q\leq 1$ and $0\leq e_0 \leq0.96$.

The speedup in evaluating the surrogate model is $2-3$ orders of magnitude faster than generating the corresponding {\it CBwaves} waveforms, reaching a factor of several thousand around $10-50M_{\odot}$.

\subsection{Effective One Body}

In \cite{PhysRevX.4.031006} accurate surrogate models for EOB waveforms of non-spinning BBH coalescences were constructed using the Reduced Basis (RB) framework, corresponding to modes $(2, 1), (2, 2), (3, 3), (4, 4)$ and $(5, 5)$ with mass ratios from $1$ to $10$. The authors benchmarked the surrogate model against a fiducial one generated with the EOB solver of the LAL software package. For a sampling rate of $2048$ Hz they found a speedup of $\approx 2,300$, about three orders of magnitude faster than the LAL waveform model.

Reference \cite{Lackey:2016krb} presented a surrogate model of a non-spinning EOB waveform model with $l=2,3,4$ tidal multipole moments that reproduces binary neutron star (BNS) numerical simulation waveforms up to merger. The authors find, within the RB framework, that $12$ amplitude and $7$ phase basis elements are sufficient to reconstruct any BNS waveform with a starting frequency of $10$ Hz. The surrogate has maximum errors of $3.8$ in amplitude ($0.04$ excluding the last 100M before merger) and $0.043$ radians in phase. Following a different trend, Ref.~\cite{PhysRevD.100.024002} implemented Gaussian process regression to build a frequency-domain surrogate version for an aligned-spin BNS waveform model using the EOB formalism. The resulting surrogate has a maximum mismatch of $4.5 \times 10^{-4}$ and a speedup $O(10^{3})$ with respect to the original model.

As an alternative to a greedy-based method for ROM, Singular Value Decomposition (SVD) can be used to generate a reduced basis for the GWs. Along this line, Refs.~ \cite{Purrer:2014fza} and \cite{Purrer:2015tud} presented two frequency-domain reduced order models for EOB models {\tt SEOBNRv1} \cite{PhysRevD.86.024011} and {\tt SEOBNRv2} \cite{PhysRevD.89.061502}, respectively. These surrogates are built upon an SVD-based method to construct reduced basis and implement tensor product splines as interpolation method. The surrogate for {\tt SEOBNRv2} is a spin-aligned model for the GW dominant $(2,2)$ mode and extends the spin range of the first surrogate to almost the entire Kerr range. It also covers the entire parameter space in which the first one is defined: symmetric mass ratios $0.01\leq \eta \leq 0.25$ and spin magnitudes $-1\leq \chi_i \leq 0.99$. In general, the mistmatches are better than $\sim 0.1 \%$ against {\tt SEOBNRv2} except in regions of parameter space in which the original model presents discontinuities, inducing mistmatches $\sim 1 \%$ in the surrogate.

\subsection{Post-Newtonian (PN)}

The first contact between the Reduced Basis method with gravitational waveform modeling occurred in \cite{Field:2011mf}. The authors built a reduced basis for 2PN waveforms corresponding to a 2-D space of non-spinning BNS inspirals with mass components in the range $[1,3]M_\odot$. They found that, remarkably, to machine precision error, only $921$ basis elements are needed to represent the full template bank used as fiducial model. Also the greedy-approach here used was compared against a metric template placement method, finding exponential decay of the greedy error with the number of bases as opposed to the approximately linear convergence rate of the metric approach. 

Later on, in \cite{Field:2012if} a reduced basis was built for the non-precessing case of BBH inspirals (4-D parameter space: $2$ masses, $2$ aligned or anti-aligned spins) using the restricted TaylorF2 PN approximation with component masses in the range $[3,30]M_\odot$ and dimensionless spin magnitudes in the full range $[-1,1]$. The authors found that, for a tolerance error of $10^{-11}$, when increasing the dimensionality of the parameter space from 2-D to 4-D the number of basis elements needed to span the whole space of waveforms increased only in $6.6\%$ with respect to the $1,725$ bases needed for the 2-D case. Furthermore, going from the 3-D to the 4-D case implied adding only $15$ more basis elements. This opened the possibility that the curse of dimensionality could be beaten for the complete 8/7-D case, as discussed in Section \ref{sec:chall}.

In \cite{Blackman:2014maa} the problem of building a reduced basis for the full 7-D case, where there are no closed-form expressions and ordinary differential equations need to be solved, was tackled. The waveforms correspond to 3.5PN precessing inspirals (mass ratio $q\in[1,10]$ and dimensionless spin magnitudes $||\chi_i||\leq 0.9$). It used a modified version of the standard greedy algorithm to randomly resample the 7-D parameter space at each iteration with a fixed number $K$ of waveforms. This was crucial to overcome the limitations imposed by the curse of dimensionality against the construction of a densely populated training space. The sampling number $K$ was increased in each run until reaching $K=36,000$ for which the number of RB waveforms became independent of $K$. Choosing to work in the binary's precessing frame, the authors exploited the fact that in this frame the waveforms have a weaker parametric dependence than they do in the inertial one. Another important ingredient was the choice of waveform parametrization in phase instead of time or frequency, taking advantage of the smooth dependence of waveforms on this variable. They find that with all these modifications, only $50$ waveforms are needed to represent the entire 7-D space with an error of $10^{-7}$.

\subsection{Ringdown} \label{sec:ringdown}

With the advent of GW astronomy since the first detection of the GW150914 event in 2015 a new era for testing general relativity and alternative theories of gravity in strong regimes \cite{PhysRevLett.123.111102,Berti:2018cxi,Zhao:2019suc,Radice:2016rys} has opened. In the case of compact binary coalescences different techniques were developed in recent years \cite{PhysRevD.88.122002,PhysRevD.98.024052,OBrien:2019hcj} to improve the extraction of information after merger. As it is well known, the account of post-merger properties is accomplished through the study of the quasinormal modes (QNMs) of the remnant Kerr black hole. Different models for ringdown waveforms have been constructed in rather recent years through several techniques. See for example \cite{London:2015ytg,London2014,London:2018gaq,London:2018nxs} for recent applications of greedy and regression methods in the construction of QNM models.

In practice, single-mode ringdown searches can limit the number of measured events in advanced ground-based detectors such as LIGO. Moreover, parameter estimation errors can become large for such single-mode searches when the actual waveform contains a second mode. Besides this, there are further motivations on multi-mode searches, such as consistency tests of GR (e.g., the no-hair theorem~\cite{Berti:2007zu,Dreyer:2003bv,Berti:2005ys}) and feature inference about the progenitors of the final black hole. Looking forward for future multi-mode ringdown searches, the Reduced Basis scheme was implemented in \cite{Caudill:2011kv} to construct a compact representation of multi-mode QNM catalogs. A single QNM waveform has the form
\be
h_{lmn}(t)={\cal A}_{lmn}(\Omega) \frac{M}{r} e^{- \pi \frac{ f_{lmn}}{Q_{lmn}}t} \cos{(2\pi f_{lmn}t)} \, , 
\ee
where the different symbols are defined as ${\cal A}_{lmn}(\Omega):=$ orientation-dependent dimensionless amplitude; $r:=$ distance to the source; $M:=$ black hole mass; $f_{lmn}:=$ central frequency; $Q_{lmn}:=$ quality factor, and geometric units $G=c=1$ are taken for granted. Starting from the one-mode case, a template bank of ringdown waveforms was constructed for the $(l,m,n)=(2,2,0)$ and $(3,3,0)$ QNMs. Asking for several minimal matches $(MM)$ between template and signal the parameter space is filled using a metric-based population criteria over the 2-D space $(f,Q)$. For example, a minimal match $MM=0.99$ corresponds to a catalog of $2,213$ waveforms with spin magnitude of the remnant black hole being in the range $[0,0.9947]$ and mass $M$ in the range $[2.9744,3025.7]M_{\odot}$. Table \ref{fig:table1} shows the number of RB elements needed for each case. 

\begin{figure}
\begin{center}
\includegraphics[width=0.4\linewidth]{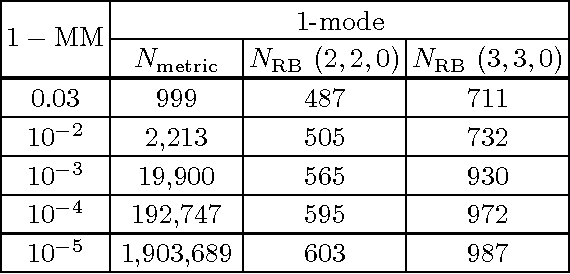}
\caption{Taken from \cite{Caudill:2011kv}. Number of reduced basis waveforms ($N_{RB}$) and metric-based templates $N_{metric}$ needed to represent one-mode QNM training spaces with $(l, m, n) = (2, 2, 0)$ and (3, 3, 0) for different minimal matches MM. The training space representation error is $\epsilon=10^{-12}$. For $\tt dim=2$, $N_{metric}$ scales with MM as $N_{metric}\propto (1-MM)^{-1}$.}
\label{fig:table1}
\end{center} 
\end{figure}

The RB representation was validated through a Monte Carlo simulation by randomly sampling the parameter space with $10^7$ waveforms. An average waveform representation error $||h_{\lambda}-\cP_N h_{\lambda}||^2\approx 4.21\times 10^{-13}$ was found, one order of magnitude better than the maximum training space representation error $\epsilon^2=10^{-12}$. For the two-mode case, the authors linearly compose the previous single modes $(2,2,0)$ and $(3,3,0)$ and find similar results than in the one-mode case, since the algorithm takes advantage of the linearity of the QNMs. Following this observation, the authors finally propose a simple method to represent unconstrained multi-mode waveforms by approximating it by the sum of individual one-mode projections.

\part{Data Analysis}

\newpage 
\section{Reduced Order Quadratures} \label{sec:roq}

We have so far presented a method to produce a reduced model which can be used as a representation for an underlying set of functions. This combines the RB and EIM frameworks to produce  accurate representations. Finally, one builds a surrogate predictive model through interpolation in parameter space at EIM nodes.  Next we discuss how this framework can be used to compute numerical approximations of integrals (quadratures).

Reduced order quadratures (ROQ) were introduced, at least in the context of GWs, in~\cite{antil2012two}. They use the EIM to build a set of nodes and weights to construct the integral, so the main difference with standard quadratures is that they are application-specific for parametrized problems. This results in fast online quadrature evaluations, which are at the core of data analysis when computing correlations, likelihoods, etc. 

\begin{comment}
Reduced Order Quadratures choose a nearly optimal subset of nodes from the training points in the physical domain (time, frequency, space) at which any given basis is known. These training points can be arbitrarily located: they can be equally spaced, randomly distributed, etc. This is an important aspect for many practical purposes, such as experimental data, where one might not be able to dictate when signals are measured, and it is in sharp contrast with fast converging Gaussian quadratures, which do dictate the time at which data should be measured. The latter might not only be impractical but also unfeasible in many experimental scenarios.   
\comment Being based on the EIM, the method naturally applies to multiple dimensions, unstructured data and meshes of arbitrary shapes. 
\comment By design, also due to being based on the EIM, very fast convergence with the number of ROQ nodes is observed, typically exponentially in the cases of interest. This would not be possible using standard methods which rely on smoothness of the functions to be integrated. More precisely, in the case of GWs, in order for Gaussian quadratures to converge fast the signal would have to be smooth as a function of time, which is never the case for GW signals due to the presence of measurement noise. In contrast, ROQ lift this requirement and in practice achieve exponential convergence {\em even for noisy data}.  
\end{comment}

The idea of ROQ is remarkably simple for its impact:
\begin{enumerate}
\item Step 1. Build a reduced basis for the problem of interest. The basis can be built, for example, using a POD or greedy approach.  
\item Step 2. Build an empirical interpolant, based on the previous basis, as discussed in Section~\ref{sec:eim}. This approximates a parametrized function $f(x, \lam)$ in the space of interest by an empirical interpolant of the form
\be
{\cal I}_n [f](x, \lam) := \sum_{i=1}^n B_i(x) f(X_i, \lam)  \approx f(x, \lam) \label{eq:EIM2}
\ee
In GW science $f$ would typically be a waveform, and $x$ time or frequency but the method is generic. 

The affine parameterization in $\lam$ and physical dimensions $x$ achieved by (\ref{eq:EIM2}) is one of the critical ingredients that allows for ROQ. By affine parameterization one means a decomposition in terms of products of functions which depend only on parameter and physical variables, as in (\ref{eq:EIM2}). 

\item Step 3. Reduced order quadratures follow what would otherwise be the standard procedure when building quadratures, but replacing standard polynomial interpolants with EIM based ones. Namely,  
\be
I_n [ f ](\lam) := \int  f(x, \lambda) dx \approx \int {\cal I}_n [f](x,\lam) dx  = \sum_{i=0}^n w_i f(X_i, \lam) \, ,  \label{eq:ROQ} 
\ee
where the ROQ weights 
\be
\omega_i := \int B_i(x) dx \label{eq:roqw}  
\ee
are computed offline, with any quadrature method of choice or availability for computing (approximating) the integrals in Eq.~(\ref{eq:roqw}).
\end{enumerate}

\begin{comment}
If the number of available (say, time or frequency) samples is $L$, then one would typically use them to precompute the weights (\ref{eq:roqw}) in the offline stage. After this offline work, the online evaluation is decreased to $n$ ROQ nodes, with no practical loss of accuracy and usually $n \ll L$, as we discuss in Section~\ref{sec:PE}, leading to dramatic speedups in likelihood computations, among other applications. 
\end{comment} 

\subsection{ROQ, other quadrature methods, dependence on dimensionality} \label{sec:roq_multidim} 

Gaussian quadratures (such as Chebyshev and Legendre) are considered some of the most efficient methods for integrating smooth generic functions. This is not just a perception: there is a whole theory of why this is generically the case. Compared to ROQ, though, they do suffer from several disadvantages, mentioned in the Discussion~\ref{disc:global_interp} of Section~\ref{sec:runge}:
\begin{enumerate}
\item The location of their nodes, at which the integrand has to be known, is dictated by the method, which is unrealistic for any experiment or application based on observations. 
\item They are not hierarchical. That is, more nodes cannot simply be added for higher accuracy but each new quadrature has to be built from scratch. 
\item Their fast convergence and near-optimality is subject to very specific conditions: smoothness of the integrand being the most important one after the imposition of the nodes location. This is an unrealistic assumption for signals taken from experiments. 
\item As with any polynomial-based approach, their efficiency {\em in general} (as when using grids based on tensor products of one-dimensional ones) decreases with the dimensionality of the problem, unlike ROQ for cases of interest in this review.  
\end{enumerate}
As we discuss next, ROQ can beat these best known generic methods while working under more relaxed conditions. This is at the expense of offline work. Obstacles (1) and (2) are well known, so there is not much need to comment on them. In Refs.~\cite{PhysRevLett.114.071104, Canizares:2013ywa} point (3) is discussed and how ROQ can lead to fast convergence even in the presence of noisy data. 

So next we discuss point (4), by considering two simple examples, in one and two dimensions (both physically and parametrically). They are taken from reference \cite{antil2012two}.

\begin{example} \label{ex:roq1d}

The example function family of interest here is (there is nothing particular about this choice)
\be
f(x,  \lam) =  \left[ \left( x - \lam_1 \right)^2 + 0.1^2 \right]^{-1/2} \label{eq:roq1d} \, , 
\ee
where both $x$ and $\lam$ are real and we want to compute an approximation to its integral, 
\begin{align} \label{eq:Int1d}
I (\lam) = \int_\Omega f(x, \lam) dx \, , \quad
\end{align}
\end{example}
with, out of arbitrariness, $\Omega = [-1,1]$ and $\lam_1 \in  [-0.1,0.1]$. We use two approaches to approximate (\ref{eq:Int1d}):
\begin{enumerate}
\item {\bf Gaussian quadratures:} Numerical integration is carried out using up to $n = 150$ nodes and a  Gauss-Legendre rule. Recall that for each value of $\lambda $ a separate quadrature for each value of $n$ has to be performed since the method is not hierarchical. 

\item {\bf Reduced Order Quadratures:} A reduced basis (in this case using a greedy approach) is first built for the family of functions (\ref{eq:roq1d}), subsequently the empirical interpolant and nodes and, finally, the ROQ. Since ROQ are hierarchical the additional cost for showing a convergence test from $n$ to $(n+1)$ is independent of $n$, only an extra node needs to be added. This, again, is in contrast to regenerating each quadrature rule as in the Gaussian case. 
\end{enumerate}
The results are shown in Figure~(\ref{fig:ROQ_2D}), as black solid and dashed lines. Even for an error of at most $10^{-4}$, ROQ provide a factor of $\sim 4$ savings, with the savings dramatically increasing with higher accuracy. We recall once again that we are comparing ROQ with one of the best general purpose quadrature rules.

\begin{figure}[ht]
\centering
\includegraphics[width=0.5\linewidth]{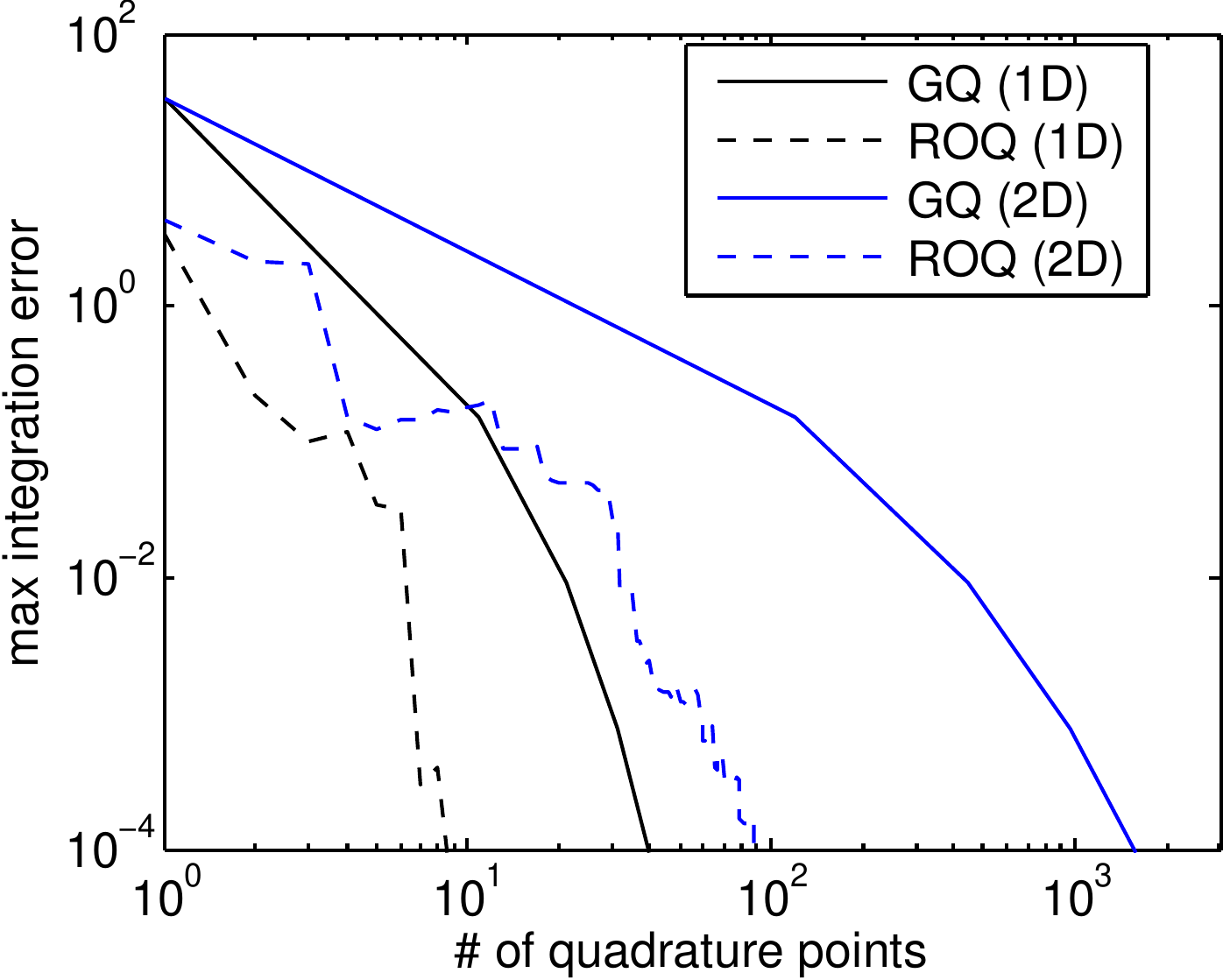}
\caption{
Error curves for the 1-D and 2-D dimensional cases of Examples~\ref{ex:roq1d} and \ref{ex:roq2d} from Ref.~\cite{antil2012two}, respectively, using Gauss-Legendre and ROQ rules. Errors are computed by taking the maximum over the entire training set. The ROQ savings increase from $\sim 4$ to $\sim 12$ as the number of spatial dimensions is increased from one to two. Further savings are expected as the number of spatial dimensions increases. Note that the non-monotonicity of the ROQ curves is because the EIM does not optimize for accuracy at each step, this is discussed in Section~\ref{sec:cond}.}
\label{fig:ROQ_2D}
\end{figure}
Approximating functions by products of functions (in particular of polynomials) in each physical dimension leads to evaluation costs of the approximants which scale exponentially with the dimentionality. As with higher accuracy, increasing the dimensionality of the problem leads to ROM providing larger savings when compared to generic methods. We next give an example.

\begin{example} \label{ex:roq2d}

The family of functions of interest is now $2$-dimensional, both in space and parameters, 
\be
f(x,\lam) = \left[ \left( x_1 - \lam_1 \right)^2 + \left(x_2 - \lam_2 \right)^2 + 0.1^2 \right]^{-1/2}  \label{eq:roq2d} \, , 
\ee
where the physical domain is $x = (x_1,x_2) \in \Omega = [-1,1]\times  [-1,1]$ and for the parametric one $\lam \in [-0.1, 0.1] \times [-0.1, 0.1] $, and we are interested in multiple evaluations of the parametrized integral
$$
I (\lam)= \int_{\Omega} f(x, \lam) dx\, . 
$$
When considering Gaussian quadratures the curse of dimensionality already in two dimensions becomes apparent. In standard multidimensional quadratures the integrand is approximated by the product of one-dimensional polynomials. Therefore we consider up to $150^2$ Gauss nodes to integrate functions from the family (\ref{eq:roq2d}). And, again, because Gaussian quadratures are not hierarchical, this requires a set of $150^2$ quadrature rules for each quadrature evaluation in a convergence test. 

Next, we build a reduced greedy basis, EIM, and ROQ. The convergence of both methods, Gaussian and ROQ, are displayed as solid and dashed blue lines in Fig.~\ref{fig:ROQ_2D}. Instead of a factor $\sim 4$ savings as in the 1-D case for a modest error of $10^{-4}$, for this same accuracy the savings of ROQ are $\sim 12$. 
As we will show in the next section, for realistic problems in GW physics, the savings are much larger than those of these test problems. 
\end{example}

\section{Accelerating Parameter Estimation with ROQ} \label{sec:PE}

Once a detection of a gravitational wave is made one would like to infer the astrophysical properties of the source which emitted it. The goal here is to do a full Bayesian analysis to compute the posterior probability distribution function (PDF) of a set of astrophysical parameters ($\lam$) which describe the source. Being able to do so in real time (or nearly so) is particularly important for rapid followups of electromagnetic counterparts enabling multi-messenger astronomy, among other motivations. For this reason we place emphasis on binary neutron stars, or mixed pairs.  

Assuming that the detector data $d$ contains the source's signal $h(\lam_{\tt true})$ and stationary Gaussian noise $n$ in an additive way \footnote{Here the noise is denoted by $n$, as is the dimensionality of a reduced basis throughout this review. Hopefully, from context this should not cause confusion.},
$$
d = h(\lam_{\tt true}) +  n\, , 
$$
the likelihood of data $d$ corresponding to a parameter $\lam$ is given by~\cite{jaranowski2009}
\be
\mathcal{L}( d | \lam ) = A \exp [ -\frac{1}{2} (d - h (\lam ), d- h (\lam ) ] \, . \label{eq:prob}
\ee
The evaluations of likelihoods for parameter estimation (PE) is a high dimensional one. For binary systems, even excluding equations of state in the presence of matter, it is already a $15$-dimensional problem, with $8$ parameters being {\it intrinsic} (mass and spin vector of each binary component) and $7$ {\it extrinsic} ones such as luminosity distance, coalescence time, inclination of the binary plane, and sky localization.  

The bulk of the computational cost in (\ref{eq:prob}) comes from two sources:
\begin{enumerate}
\item Computing the gravitational wave candidates $h(\lam)$. 
\item Evaluating the inner products in (\ref{eq:prob}) . 
\end{enumerate}
Surrogate models can decrease the computational cost of (1), and ROQ those of (2). In this section we briefly review (2); that is, the use of ROQ and ROM in general for accelerating parameter estimation (PE). Therefore, this section is intendently limited in scope, and it is not intended to be a general review of approaches for accelerating PE, which is a field by itself given its importance.  

Our presentation follows closely that one of Ref.~\cite{Canizares:2013ywa}. 

In Eq.~(\ref{eq:prob}) $(a | b)$ is a weighted inner product for discretely sampled data. More explicitly, and working in the frequency domain, of the form
\be
(d|h(\lam)) = 4\mathbb{R} \ \Delta f \sum_{i=1}^{L} \frac{\bar{d}(f_i) h(f_i, \lam)}{S_{n}(f_i)}\, , \label{e:inner}
\ee
where $d(f_i)$ and $h(f_i, \lam)$ are the (discrete) Fourier transforms at frequencies $\{ f_i \}_{i=1}^L$, a bar denotes complex conjugation, and the power spectral density (PSD)  $S_{n}(f_i)$ characterizes the detector's noise. 

For a given observation time $T=1/\Delta f$ and detection frequency window $(f_{\tt high} - f_{\tt low})$ there are 
\be
L = 
{\tt int}\left( \left[ f_{\tt high} - f_{\tt low} \right] T \right) \label{eq:Ldef}
\ee
sampling points in the sum (\ref{e:inner}). Usually $L$ is large, which determines the second bulk of the computation when evaluating the likelihood (\ref{eq:prob}). 

\subsection{Constructing the Reduced Order Quadrature} \label{sec:roq-construction}

The ROQ scheme for parameter estimation requires three steps, as described in Section~\ref{sec:roq}. Here we slightly reformulate them in order to bring up some practical issues. These steps are:
\begin{enumerate}
\item Construct a reduced basis, i.e. a set $n$ elements whose span reproduces the GW model with a desired precision. 
\item Using the EIM, construct the empirical interpolant and its nodes. 
\item The ROQ weights (\ref{eq:wk}) are computed, and are used to replace, without loss of accuracy, inner product evaluations~\eqref{e:inner} by ROQ compressed ones.\\

\end{enumerate}

\begin{comment}
\begin{itemize}
\item \noindent {\it Step 1.} 
The reduced basis only needs to be built over the space of intrinsic parameters for the waveform family. Furthermore, if the basis is generated using a PSD $\equiv 1$ the representation of the waveform family can be used with any PSD whenever the weights are built as in Eq.~(\eqref{eq:wk}).
 
In practice, for the purposes of this Section, basis generation proceeds in two stages. A reduced basis is first constructed, which can be evaluated for any value of. Next, this basis is evaluated at $L$ equally spaced frequency samples appropriate for the detector. 

\item \noindent {\it Step 2.}  Given an accurate basis of dimension $n$, it is possible to uniquely and accurately reconstruct any waveform $h(\lam)$ from only $n$ subsamples $\{ h(F_k, \lam)\}_{k=1}^n$ using the EIM. The frequency nodes $\{F_k\}_{k=1}^n$, selected from the full set $\{f_i\}_{i=1}^L$, are also obtained by the EIM. 

This step provides a near-optimal compression strategy in frequency which is complimentary to the parameter one of Step (1). The model's empirical interpolant, valid for all parameters, can be written as (cf. Eq.~(19) of Ref.~\cite{PhysRevX.4.031006})
\be
{\cal I} [f](f_i, \lam) =  \exp(-2 \pi \mathrm{i} t_{c} f_i ) \sum_{j=1}^n B_j (f_i) h (t_c=0, F_j, \lam ) \approx  h(f_i, \lam)\, ,
\label{eq:EIM}
\ee 
where, for the sake of the discussion below, we have temporarily isolated the coalescence time $t_{c}$ from the other
extrinsic parameters. 

\item \noindent {\it Step 3.} Except for $t_c$, no extrinsic parameter affects the construction of the ROQs. The coalescence time, however, requires special treatment. One can see this by substituting Eq.~(\ref{eq:EIM})  into Eq.~(\ref{e:inner}), 
\be
(d|h(t_{c}, \lam)) 
= \sum_{k=1}^n \omega_k(t_{c}) h(t_c=0, F_k, \lam) \, ,
\label{e:roq}
\ee
with the ROQ weights given by 
\be
\omega_k(t_{c}) = 4\mathbb{R} \Delta f \sum_{i=1}^L \frac{ \bar{d}(f_i) B_k (f_i)}{ S_n(f_i) } \exp(-2 \pi \mathrm{i} t_{c} f_i ) \, .
\label{eq:wk}
\ee

\end{itemize}
\end{comment}
In practice, the dependence of (\ref{eq:wk}) on $t_c$ can be achieved through a simple domain decomposition, using that an estimate for the time window $W$ centered around the coalescence time $t_{\tt trigger}$ is given by the GW search pipeline. This suggests a prior interval $[ t_{\tt trigger} - W, t_{\tt trigger} + W] $ to be used for $t_c$. This prior interval is then split into $n_c$ equal subintervals of size $\Delta t_c$. The number of subintervals is chosen so that the discretization error is below the measurement uncertainty on the coalescence time.  Finally, on each subinterval a unique set of ROQ weights is constructed. 

Step (3) is currently implemented in the LALInference pipeline as summarized in Algorithm~(\ref{alg:Weights}). The offline steps (1) and (2) are carried out independently. By construction, the approach guarantees that these offline steps need be to carried out only once for each waveform family model. \\

{\scriptsize
\begin{algorithm}[H]
\caption{Computing the ROQ weights for parameter estimation} 
\label{alg:Weights}
\begin{algorithmic}[1]
\vspace{0.2cm}
\State {\bf Input:} $d, S_n, \{ B_j \}_{j=1}^{N}, \Delta f, t_{\tt trigger}, W, \Delta t_c$.
\vskip 10pt
\State Set $n_c=  {\tt int} \left( \left( 2 W \right)/\Delta t_c \right) + 1$ 
\For{$j = 1 \to n_c$} 
\State $T_j = t_{\tt trigger} - W + \left(j - 1 \right) \Delta t_c$
\For{$k = 1 \to N$}
\State Compute $\omega_k(T_j)$ via Eq.~(\ref{eq:wk})
\EndFor
\EndFor
\vskip 10pt
\State {\bf Output:} $\{ T_j \}_{j=1}^{n_c}, \{\{ \omega_k(T_j) \}_{k=1}^N\}_{j=1}^{n_c}$.
\end{algorithmic}
\vspace{0.2cm}
\end{algorithm}
}

\subsubsection*{Compressed Norm Evaluations} \label{sec:norms}

There are two more terms to consider for fast computations of inner products, which can be seen from Eq.~(\ref{eq:prob}). One of them is $(h(\lam) |h(\lam ) )$. 
Unless the GW model is a closed-form expression, one needs to build a fast online evaluation for this norm for values of $\lam$ that are only known at run time. This can be achieved by constructing a reduced basis for $(h, h )$, then its empirical interpolant, and finally its ROQ.  
The other term, $(d|d)$, only needs to be computed once per parameter estimation or search analysis, so it does not require any special treatment to speed up its calculation. 

\subsection*{Total speedup}
Notice that even though the ROQ weights (Alg.~\ref{alg:Weights}) are computed in the online stage, they only depend on the detection-triggered data $d$ and not on any $h(\lam )$. Therefore, they can be computed in what can be referred to as the  {\it startup stage}, which requires $n$ full (of size $L$) inner product (\ref{e:inner}) evaluations for each $t_c$ interval. As discussed below, in practice this cost is negligible, while each likelihood is subsequently calculated millions of times, leading to significant speedups in parameter estimation studies, resulting in observed speedups in the whole PE study equal to $L/n$, which is the reduction in the number of terms needed to compute  (\ref{e:roq}) instead of  (\ref{e:inner}). This assumes that $n<L$ but, as we see below, this is indeed the case in problems of interest (furthermore, usually $n \ll L$).  

This speedup comes from operation counts, but it has also been observed in practice in actual implementations in the LIGO Algorithm Library (LAL) pipeline \cite{LAL_software}, as discussed below. For BNS,  for the early advanced detectors' configuration \cite{2013arXiv1304.0670L} ROQ showed to provide a factor of $\sim30\times$ speedup in PE for low-frequency sensitivity of $40$Hz, and $\sim 70\times$ and $\sim150\times$ as the sensitivity band is lowered to $20$Hz and $10$Hz, respectively; in all cases without practical loss of accuracy or systematic biases. 

\begin{example}
The material for this example is taken from Ref.~\cite{PhysRevLett.114.071104}. 

The majority of a binary neutron star's GW signal can be expected to be in the inspiral regime \cite{lrr-2009-2}, which can be described by the closed-form TaylorF2 approximation~\cite{Buonanno:2009zt}. While TaylorF2 does not incorporate spins or the merger-ringdown phases of the binary's evolution, these might not be important for BNS parameter estimation and can therefore be neglected, at least in a first approximation~\cite{Singer:2014qca}. For a thorough study on this point using the SpinTaylorF2 approximation, see~\cite{PhysRevD.92.044056}. 

Even for this simple to evaluate waveform family, inference on a single data set used to require significant computational wall-time with standard parameter estimation methods~\cite{2013PhRvD..88f2001A}. 

In Ref.~\cite{PhysRevLett.114.071104} the authors first computed the 
observation time $T$ required to contain a typical BNS signal. Next, a reduced basis of dimensionality $n$ needed to represent this model for any pair of BNS masses was constructed. The upper frequency $f_{\tt high}$ was fixed to $1024$Hz while $f_{\tt low}$ varied between $10$Hz and $40$Hz. 

The time taken for a BNS system with an initial GW frequency of $f_{\tt low}$ to inspiral to
$1024$Hz,
\begin{align}
\label{eq:fitL}
T_{\tt BNS} =  \left[ 6.32 + 2.07\times \frac{10^6}{ \left( f_{\tt low}/\text{Hz} \right)^3 + 5.86\left( f_{\tt low}/\text{Hz} \right) ^2  } \right ] {\text s}\, ,
\end{align}
was empirically found by generating a $\left(1+1\right)\,M_{\odot}$ waveform (directly given in the frequency domain) and Fourier transforming it to the time domain where the duration up to when the waveform's evolution terminates is measured.
Equation~(\ref{eq:fitL}) and subsequent fits were found using a genetic algorithm-based symbolic regression software, {\tt Eureqa}~\cite{eureqa,Schmidt03042009,Schmidt2010}. The length $L$, as implied by Eq.~(\ref{eq:Ldef}), is plotted in the top panel of Fig.~\ref{fig:Nbasis_speedup}.
\begin{figure}[ht]
\begin{center}
\includegraphics[width=0.7\linewidth]{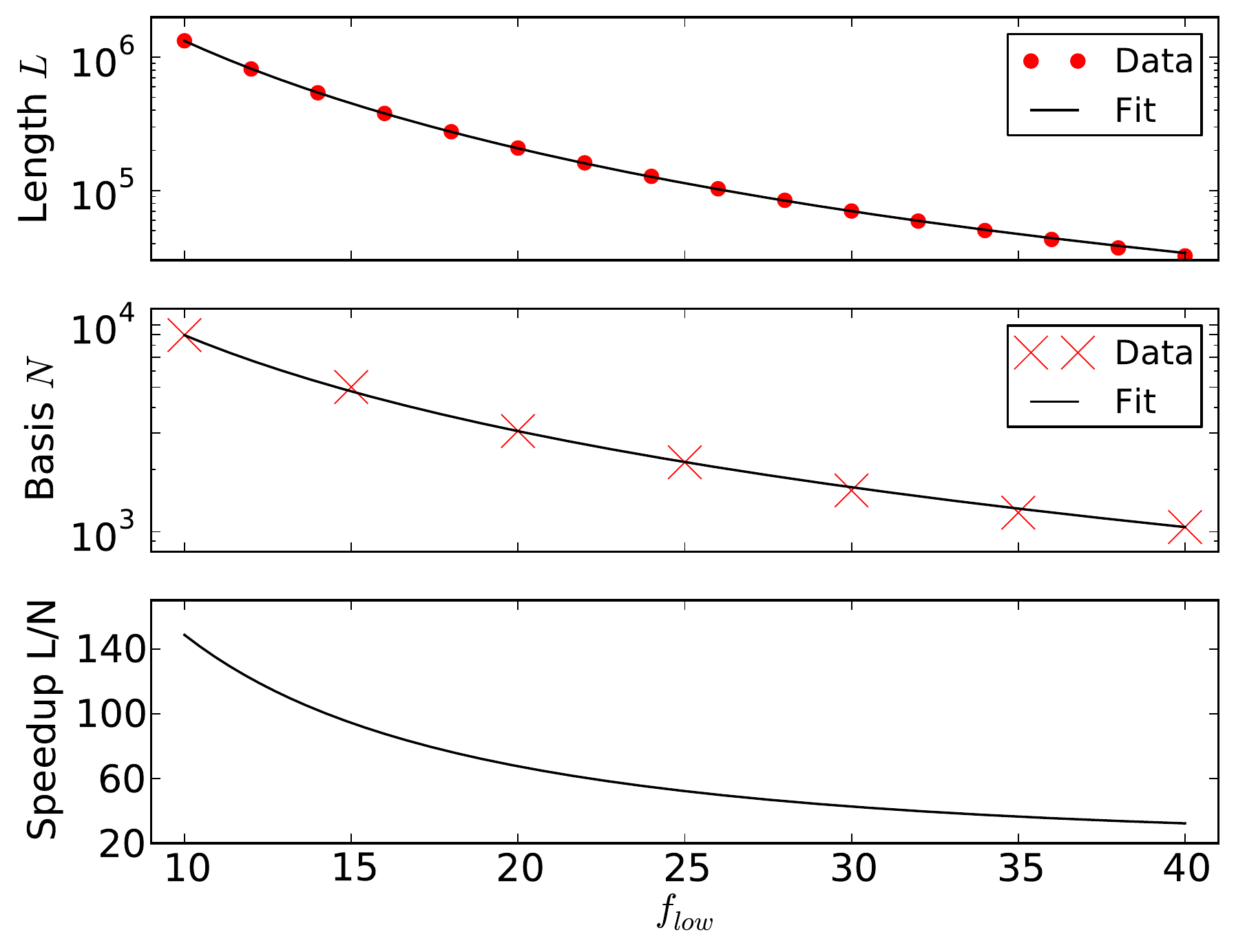}
\caption{{\bf Top}: Length $L$ (red dots) of a typical   
binary neutron star inspiral waveform, with the solid black curve connecting this data implied by the fit~(\ref{eq:fitL}). {\bf Middle}: Number of reduced basis waveforms (red crosses), with the solid black curve given by the fit~(\ref{eq:fitNrb}). {\bf Bottom}: Speedup implied by operation counts, as given by equation~(\ref{eq:Cr_spa}). Figure from \cite{PhysRevLett.114.071104}, where what here is denoted as $n$ (the number of basis) in that reference is $N$ (here used for the size of the initial training set).}
\label{fig:Nbasis_speedup}
\end{center}
\end{figure}

As discussed, each basis only needs to be constructed over the space of intrinsic parameters -- in this case the two-dimensional space of 
component masses, chosen to be in the range $\left[1,4\right]M_{\odot}$.  This range is wider than expected for neutron stars, but ensures that the resulting PDFs do not have sharp cut-offs \cite{Mandel:2014tca}. The number of reduced basis required to represent the TaylorF2 model within this range with a representation error around double precision ($\sim 10^{-14}$) can be fit by 
\begin{align}
\label{eq:fitNrb}
n_{\tt BNS} = 3.12\times 10^5 \left( f_{\tt low}/\text{Hz} \right)^{-1.543} \, , 
\end{align}
and is shown in the middle panel of Fig.~\ref{fig:Nbasis_speedup}. It was found that increasing the high-frequency cutoff to
$4096$ Hz only adds a handful of basis elements, while $L$ changes by a factor of $4$, indicating that the speedup for an inspiral-merger-ringdown model might be higher, especially given that not many EIM nodes are needed for the merger and ringdown regimes \cite{PhysRevX.4.031006}. This was indeed shown in \cite{PhysRevD.94.044031}, as discussed below.

Recalling equation (\ref{eq:Ldef}), the expected speedup from standard to ROQ-compressed likelihood evaluations is given by 
\begin{align}
\label{eq:Cr_spa} 
\frac{L}{n} \approx \left( 1024 \text{Hz} - f_{\tt low} \right)\frac{T_{\tt BNS}}{n_{\tt BNS}} \, ,
\end{align}
with $T_{\tt BNS}$ and $n_{\tt BNS}$ given by Eqs.~(\ref{eq:fitL}) and~(\ref{eq:fitNrb}), respectively. As reported in~\cite{PhysRevD.94.044031}, this speedup is indeed observed using LALInference, and is shown in Fig.~\ref{fig:Nbasis_speedup} (bottom), with a reduction in computational cost and time of  $\sim 30$ for the initial detectors (with a cutoff of $f_{\tt low} = 40\,$Hz) and $\sim 150$ once the advanced detectors reach $f_{\tt low} \sim 10\,$Hz. 

\end{example}

\subsection{Implementation in LALInference} 
Reduced order quadratures for compressed likelihood evaluations and Algorithm~(\ref{alg:Weights}) were originally implemented in the LAL parameter estimation pipeline, known as LALInference \cite{LAL_software, 2013PhRvD..88f2001A}, by Vivien Raymond and Rory Smith. The resulting variation is called LALInference$\_$ROQ. This section provides unpublished details which were not included in \cite{PhysRevLett.114.071104}, courtesy also of VR and RS. Below is a comparison between MCMC parameter estimation results using the standard version of LALInference and ROQ accelerated studies using LALInference$\_$ROQ for the previous example, where TaylorF2 is the waveform model. Synthetic signals embedded in simulated Gaussian noise were injected into the LAL pipeline, for settings anticipating at the time the initial configuration of aLIGO, using the zero detuned high power PSD \cite{techrep:aLIGOsensitivity} and $f_{\tt low} = 40$Hz.

The time window was taken to be $W= 0.1$s about the coalescence time $t_c$ of a binary neutron star signal~\cite{Smith:2012du, 2013PhRvD..88f2001A}. Following the discussed procedure, LALInference$\_$ROQ discretizes this prior into $n_c=2,000$ sub-intervals, each of size $\Delta t_c=10^{-5}$s, for which it constructs a unique set of ROQ weights on each sub-interval. A width of $10^{-5}$s ensures that this discretization error is below the measurement uncertainty on the coalescence time, which is typically $\sim 10^{-3}$s \cite{2013PhRvD..88f2001A}. 

As expected, the ROQ and standard likelihood approaches produce statistically indistinguishable results for posterior probability density functions over the full $9$-dimensional parameter space ($2$ intrinsic dimensiones and the full $7$ extrinsic ones). Figure~\ref{fig:MCMC} and Table~\ref{tab:parameters} describe results for the nine parameters obtained in one particular MCMC simulation; other simulations were qualitatively similar. 

\begin{landscape}
\begin{table}
\begin{tabular}{ l l l l l l l l l l l l}
  \hline
   & $\mathcal{M}_c\,(M_{\odot})$ & $\eta$ & $D\,$(Mpc) & $t_c$(s) & $\alpha$ (rad) & $\delta$ (rad) & $\phi_c$ (rad) & $\psi$ (rad) & $\iota$ (rad) & SNR \\
  \hline
  injection & 1.2188 & 0.25 & 172 & 0 & 4.91 & -0.981 & 1.46 & 0 & 1.89 & 11.4 \\
  standard & $1.2188^{1.2189}_{1.2184}$  & $0.249^{0.250}_{0.243}$  &  $153^{107}_{334}$  & $0^{0}_{0}$ & $4.89_{1.79}^{4.98}$ & $-0.978_{-1.01}^{1.20}$ & $3.35_{0.309}^{6.07}$ & $1.58_{1.54\times10^{-2}}^{3.13}$ & $1.82_{1.32}^{2.91}$ & 12.9 \\
  ROQ & $1.2188^{1.2189}_{1.2184}$ & $0.249^{0.250}_{0.243}$  &  $149_{100}^{352}$ & $0_{0}^{0}$ & $4.89_{1.82}^{4.98}$ & $-0.978_{-1.01}^{1.19}$ & $3.01_{0.543}^{5.97}$ & $1.58_{2.85\times 10^{-2}}^{3.11}$ & $1.79_{1.39}^{2.82}$ & 12.9 \\
  \hline
\end{tabular}
\caption{
Chirp mass $\mathcal{M}_c$, symmetric mass ratio $\eta$, source distance (from Earth) $D$, coalescence time $t_c$, right ascension $\alpha$, declination $\delta$, coalescence phase $\phi_c$, polarization phase $\psi$, inclination $\iota$ and Signal-to-Noise Ratio (SNR) of the analysis from Figure~\ref{fig:MCMC}.  
Median value and 90\% credible intervals are provided for both the  standard likelihood (second line) and the ROQ compressed likelihood (third line). 
The SNR is empirically measured from $\mathrm{Likelihood_{max}} \approx \mathrm{SNR}^2 / 2$. The differences between the two methods are dominated by statistics from computing intervals with a finite number of samples. \textit{Credit: Vivien Raymond}}
\label{tab:parameters}
\end{table}
\end{landscape}
It is also useful to quantify the fractional difference in the 9-D likelihood function computed using ROQ and the standard approach. This fractional error has found to be
$$
\Delta\log\,\mathcal{L} = 1 - \left( \frac{ \log\,\mathcal{L}}{\log\,\mathcal{L}_{\tt{ROQ}} } \right) \lesssim 10^{-6}
$$ 
in all cases. That is, both approaches are indistinguishable for all practical purposes. 

In addition to providing indistinguishable results, ROQ accelerated inference is significantly faster: The ROQ-based MCMC study with the discussed settings takes $\sim 1$ hour, compared to $\sim 30$ hours using the standard likelihood approach, in remarkable agreement with the expected savings based on operation counts. The wall-time of the analysis is proportional to the total number of posterior samples of the MCMC simulation, which in this case was $\sim 10^{7}$. The startup stage required to build the ROQ weights has negligible cost and is completed in near real-time, $\sim30$s, which is equivalent to $\sim0.028 \%$ of the total cost of a standard likelihood parameter estimation study.

\begin{figure}
\includegraphics[width=0.98\linewidth]{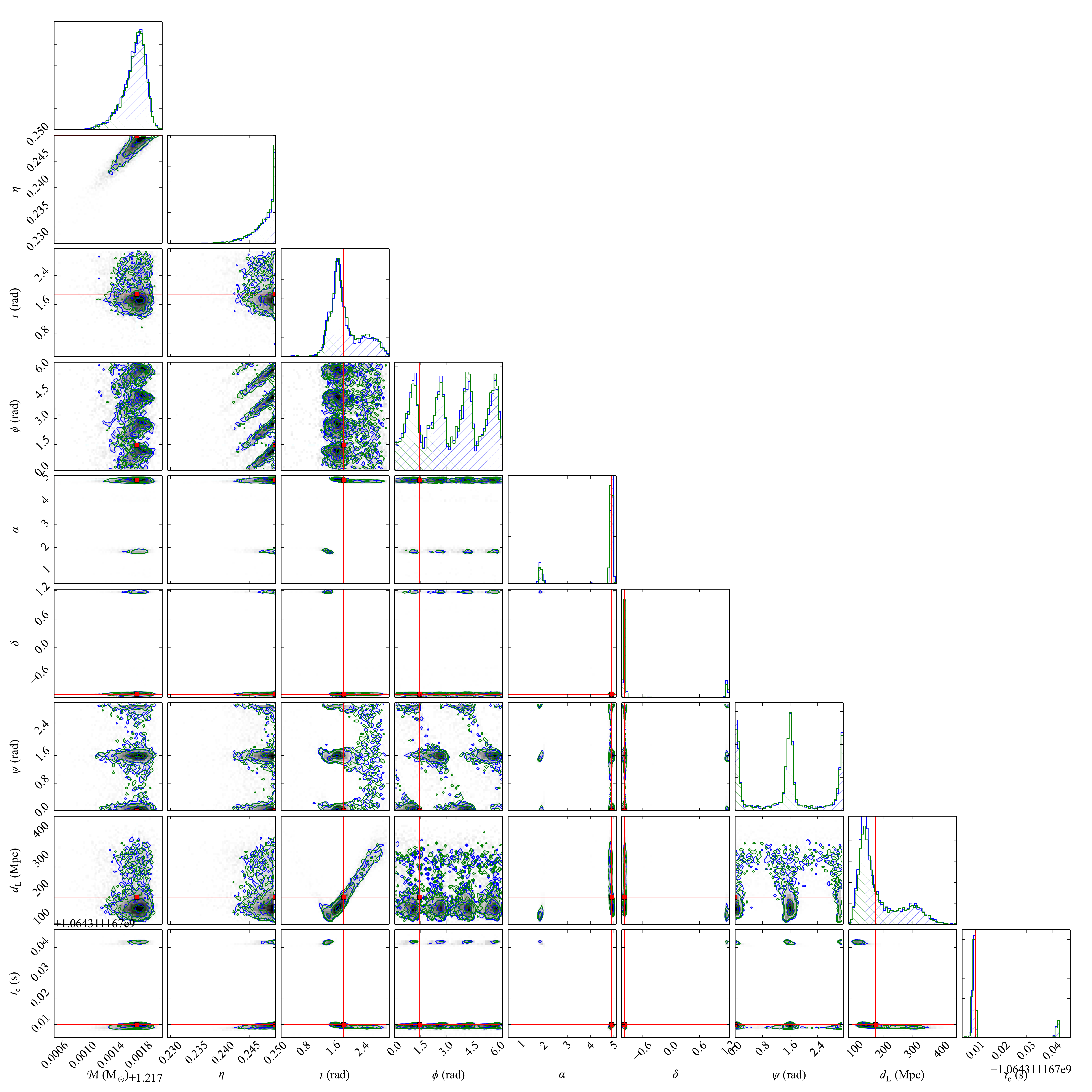}
\caption{Probability density function for the chirp mass $\mathcal{M}_c$, symmetric mass ratio $\eta$, inclination $\iota$, coalescence phase $\phi_c$, right ascension $\alpha$, declination $\delta$, polarization phase $\psi$, source distance (from Earth) $D$ and coalescence time $t_c$ of a simulated event in LIGO/Virgo data. In green as obtained in $\sim 30$ hours by the standard likelihood, and in blue as obtained in $1$ hour with the ROQ. The injection values are in red, and are listed in Table~\ref{tab:parameters}. The overlap region of the sets of PDFs is the hatched region. \textit{Credit: Vivien Raymond.}}
\label{fig:MCMC}
\end{figure}

For a lower cutoff frequency of $f_{\tt low} \sim 20$Hz, the speedup reduction is from a couple of weeks to hours. For advanced detectors with $f_{\tt low} \sim 10$Hz, the longest BNS signals last around $2048$s in duration. Assuming a fiducial high frequency cut-off of $1024\,$Hz, which is approaching the upper limit of the sensitivity of aLIGO/AdV, datasets can be as large as $L\sim 1024\text{Hz}^{-1}\times2048\text{s} \sim 10^6$. Assuming that the advanced detectors require at least $\sim 10^7$ posterior samples, this implies runtimes upwards of $\sim 100$ days and one Petabyte worth of model evaluations using the standard approach. On the other hand, ROQ reduces this to hours. Remarkably, this approach when applied to the advanced detectors operating at design sensitivity is faster than even the standard likelihood one used for the initial detectors. Additionally, with parallelization of the sum in each likelihood evaluation essentially real-time full Bayesian analysis {\em might} be achieved.  More details can be found on the website \cite{LAL-ROQ}. 

\subsection*{Further reading}
Even though ROQs decrease the computational cost of likelihood evaluations for binary neutron stars for advanced detectors by about two orders of magnitude, this is still in the order of hours, which does not meet the need for real-time followup of GW detections through electromagnetic counterparts. Therefore, other ideas, beyond or on top of ROQ are needed. One approach, named Focused Reduced Order Quadratures (FROQ) \cite{Morisaki_2020}, is to restrict the parameter estimation search based on trigger values from the detection pipeline, which further reduces the cost to around 10 minutes, which is a good target from an astrophysical point of view for searches of electromagnetic counterparts. Even though the study of \cite{Morisaki_2020} uses Post-Newtonian closed-form approximations, if NR surrogates are used for other scenarios, waveform evaluation cost should not be an issue for extending FROQ. 

In Reference \cite{PhysRevD.94.044031} the authors built the first ROQ for precessing inspiral-merger-ringdown compact binaries, using the IMRPhenomPv2 model, which includes all $7$ intrinsic parameters. For low cutoff frequencies of $f_{\tt low} =20$Hz, the authors found speedups of up to between $\sim 4\times$ and $\sim 300 \times$ (for short, BBH, and long, BNS systems, respectively), leading to an estimate of $6-12$ hours for a full PE study, as opposed to $\sim 1$day instead of $\sim 6$ months.  The resulting code is also available from LALInference. This study was extended in \cite{PhysRevD.97.044033} adding a parametrization for deviations from General Relativity using IMRPhenomPv2. 

We mentioned that one of the computational burdens of PE is the computation of the gravitational waveforms and that this can be solved using surrogate models. This speedup was analyzed in detail using SVD interpolated waveforms in \cite{Smith:2012du}. 
 
Even though in this Section we have focused on the use of ROQ for parameter estimation, the procedure for GW searches is similar. 

As a proof of the generality of ROQ, it was used in Ref.~\cite{Brown_2016} in the context of studies of laser light scattering with a speedup of $\sim 2,750$ with essentially no loss in precision. 

As emphasized in Section \ref{sec:chall}, an important topic that we have not touched in this review is that one of statistical machine learning approaches (as opposed to algorithmic ones as in this review) for GWs prediction and analysis. An exception in order here is that one of \cite{PhysRevLett.122.211101}, since it bridges ROM with Artificial Neural Networks (ANN) and addresses an important problem. Namely, the challenge of interpolating the projection coefficients in a Reduced Basis approach. Instead of algorithmically interpolating them, the authors use ANN to map the relationship from parameters to projection coefficients. Though the focus of the reference is on parameter estimation of GWs, the fundamental idea is applicable to the predictive step itself; in fact it was proposed around the same time in a different context in \cite{WANG2019289}. 

\section{Challenges, open issues and comments} \label{sec:chall}
To discuss one of the main challenges left in reduced order and surrogate models for gravitational waves we review the case of inspiral spinning non-precessing binary PN black hole waveforms, taken from Ref.~\cite{Field:2012if}. The parameter space is 4-D and the sector covered corresponds to $m_i\in[3-30]\msun$, with $\chi _i\in [-1,1]$ ($i=1,2$). The authors study, in the context of Reduced Basis, the {\tt rb} for parameter dimensions ${\tt dim}=2,3,4$, which correspond to the non-spinning, spinning-aligned, and spinning non-precessing cases, respectively. In Figure~\ref{fig:greedy_counter} there is a histogram corresponding to the 4-D case showing the greedy parameters count for mass ratio $m_1/m_2\in[0.1, 10]$ for a representation error of $\sim 10^{-14}$ of {\em any} waveform in the considered parameter range. In Figure~\ref{fig:sparse}, for visual representation, we show the distribution of the greedy points in parameter space for the 3-D case where the spins are aligned or anti-aligned, with $\chi_1=\chi_2=\chi$. The lower cutoff frequency being considered at the time was $40$Hz.  

\begin{figure}[htp]
\begin{center}
  \includegraphics[width=0.6\linewidth]{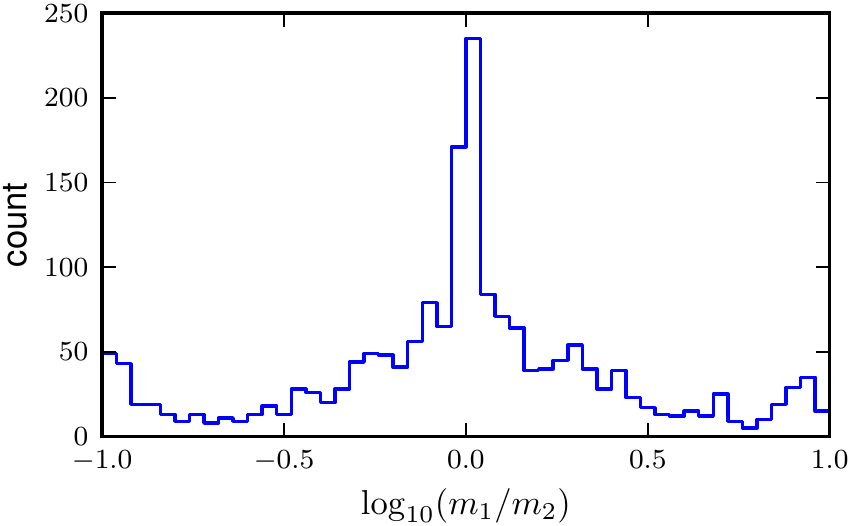}
  \caption{The greedy parameter choices for the mass ratio $m_1/m_2\in[0.1, 10]$. Notice that the greedy points cluster in the near-equal mass sector. Figure taken from~\cite{Field:2012if}.}
  \label{fig:greedy_counter}
\end{center}
\end{figure}

\begin{figure}[htp]
\begin{center}
  \includegraphics[width=0.6\linewidth]{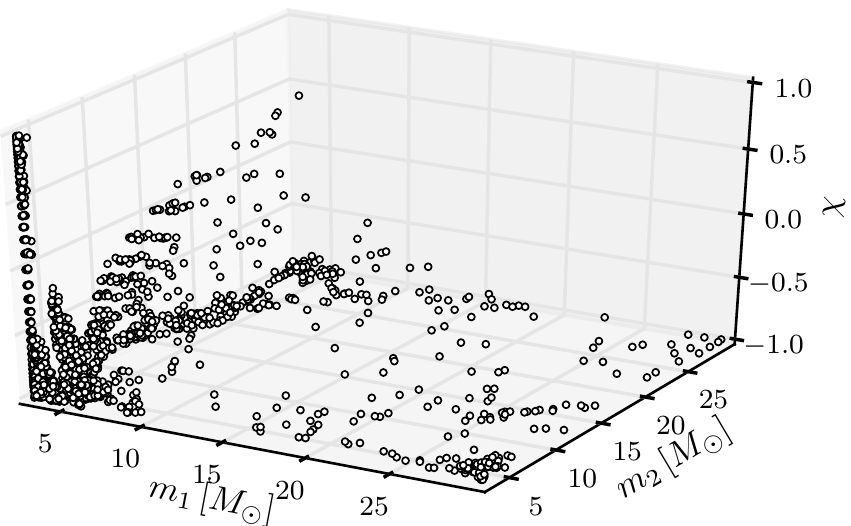}
  \caption{The greedy parameter choices for the mass ratio $m_1/m_2$ in the 4-D case. Right: the greedy parameter choices for the 3-D $(m_1, m_2, \chi)$ for the aligned spin case. Figure taken from~\cite{Field:2012if}.}
  \label{fig:sparse}
\end{center}
\end{figure}

\begin{figure}[htp]
\begin{center}
\includegraphics[width=0.6\linewidth]{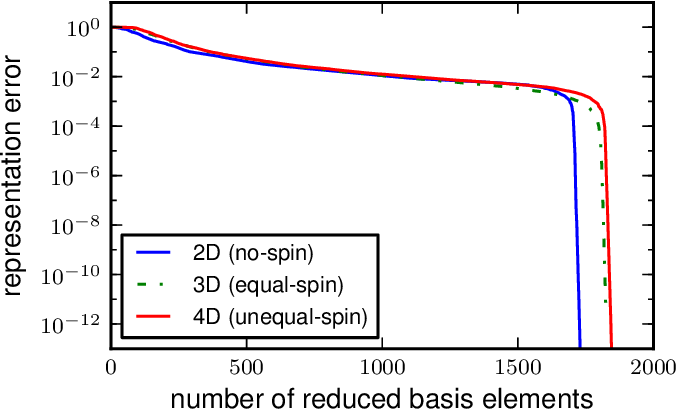}
\caption{The reduced basis representation error as a function of the dimensionality of the reduced bases for ${\tt dim}=2,3,4$. The number of reduced basis elements barely grows with the dimensionality of the problem $\tt dim$, suggesting that at least in the context of GWs from compact binaries, the curse of dimensionality might be beaten by the reduced basis approach. Figure taken from~\cite{Field:2012if}.}
\label{fig:dimensionality}
\end{center}
\end{figure}

\begin{comment}
\begin{enumerate}
\item From Fig.~\ref{fig:greedy_counter} it can be seen that the largest number of selected greedy points is near the equal mass case. The reason for this is that since a fixed frequency range is considered, the longer waveforms correspond to nearly equal mass components, have a larger number of cycles and therefore more structure and representative information. 

\item In the non-spinning case around $1,700$ basis waveforms are needed for machine precision representation, see Fig.~\ref{fig:dimensionality}. This is a rather large number of data points to fit each waveform phase and amplitude at the EIM nodes for the considered mass range when building a surrogate as proposed in Section~\ref{sec:surrogate_eim}. Therefore, a low order local polynomial or spline interpolation would probably be accurate enough for practical purposes when the fiducial model is fast enough to evaluate -- as it is in this example -- while avoiding Runge's phenomenon (discussed in Section~\ref{sec:interp}). 

\item The reason why as increasing the dimensionality of the problem the reduced basis approach seems to beat the curse of dimensionality can be seen from the right panel of Fig.~\ref{fig:sparse}. Namely, the data becomes extremely sparse, and the number of basis elements remains approximately constant while the volume of the parameter space grows exponentially with the number of dimensions (Figure~\ref{fig:dimensionality}). This sparseness is a very much desired condition indeed for any ROM approach, but it introduces difficult challenges, as discussed below. 
\end{enumerate}
\end{comment}

In the case of projection and when the waveforms are known in closed form, or can be computed in a relatively inexpensive way, there are no major issues: in order to evaluate the representation of the waveform one simply computes the projection coefficients onto the basis. Similarly for predictive models, since in those cases there are an ``arbitrary'' number of local waveforms to use for fitting in Step 3 of an EIM-based interpolant (Section~\ref{sec:surrogate_eim}): a local fit of low order (be it in the form of interpolation, least squares, splines, etc.) can be used with high accuracy and while avoiding Runge's phenomena. 

A very different situation is that one in which the training set is computationally expensive to build, as when it involves solving the full Einstein equations through numerical relativity and only a sparse training set can be constructed and there are not  enough points nearby in parameter space to perform local fits with enough accuracy. 

One solution is to keep running NR simulations to enrich the training set. This approach {\em might} be affordable, with enough computational resources, for binary black holes. For binary neutron stars or mixed pairs, on the other hand, where the parameter dimensionality grows considerably, this might be impractical if not just impossible in reality. 

This challenge follows directly from the agnostic, data-driven, purposely design of the surrogate approach discussed in this review, in which no differential equations are invoked. The motivation for this design is that it is a major effort to build a production Einstein solver for realistic problems of interest, and an intrusive approach would require major algorithmic and software changes and as a consequence the NR community might be reluctant to it. Yet, that might be the only feasible approach in the long run. An intermediate solution {\em might} be doing further research in state of the art of {\em global} approaches to approximations in multiple dimensions which are fast to evaluate and do not suffer from Runge's phenomena, be it in the form of deterministic algorithms or statistical machine learning/artificial intelligence type. 

In this review we have covered mostly algorithms of the first kind and tangentially mentioned some results of the second kind. But there is a lot of ongoing activity and interesting results on the ML/AI side, though it appears too premature to cover them here, so we will defer their discussion to a future, updated version of this Living Review. For a recent preliminary comparative study, though, on ML/AI surrogates, see \cite{Setyawati_2020}.  
In terms of parameter estimation and within reduced order modeling, the use of Focused Reduced Order Quadratures, discussed at the end of Section~\ref{sec:PE}, might be a good enough solution for near real-time or fast-enough followups of electromagnetic counterparts.

A related active field but not in GW science is that one of Uncertainty Quantification (UQ) ~\cite{10.5555/2568154} and, in particular, generalized polynomial chaos~\cite{10.2307/j.ctv7h0skv}, which we have not discussed in this review but are closely related to Reduced Basis and can remove the usual assumption (and its consequences) of the noise being Gaussian and stationary noise, which is not the case in GW science. The application of these areas of research in GW science are unknown to us, so we also defer them to a further version of this review. However, it has to be highlighted that other techniques to deal with non-Gaussian noise have found new detections from public LIGO data~\cite{Zackay:2019kkv}, so the importance of the topic cannot be overemphasized. 

\section{Acknowledgments} \label{sec:ack}
We thank, for valuable discussions about topics covered in this review through collaborations and/or interactions throughout the years:
Harbir Antil, 
Peter Binev, 
Jonathan Blackman, 
Priscilla Canizares, 
Sarah Caudill, 
Wojciech Czaja, 
Albert Cohen, 
Scott Field, 
Jonathan Gair, 
Chad Galley, 
Chad Hanna, 
Frank Herrmann, 
Jan Hesthaven, 
Stephen Lau, 
Jason Kaye, 
David Knezevic, 
Tom Loredo, 
Akil Narayan, 
Ricardo Nochetto, 
Evan Ochsner, 
Dianne O'Leary, 
Michael P\"urrer, 
Vivien Raymond, 
Gianluigi Rozza, 
Rory Smith, 
Benjamin Stamm, 
B\' ela Szil\' agyi, 
Eitan Tadmor, 
Michele Vallisneri, 
and Alan Weinstein. 
We apologize in advance for any omissions.

We especially thank Scott Field, Chad Galley and Rory Smith for initial contributions to an initial draft of this review years ago, which we have attempted to write from scratch, given the results over the last years in the field, but some imprints from the original conception might still be left. Special credits are due to Frank Herrmann, who played a key role in the foundational efforts of reduced basis in gravitational wave science since 2009, bringing it from exploratory analyses and initial proofs of concept to a production-level stage. Special credits are also due to Jan S. Hesthaven, who introduced one of us (MT) to the idea of using the Reduced Basis framework for GW science back in 2009 at Brown University, and Ben Stamm for a short but intensive course at UMD that he gave us to get started. 

We also thank Jorge Pullin, the Horace Hearne Institute and the Center for Computation and Technology at LSU, for hospitality while part of this review was written. This work was supported partially by CONICET and EVC-CIN (Argentina).

\bibliography{references}

\end{document}